\documentclass{aa}  

\usepackage{graphicx}
\usepackage{svg}
\usepackage{adjustbox}
\usepackage{txfonts}
\usepackage{lipsum}
\usepackage{subcaption}
\usepackage{lscape}
\usepackage{placeins}

\usepackage[colorlinks=true,linkcolor=blue,citecolor=blue]{hyperref}
\usepackage{orcidlink}
\usepackage{multirow}

\begin{document}

   \title{Performance of morphological classifiers for galaxy mergers compared to current machine learning methods}

   \author{A. P. Cotter\orcidlink{0009-0004-1910-4990}\inst{1}
        \and W. J. Pearson\orcidlink{0000-0002-7300-2213}\inst{1}
        \and S. Dey\orcidlink{0000-0002-4679-0525}\inst{1}
        \and B. Margalef-Bentabol\inst{2}
        \and A. Guzmán-Ortega\orcidlink{0000-0002-0027-9692}\inst{3}
        \and V. Rodriguez-Gomez\inst{3}
        }

   \institute{National Centre for Nuclear Research, Pasteura 7, 02-093 Warszawa, Poland\\
             \email{Aidan.Cotter@ncbj.gov.pl}
            \and SRON Netherlands Institute for Space Research, Landleven 12, 9747 AD Groningen, The Netherlands
            \and Instituto de Radioastronomía y Astrofísica, Universidad Nacional Autónoma de México, Apdo. Postal 72-3, 58089 Morelia, Mexico
            }

   \date{Received Month YY, XXXX}

  \abstract
   {}
   {Non-parametric morphological statistics can be useful in efficiently determining galaxy merger classifications. This work is aimed at comparing the performance of morphological merger classifiers to state-of-the-art machine learning (ML) models. A secondary aim is to produce updated criteria for mergers based on non-parametric morphological statistics.}
   {The Gini coefficient (G), $M_{20}$ statistic, and concentration ($C$) were calculated for mock Hyper Suprime-Cam Subaru Strategic Program (HSC-SSP) images based on the IllustrisTNG and Horizon-AGN simulations, along with observations from HSC-SSP. IllustrisTNG images were used to find the line that best separates mergers and non-mergers in 2D morphological space with a Markov chain Monte Carlo (MCMC) method.}
   {Based on the MCMC results, we classified galaxies with $G>(-0.267\pm0.081)M_{20}+(0.143\pm0.012)$ or $G>(0.162\pm0.048)C+(0.149\pm0.12)$ as mergers. These criteria had precisions levels of 69.5\% and 72.3\% when applied to previously unseen IllustrisTNG mock HSC-SSP images, respectively. The precision of the morphological classifications are consistent with state-of-the-art ML methods. The morphological classifiers were found to be effective at selecting only pre-mergers. On the other hand, post-merger galaxies are indistinguishable from non-mergers in terms of their $G$, $M_{20}$, and $C$ values. Morphological classifiers displayed a similar robustness to new data to ML methods up to a redshift of $\sim0.52$ and maintained the robustness better than ML methods based on convolutional neural networks (CNNs) in the redshift range $0.52<z<1$.}
   {This work presents updated morphological classifiers with the capacity to achieve a similar level of precision to that of ML-based merger classifiers that have a high robustness with regard to new data. Updated morphological statistics are needed to identify the features of post-merger galaxies.}

   \keywords{methods: numerical --
                methods: data analysis --
                galaxies: evolution --
                galaxies: interactions --
                galaxies: statistics --
                galaxies: fundamental parameters
               }

   \maketitle
   \nolinenumbers

\defcitealias{Margalef-Bentabol2024}{MB24}

\section{Introduction}
Mergers are an important aspect of galaxy evolution. In the Lambda cold dark matter ($\Lambda$CDM) cosmological model, galaxies assemble mass hierarchically \citep{White1978, White1991, Fakhouri2008} by merging with their neighbours. These interactions trigger key astrophysical phenomena such as enhanced star formation rates \citep{Barnes2004, Saitoh2009, Pearson2019} and active galactic nuclei \citep{Alonso2007, Ellison2025}. Mergers have also been linked to morphological transformations and structure formation within galaxies \citep{Bois2011, Simmons2014, Sparre2017, Smethurst2025}. 
To effectively study the role of mergers in the evolution of galaxies, accurate and reliable merger classification methods are essential. As increasingly large surveys such as Euclid \citep{EuclidCollaboration2025may} and the Legacy Survey of Space and Time (LSST) \citep{Ivezic2019} begin their operations, more efficient methods of merger classification are required to select mergers from large sets of data. The classifiers presented in this work are part of ongoing work to develop reliable methods for identifying merging galaxies in large surveys.

One simple technique for identifying mergers is the close-pairs method \citep{Zepf1989, Patton2002, Lopez-Sanjuan2015, Conselice2022, Puskas2025}. This technique identifies mergers as pairs of galaxies with small projected separation and a small relative velocity, typically < 150 kpc and < 300 kms$^{-1}$, respectively \citep{Bustamante2020}. These criteria vary throughout the literature. This method is well suited for poorly resolved galaxies since the morphology does not need to be included; however, spectra are indeed required for kinematic data and so this method is not applicable to most large datasets. Additionally, the close-pairs method is only effective for identifying pre-mergers. 

Another simple method for classifying mergers is visual classification \citep{Schawinski2007, Darg2010, Pearson2022, EspejoSalcedo2025, LaMarca2026}. This technique can be used to produce very detailed classifications and it is a good method for new types of data. Visual classification does not scale well to large datasets because it is a very time-intensive process and is not easily reproducible due to the subjective nature of visual classification.

Modern merger classification is dominated by machine learning (ML), with convolutional neural networks (CNNs) and vision transformers commonly used to produce highly accurate merger classifiers \citep{Ackermann2018, Pearson2019, Bickley2021, Pearson2022, Margalef-Bentabol2024, Ferreira2024}. The training process for ML methods require large datasets and can be computationally expensive in terms of both components and processor time. Additionally, ML classifiers often do not transfer well to new data which can make them less reliable. Merger samples produced exclusively using ML methods can be highly contaminated by non-mergers due to their poor reliability on new data \citep{Pearson2019, Bickley2021} and require visual classification to to produce pure samples of mergers \citep{Pearson2022}, which results in an even more time intensive classification. Alternatively, ensemble classifiers can be used to remove contamination and produce high purity merger samples \citep{Ferreira2024}; however, this requires substantially more computational power to be implemented.

Morphological classifiers select merging galaxies using a set of statistics that have the capacity to trace their key features. These features are described by parametric and non-parametric morphological statistics; there are many such statistics, but for merger classification, the most popular statistics employed include: concentration ($C$), asymmetry (A), smoothness (S) system \citep{Kent1985, Bershady2000, Conselice2000, Conselice2003}, and the Gini coefficient (G), and $M_{20}$ (the second-order moment of the brightest 20\%) system \citep{Lotz2004, Lotz2008}. Merger criteria have been defined for these statistics in \citet{Conselice2003}, \citet{Lotz2004}, and \citet{Lotz2008}.

The merger criteria found in \citet{Conselice2003}, \citet{Lotz2004}, and \citet{Lotz2008} were based on samples of luminous infrared galaxies (LIRGs) \citep{Sanders1996, Borne2000, Canalizo2001}. These are a subclass of mergers that radiate brightly in the infrared due to the heating of dust during a starburst commonly caused by a merger. They do not include all types and stages of mergers and, thus, LIRGs make an incomplete, biased merger sample.

Random forest (RF) models have also been used to classify mergers based on morphological statistics. In \citet{Guzman-Ortega2023} and \citet{Snyder2019} up to ten morphological statistics were included in RF classifiers. A major weakness in morphological classifiers is that in decomposing images into simple statistics, information about the galaxy is lost and so each classifier is only sensitive to a small number of feature. By including many morphological statistics, a classifier can be sensitive to a wider range of merger signatures.
        
This work compares the performance of non-ML merger classifiers, relying on morphological statistics, against state-of-the-art ML techniques. We define the updated morphological criteria for mergers for this comparison. These criteria were optimised based on a set of mock images to remove the bias present in the merger criteria seen in the literature. We have extended the merger challenge from \citet{Margalef-Bentabol2024} (hereafter \citetalias{Margalef-Bentabol2024}) to include these morphological classifiers to enable a direct comparison between the various methods. \citetalias{Margalef-Bentabol2024} provides a good comparison for merger classifier performance because they include dsix ML methods trained on the same set of mock images, which effectively removed any training data biases as a factor in their relative performance.

The structure of the \citetalias{Margalef-Bentabol2024} dataset is discussed in Section \ref{sec:Data}. In Section \ref{sec:Methodology}, the non-parametric morphological statistics are defined and our method is discussed. In Section \ref{sec:Results}, the performance of the morphological classifiers is presented and in Section \ref{sec:Discussion} the performance is compared to state-of-the-art ML methods. Section \ref{sec:Conclusions} summarises the conclusions of this work.

\section{Data}\label{sec:Data}
The merger challenge presented in \citetalias{Margalef-Bentabol2024} features $882\,214$ images split across three datasets. The first dataset, used for training and testing, contains mock Hyper Suprime-Cam Subaru Strategic Program \citep[HSC-SSP;][]{Aihara2018} images based on the IllustrisTNG simulations. A second dataset, made up of mock HSC-SSP images based on the Horizon-AGN simulation, was intended to test how well the classifiers transfer to new data. Observations from the HSC-SSP were included as the third dataset intended to compare the performance of the methods on real images. The ML methods developed for \citetalias{Margalef-Bentabol2024} were trained on a subset of 504741 mock images from IllustrisTNG and so for consistency, the same training set was used throughout this work when optimising the merger classifiers.

\subsection{Simualtions}\label{subsec:simulations}
\subsubsection{IlustrisTNG}\label{subsec:IllustrisTNG}
The IllustrisTNG project \citep{Marinacci2018, Naiman2018, Nelson2018, Pillepich2018, Springel2018} is a set of magnetohydrodynamical cosmological simulations which build on the Illustris simulations \citep{Genel2014, Vogelsberger2014may, Vogelsberger2014oct, Sijacki2015}. IllustrisTNG features three simulations in varying sizes of comoving boxes: TNG300 has the largest box with comoving length at 300 Mpch$^{-1}$, TNG100 has a comoving length of 100 Mpch$^{-1}$, and TNG50 has the smallest box at a comoving length of 50 Mpch$^{-1}$. Galaxies from snapshots 50-91 in TNG100 and TNG300 were used to produce the mock images in \citetalias{Margalef-Bentabol2024}, these snapshots correspond to a redshift range of 0.1 < z < 1.

TNG300 used 2500$^3$ dark matter particles with a mass resolution of 5.9$\times$10$^7$M$_\odot$ and 2500$^3$ baryonic matter particles at a resolution of 1.1$\times$10$^7$M$_\odot$. TNG100 was run with 1820$^3$ dark matter particles with a mass resolution of 7.5$\times$10$^6$M$_\odot$ and 1820$^3$ baryonic matter particles at a resolution of 1.4$\times$10$^6$M$_\odot$. The cosmological parameters for IllustrisTNG come from the Planck 2015 results \citep{PlanckCollaboration2016}.

Galaxies are identified in IllustrisTNG using the \texttt{SUBFIND} algorithm \citep{Springel2001, Dolag2009} which is based on the friends-of-friends approach \citep{Davis1985}. Merger trees are constructed using the \texttt{SubLink} algorithm \citep{Rodriguez-Gomez2015}, this also allows for merger times and mass ratios to be calculated. Truth labels for mergers in simulations can be made with certainty using the merger trees, this enabled \citetalias{Margalef-Bentabol2024} to construct an unbiased and pure training sample with mock images. \citetalias{Margalef-Bentabol2024} classified mergers as being between 0.8Gyrs before and 0.3Gyrs after coalescence with another galaxy. Only major mergers, with a mass ratio of $>1:4$ in the snapshot in which the secondary mass is at its maximum \citep{Rodriguez-Gomez2015}, were considered in \citetalias{Margalef-Bentabol2024}

\citetalias{Margalef-Bentabol2024} selected galaxies from TNG100 with stellar mass > 10$^9$M$_\odot$ and consisted of at least 714 particles. From TNG300, galaxies with stellar mass > 8$\times$10$^9$M$_\odot$ and containing at least 727 particles were selected for \citetalias{Margalef-Bentabol2024}. The sample that \citetalias{Margalef-Bentabol2024} selected from IllustrisTNG was balanced between mergers and non-mergers.

\subsubsection{Horizon-AGN}\label{subsec:Horizon-AGN}
Horizon-AGN \citep{Dubois2014} is a cosmological simulation with a comoving box size of 100 MPc$h^{-1}$ containing 1024$^3$ dark matter particles at a mass resolution of 8$\times$10$^7$M$_\odot$. The Horizon-AGN simulation used the cosmological parameters from the WMAP 7 results in \citet{Komatsu2011} which are within 10\% relative variation of the Planck results from \citet{PlanckCollaboration2014} from which the initial conditions were produced using the \texttt{MPGRAFIC} package \citep{Prunet2008}.

Galaxies in Horizon-AGN are identified using the \texttt{AdaptaHOP} algorithm \citep{Aubert2004, Tweed2009} which works by computing the local density of each stellar particle. Merger trees are constructed with the \texttt{TREEMAKER} algorithm \citep{Tweed2009}. For \citetalias{Margalef-Bentabol2024}, as with IllustrisTNG mergers were restricted to mass ratios $>1:4$ and only mergers up to 0.8Gyrs before or 0.3Gyrs after coalescence were considered. Galaxies with a stellar mass of $>10^9$M$_\odot$ and with at least 500 particles were selected for \citetalias{Margalef-Bentabol2024}. This selection ensured that the sources were well-resolved.

\subsubsection{Mock images}\label{subsec:Mock images}
\citetalias{Margalef-Bentabol2024} constructed mock images of galaxies from the simulations, ensuring that all surrounding objects were included so that close pairs would be visible in the final images. Each raw image was convolved with the HSC-SSP i-band point spread function (PSF). Poisson noise was then added to the images and finally they were injected into cut outs from HSC-SSP to add background sources.

To select cutouts from HSC-SSP, \citetalias{Margalef-Bentabol2024} required:
\begin{itemize}
    \item No catalogued bright or low-z sources within 21" of the image centre;
    \item No bad pixels, saturated pixels, pixels with invalid values, or possible missed bright sources, indicated by flags in the HSC pipeline \citep{Bosch2018}.
\end{itemize}

\citetalias{Margalef-Bentabol2024} divided the images into four redshift bins chosen to contain a similar number of galaxies and span a similar redshift range: 0.1 < z$_1$ < 0.31, 0.31 < z$_2$ < 0.52, 0.52 < z$_3$ < 0.76, and 0.76 < z$_4$ < 1. A physical size of 160 $\times$ 160 kpc was imposed on all images, the size of the image was chosen to correspond to a physical size of 160 kpc at the midpoint of each redshift bin. This resulted in images with side lengths of 320, 192, 160 and 128 pixels for \citetalias{Margalef-Bentabol2024} which are also used throughout this work.

The images generated from the IllustrisTNG simulations were then divided further into a training and testing dataset. The training set was constructed to include 90\% of the images leaving 10\% for testing, these same subsets are used throughout this work for consistency with \citetalias{Margalef-Bentabol2024}. Since these datasets were designed for training deep learning methods, all images of the same merger tree had to be in a single dataset to prevent learning by interpolation \citep{Eisert2023}. Separating merger trees may not be necessary for non-visual classifiers but to make a fair comparison to \citetalias{Margalef-Bentabol2024}, the same testing and training datasets were used. 

\subsection{Observations}\label{subsec:Observations}
Hyper Suprime-Cam (HSC) \citep{Aihara2018} is an imaging camera for the Subaru telescope with a 1.5$^\circ$ field of view and resolution of 0.168 arcsec/pixel. HSC-SSP is a survey conducted over $\sim1200$deg$^2$ in five photometric bands (grizy), \citetalias{Margalef-Bentabol2024} used only the i-band because of its depth of $\sim26$ mag for detection of point sources at $5\sigma$ and its seeing of 0.61".

Approximately 2000 visually classified images from \citet{Goulding2018} were included. These visual classifications were made for an initial sample of 5900 star forming galaxies and were intended to provide a ground truth for training a RF classifier. The visual classification was undertaken by seven experts who were originally tasked with identifying five classes (three classes linked to mergers, a non-merger class and an inconclusive class) though this was reduced to a binary merger and non-merger classification later in their analysis. \citetalias{Margalef-Bentabol2024} added a set of $\sim1000$ visually classified non-mergers, these were selected for having no visible merging features.

Photometric redshifts and stellar masses are based on the KiDS-VIKING nine-band photometry \citep{Edge2013, Kuijken2019, Wright2019} using the Bayesian photometry redshift algorithm from \citet{Benitez2000}. The redshift and mass of the sample were limited to stellar masses of >10$^9$M$_\odot$ and photo-z in the range (0.1, 1) to match the simulations.

\section{Methodology}\label{sec:Methodology}

\subsection{Source segmentation}\label{subsec:Segmentation}
To properly detect the sources in the mock images, the segmentation method from \citet{Guzman-Ortega2023} was adapted. The \texttt{SEP} library \citep{Barbary2016} was used to perform background subtraction, segmentation, and deblending with a segmentation threshold of one times the local root mean squared background and a deblending threshold of 0.00001 with a minimum segment size of 8 pixels.

The low deblending threshold is used to identify individual sources however it caused over-segmentation in some images. Mergers are particularly vulnerable to over-segmentation because they can have multiple nuclei, bright clumps where recent star formation has occurred, and tidal features. To correct for this, the distance between the brightest pixels in adjoining segments was calculated and any with sufficiently small distances were combined into a single segment. The distance used in this work was 32 pixels. This segmentation method is available on \texttt{GitHub}\footnote{\href{https://github.com/APCotter/Galaxy-phOtOmetry-with-Deblended-Segmentation-Image-Reconstruction}{github.com/APCotter/Galaxy-phOtOmetry-with-Deblended-Segmentation-Image-Reconstruction}}.

\begin{figure*}
    \centering
    \includegraphics[width=17cm]{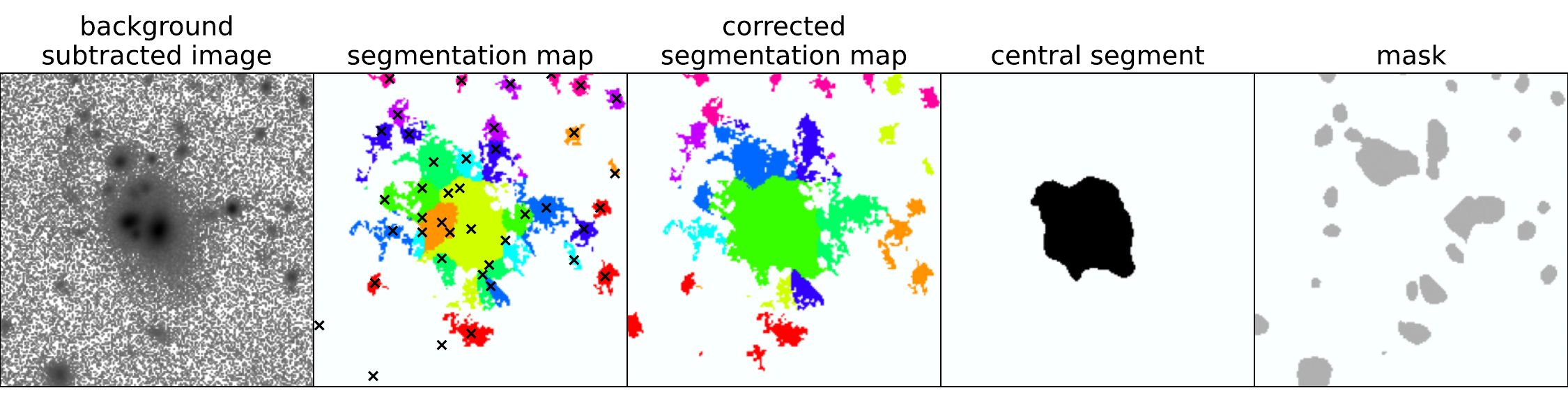}
    \caption{Example of the segmentation of a mock image of a merger from the IllustrisTNG simulation. Left most panel shows the background subtracted image with logarithmic scaling. Second panel shows an initial segmentation map with deblending applied at a threshold of 0.00001, the black crosses indicate the brightest pixel in each segment. In the middle panel, a corrected segmentation map is displayed with adjacent segments combined where the brightest pixels are closely separated. The two right most panels show the central segment and mask covering all other sources.}
    \label{fig:segmentation}
\end{figure*}

After segmentation, the central source was selected to be used and all other sources were selected as a mask. The central source and mask were then smoothed with a uniform kernel of shape (10, 10). An example of this segmentation method working can be seen in Figure \ref{fig:segmentation} where a clumpy merging system is successfully segmented.

\subsection{Non-parametric morphological statistics}\label{subsec:Parametric and non-parametric morphologies}
The morphological statistics of the \citetalias{Margalef-Bentabol2024} images were generated; for the classifiers in this work, the $C$, $G$, and $M_{20}$ statistics \citep{Conselice2003, Lotz2004} were used. The \texttt{statmorph} package (V\_0.6.1) for Python \citep{Rodriguez-Gomez2019} was used to calculate the morphological statistics, the resulting statistics are summarised in Appendix \ref{app:Histograms of the sample morphologies}.

In brief, C can be calculated from the ratio of the radii enclosing 20\% ($r_{20}$) and 80\% ($r_{80}$) of the flux from the galaxy \citep{Bershady2000, Conselice2003}. The total flux of the galaxy is taken to be within 1.5 Petrosian radii of point which minimises the asymmetry statistic \citep{Lotz2004, Rodriguez-Gomez2019}. The ratio is then normalised by a logarithm to give $C$ as
\begin{equation}
    C = 5\log_{10}\bigg(\frac{r_{80}}{r_{20}}\bigg)
    \label{eq:concentration},
\end{equation}
which has a higher value for bulge dominated galaxies and a lower value for disk dominated galaxies.

G is a statistic adapted from econometrics by \citet{Abraham2003} to describe the inequality in pixel brightness in an image. \citet{Lotz2004} defined $G$ for a 1D flattened array of pixels X as
\begin{equation}
    G = \frac{1}{|\Bar{X}|n(n - 1)}\sum_i^n(2i - n - 1)|X_i|
    \label{eq:gini definition},
\end{equation}
for n pixels which are above a threshold brightness and $\Bar{X}$ the average pixel brightness. This definition was adopted by \citet{Rodriguez-Gomez2019} in the \texttt{statmorph} package for Python.

The $M_{20}$ statistic was introduced in \citet{Lotz2004} to measure the spatial distribution of bright features such as bars, nuclei, spiral arms, and star-forming regions. The second-order moment of each pixel is defined as
\begin{equation}
    M_i=X_i\bigg((x_i-x_c)^2+(y_i-y_c)^2\bigg),
    \label{eq:Mi definition}
\end{equation}
where  X$_i$ is the pixel brightness, (x$_c$, y$_c$) is the position of the galaxy centre and (x$_i$, y$_i$) is the position of the pixel. The centre is taken to be the point that minimises the total second-order moment (M$_{tot}$) as
\begin{equation}
    M_{tot}=\sum_i^nM_i,
    \label{eq:Mtot definition}
\end{equation}
for an image with n pixels. $M_{20}$ is then given by
\begin{equation}
    M_{20}=\log_{10}\Bigg(\frac{\sum_iM_i}{M_{tot}}\Bigg),
    \label{eq:M20 definition}
\end{equation}
with the condition that $\sum_iX_i=0.2X_{tot}$. Normalising by the total second-order moment ensures that $M_{20}$ is independent of galaxy size and total flux.

\subsection{Evaluation metrics}\label{subsec:Evaluation metrics}
To quantify the performance of merger classifiers and compare them to the deep learning techniques in \citetalias{Margalef-Bentabol2024}, four metrics are defined; accuracy, precision, recall and F1 score. These metrics are calculated from the number of classifications which are true positive (TP), false positive (FP), true negative (TN), and false negative (FN), whereby positive labels are classified as mergers and negative are classified as non-mergers. 

The metrics are defined the same as in \citetalias{Margalef-Bentabol2024} to ensure consistency. Accuracy (A) is given by
\begin{equation}
    A = \frac{TP+TN}{TP+FP+TN+FN},
    \label{eq:accuracy definition}
\end{equation}
which is the ratio of true classifications to total classifications. Precision gives a measure of purity and is calculated as
\begin{equation}
    P = \frac{TP}{TP+FP},
    \label{eq:precision definition}
\end{equation}
which is the ratio of true positive classification to the total number of positive labels. Recall is a measure of completeness and is calculated as
\begin{equation}
    R = \frac{TP}{TP+FN},
    \label{eq:recall definition}
\end{equation}
which is the ratio of true positive classifications to the number of true positive labels. F1 score is calculated as
\begin{equation}
    F_1 = \frac{2}{R^{-1}+P^{-1}},
    \label{eq:f1 definition}
\end{equation}
which is the harmonic mean of recall and precision.

These metrics assume a balanced dataset, that is a dataset containing an equal number of mergers and non-mergers. To calculate the metrics for an unbalanced dataset, the number of TP, FP, TN, and FN classifications are weighted as
\begin{equation}
    \begin{split}
        TP &\longrightarrow TP\cdot W_{mergers} \\
        FP &\longrightarrow FP\cdot W_{nonmergers} \\
        TN &\longrightarrow TN\cdot W_{mergers} \\
        FN &\longrightarrow FN\cdot W_{nonmergers},
    \end{split}
    \label{eq:weighting}
\end{equation}
with the factors $W_{mergers}$ and $W_{nonmergers}$ given by
\begin{equation}
    \begin{split}
        W_{mergers}&=\frac{2(TP+FN)}{TP+FP+TN+FN}, \\
        W_{nonmergers}&=\frac{2(TN+FP)}{TP+FP+TN+FN}.
    \end{split}
    \label{eq:weighting factors}
\end{equation}

\subsection{Classifier optimisation}\label{subsec:Classifieroptimisation}
A Markov chain Monte-Carlo (MCMC) method, implemented using the \texttt{emcee} package for Python \citep{Goodman2010, Foreman-Mackey2013}, was used to optimise the lines which divide mergers and non-mergers in 2D morphological space. The training data from \citetalias{Margalef-Bentabol2024} was used for this optimisation so any biases in the training data would not be a variable in the comparison to \citetalias{Margalef-Bentabol2024}.

A visually fitted cut was used as the median of a truncated Gaussian distribution to generate 150 sets of initial parameters for the cuts. Selecting a good set of initial parameters, that is parameters close to the ideal position, is key to ensuring the MCMC is well constrained. The process used to select the initial conditions is detailed in Appendix \ref{app:MCMC initial condition and prior selection}.

The MCMC routine works by sampling the posterior distribution given by Bayes' theorem,
\begin{equation}
    P(\theta|m) = \frac{P(\theta)P(m|\theta)}{P(m)},
    \label{eq:posterior definiton}
\end{equation}
where the parameters of the cut are $\theta$, and $m$ is the distribution of morphological statistics. The prior is given as $P(\theta)$ and the likelihood function, $P(m|\theta)$, was set up as the inverse of the accuracy. To increase the gradient of the likelihood function it was calculated as
\begin{equation}
    P_{likelihood}(TP,~FP,~TN,~FN) = e - e^{\frac{TP+FP+TN+FN}{TP+TN}},
    \label{eq:likelihood function}
\end{equation}
where the classifications $TP$, $FP$, $TN$, and $FN$ depend on the parameters sampled. The constant value of e was added to make the optimal value of the function 0. The sampling was performed over 10000 steps with the optimised criteria taken from the median of the second half of the sampling.

Using a validation set prevents the optimisation routine from fitting to outlying points in the training data. A validation dataset was randomly sampled from the training data to contain 20\% of the images in the training dataset as in \citetalias{Margalef-Bentabol2024}. At each step of the MCMC, the likelihood function was evaluated for both the training dataset and the validation dataset. If the absolute value of the validation likelihood was larger than the absolute value of the training likelihood then an arbitrarily large value was subtracted from the likelihood function. This procedure is effective because it forces the chain to reject any sets of parameters which result in poor likelihood values when applied to previously unseen data, parameters which trigger this may be overfitted to the training dataset.

Setting priors also helped to prevent fitting to outliers, this was achieved by setting limits on the parameters and subtracting an arbitrarily large value from the likelihood function if the parameters exceeded the range. The priors were selected using the same method as for the initial parameters (see Appendix \ref{app:MCMC initial condition and prior selection} and also used as the limits on the truncated Gaussian distribution used to generate the initial parameters. The initial conditions and priors for the sampling are detailed in Table \ref{tab:mcmc setup}.

\begin{table*}
    \caption{\label{tab:mcmc setup}Initial conditions and priors set for the MCMC optimisation of the merger criteria.}
    \centering
    \begin{tabular}{c c c c c}
        \hline\hline
        \rule{0pt}{2.5ex}morphological statistics & initial gradient & gradient priors & initial constant & constant priors \\
        \hline
        \rule{0pt}{2.5ex}$G$, $M_{20}$ & -0.24 & -0.38, 0.13 & 0.15 & -0.06, 0.32 \\
        $G$, $C$ & 0.15 & 0.075, 0.225 & 0.12 & -0.08, 0.32 \\
        \hline
    \end{tabular}
    \tablefoot{The criteria are constructed as $y>mx+c$ with the morphological statistics as y and x (G is the y variable in both cases). Priors are used to set limits on the upper and lower values of each parameter.}
\end{table*}

\section{Results}\label{sec:Results}
\subsection{Optimised merger criteria}\label{subsec: optimised merger criteria}
\begin{figure*}
    \centering
    \includegraphics[width=8.5cm]{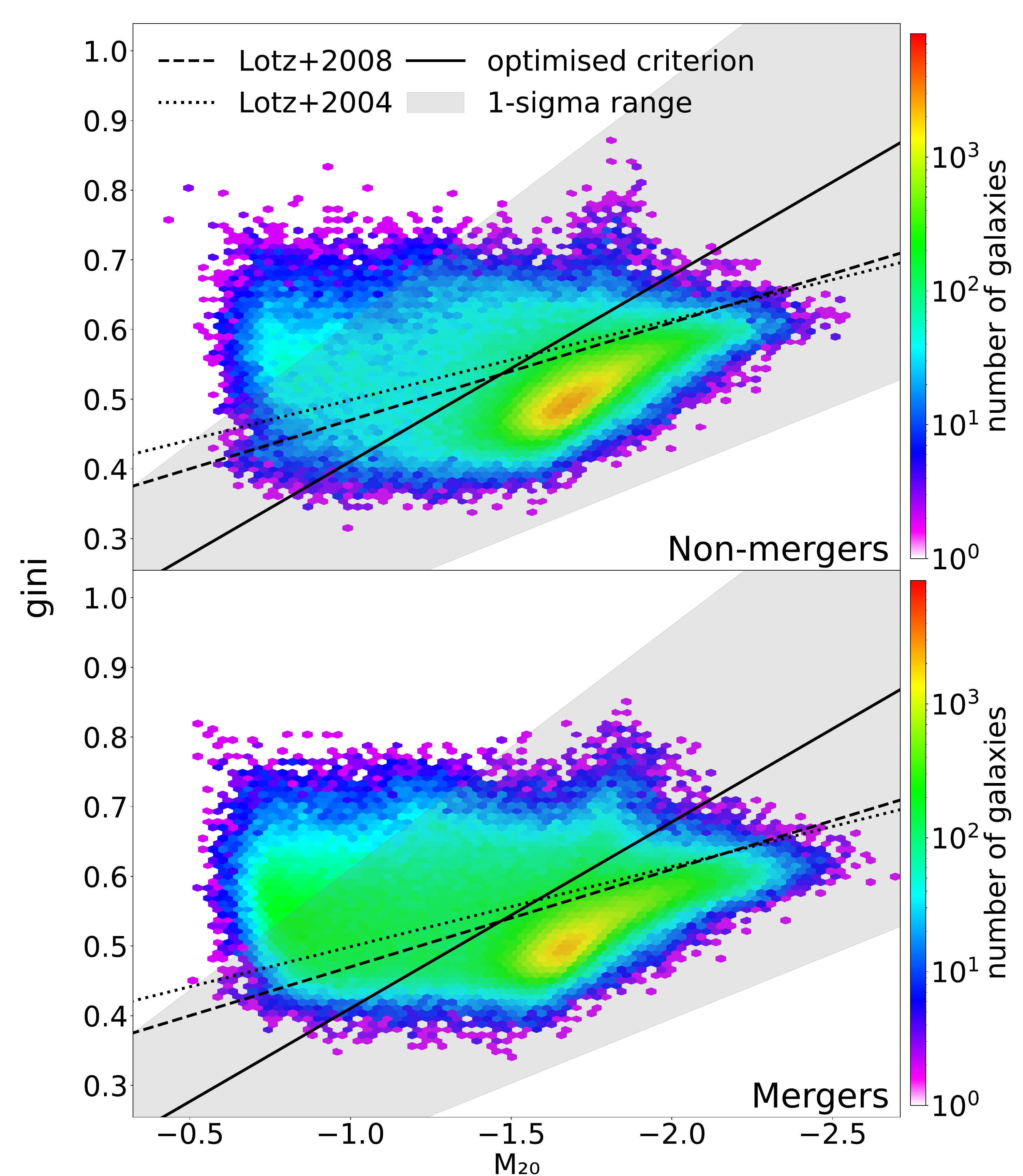}
    \includegraphics[width=8.5cm]{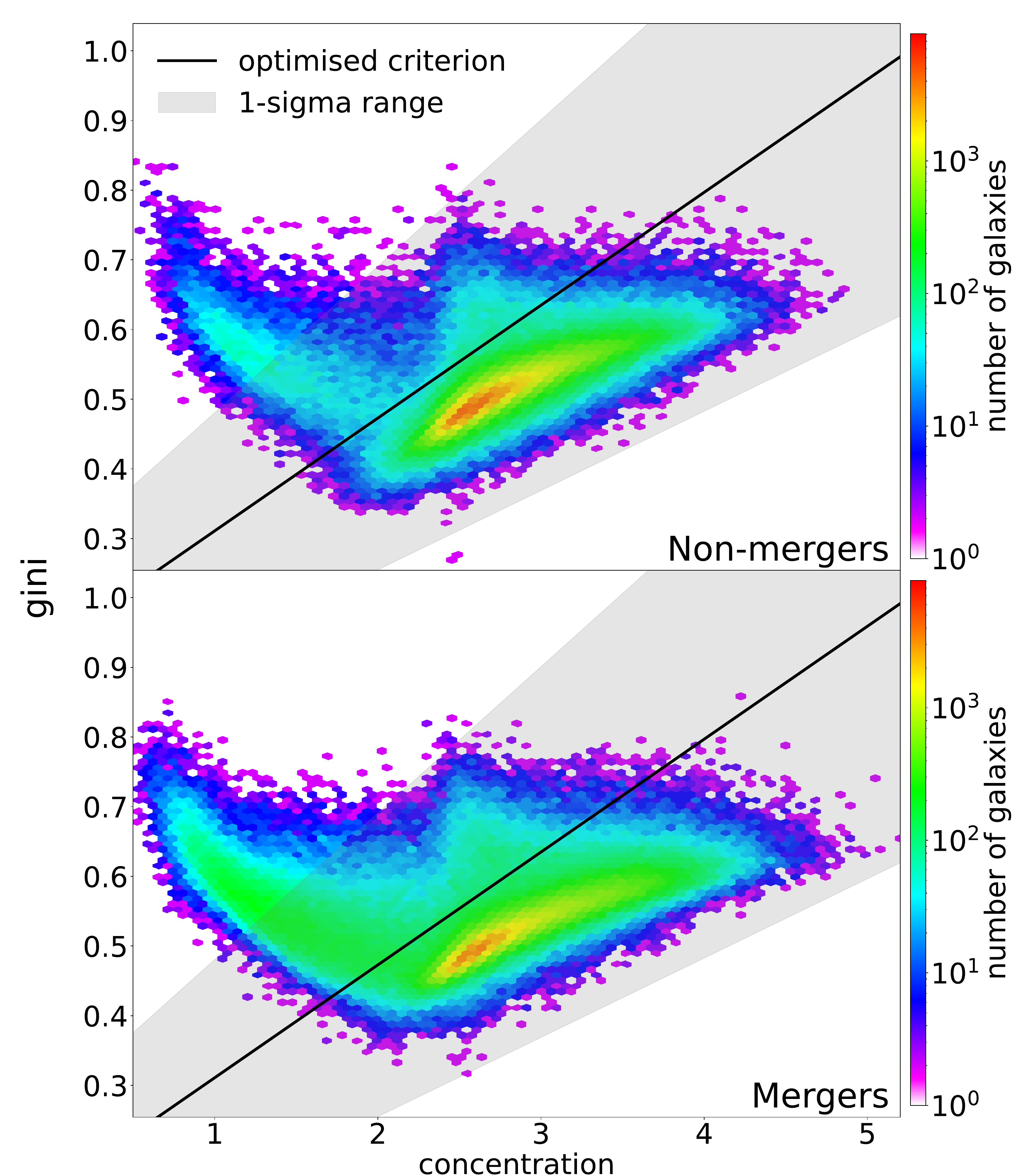}
    \caption{Distribution of the training data in $G$-$M_{20}$ (left) and $G-C$ (right) with various merger criteria. The top panels show the distribution of non-merging galaxies and the bottom panels show the distribution of merging galaxies. Colouring shows the number of galaxies per cell in morphology space from low (in purple) to high (in red). The dashed and dotted lines represent cuts from the literature. The solid lines show the result of the MCMC routine. The 1-$\sigma$ range of the optimised criterion is shown as the grey region.}
    \label{fig:morph_dist_main}
\end{figure*}

We find a criteria for merger classification using $G$ and the $M_{20}$ statistic of 
\begin{equation}
    G>(-0.267\pm0.081)M_{20}+(0.143\pm0.012),
    \label{eq:optimised gm20}
\end{equation}
using the MCMC routine. This criterion is shown as the solid black line in the left panels of Figure \ref{fig:morph_dist_main} for which all points above the line are classified as mergers.

\begin{figure}
    \centering
    \resizebox{\hsize}{!}{\includegraphics{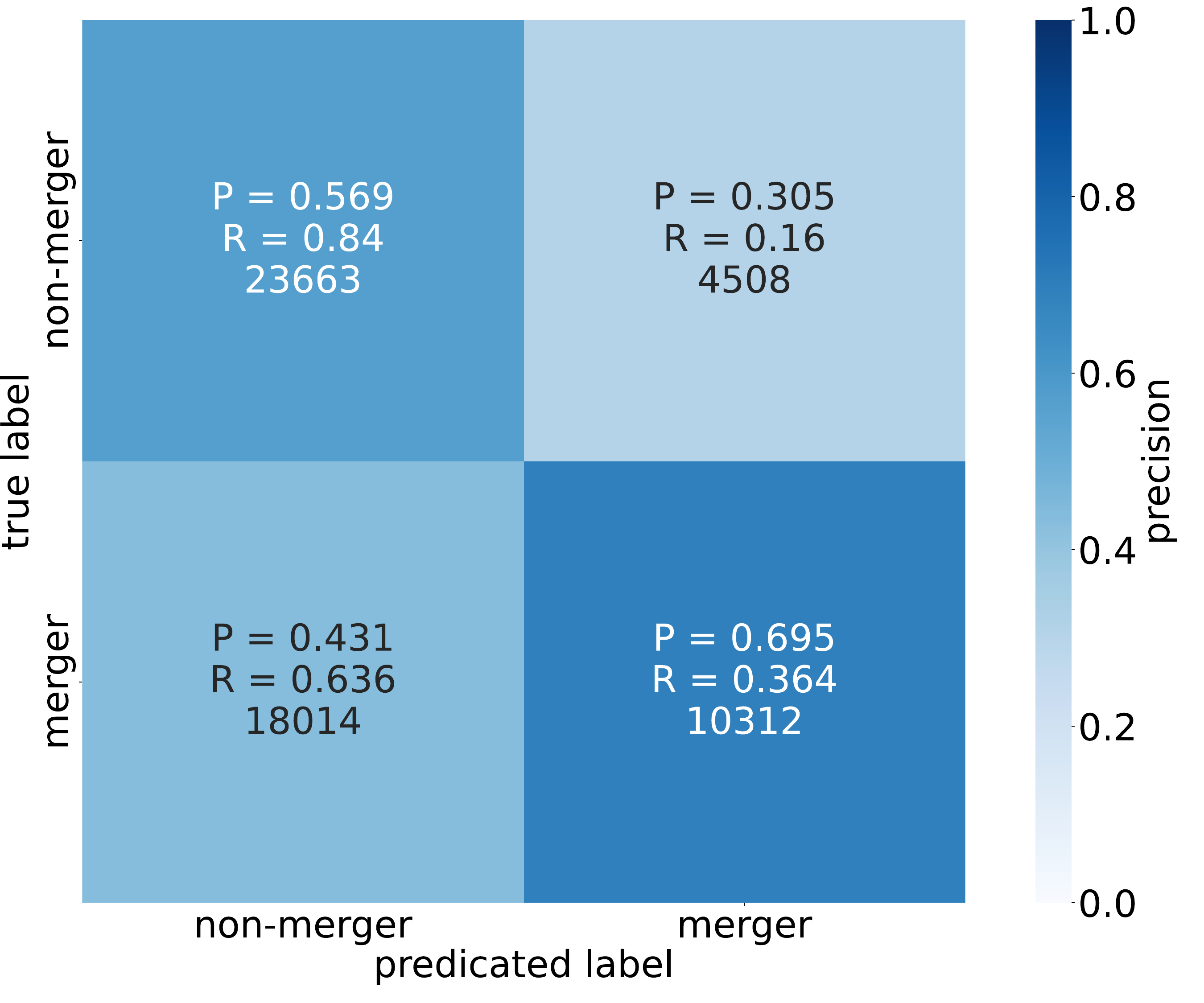}}
    \resizebox{\hsize}{!}{\includegraphics{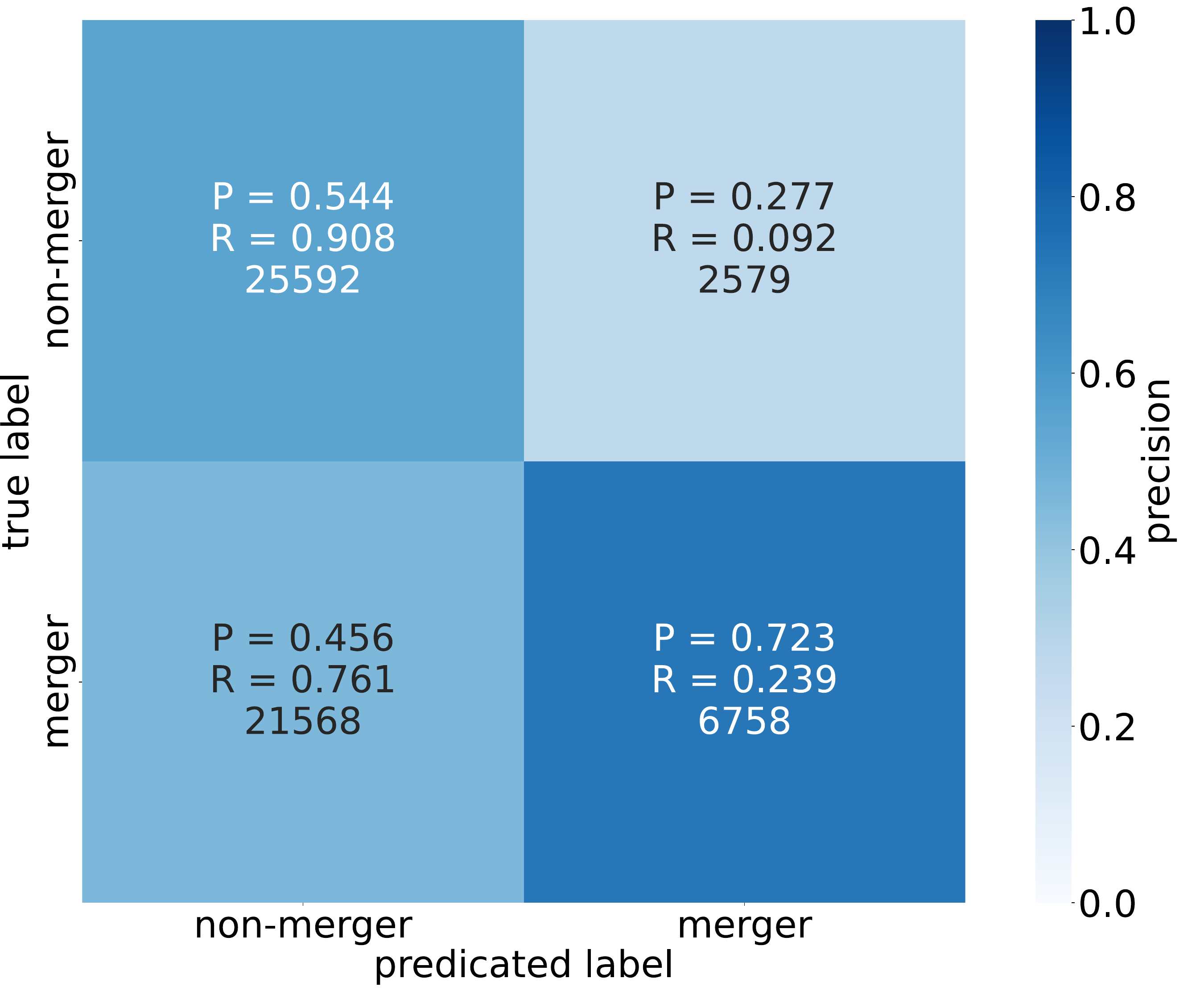}}
    \caption{Confusion matrices displaying the performance of $G$-$M_{20}$ (top) and $G-C$ (bottom) merger criteria resulting from the MCMC optimisation. Upper value in each cell is the fraction of predicted labels in each category equivalent to the precision. The middle value is the fraction of true labels in each category equivalent to the recall. The lower value is the number of images in each class.}
    \label{fig:mcmc_conf}
\end{figure}

The performance of the optimised cut is shown in the confusion matrix in the top panel of Figure \ref{fig:mcmc_conf}. A confusion matrix summarises the performance of a classifier by dividing the images into cells based on their true label (on the y axis) and predicted label (on the x axis). Each cell is coloured based on the precision of the prediction represented. The numbers in the cells represent the evaluation metrics for that cell; the precision (P) is the number of images in the class normalised vertically, the recall (R) is the number of images in the class normalised horizontally, and the bottom number is the total number of images in that class after the weighting described in Section \ref{subsec:Evaluation metrics} has been applied. This criterion identified mergers with a precision of 69.5\% and a recall of 36.4\%. Mergers and non-mergers were distinguished with an accuracy of 60.2\%.

A merger classifier using $G$ and $C$ was developed using MCMC optimisation as 
\begin{equation}
    G>(0.162\pm0.048)C+(0.149\pm0.12),
    \label{eq:optimised gc}
\end{equation}
which is displayed as a solid black line in the right panels of Figure \ref{fig:morph_dist_main}. These statistics trace a similar feature ($G$ traces the inequality of light in a pixel-wise manner and $C$ traces the inequality of light in a radial manner) and so in Figure \ref{fig:morph_dist_main} most galaxies appear in a similar region with the exception for a region with low $C$ populated mostly by mergers. The 1-$\sigma$ range for this cut is displayed as the gray region.

The performance of the $G$-$C$ classifier is summarised in the confusion matrix shown in the bottom panel of Figure \ref{fig:mcmc_conf}. Mergers were identified with a precision of 72.3\% and recall of 23.6\% giving an accuracy of 57.4\%.

\subsection{Classification of merger subclasses}\label{subsec:Classification of merger subclasses}
Mergers can be classified by the time before or after coalescence. In \citetalias{Margalef-Bentabol2024}, pre-mergers are defined as systems between 0.1Gyrs and 0.8Gyrs before coalescence, ongoing mergers are from 0.1Gyrs before coalescence to 0.1Gyrs after, and post-mergers are are set between 0.1Gyrs and 0.3Gyrs after coalescence. 

\begin{figure*}
    \centering
    \includegraphics[width=8.5cm]{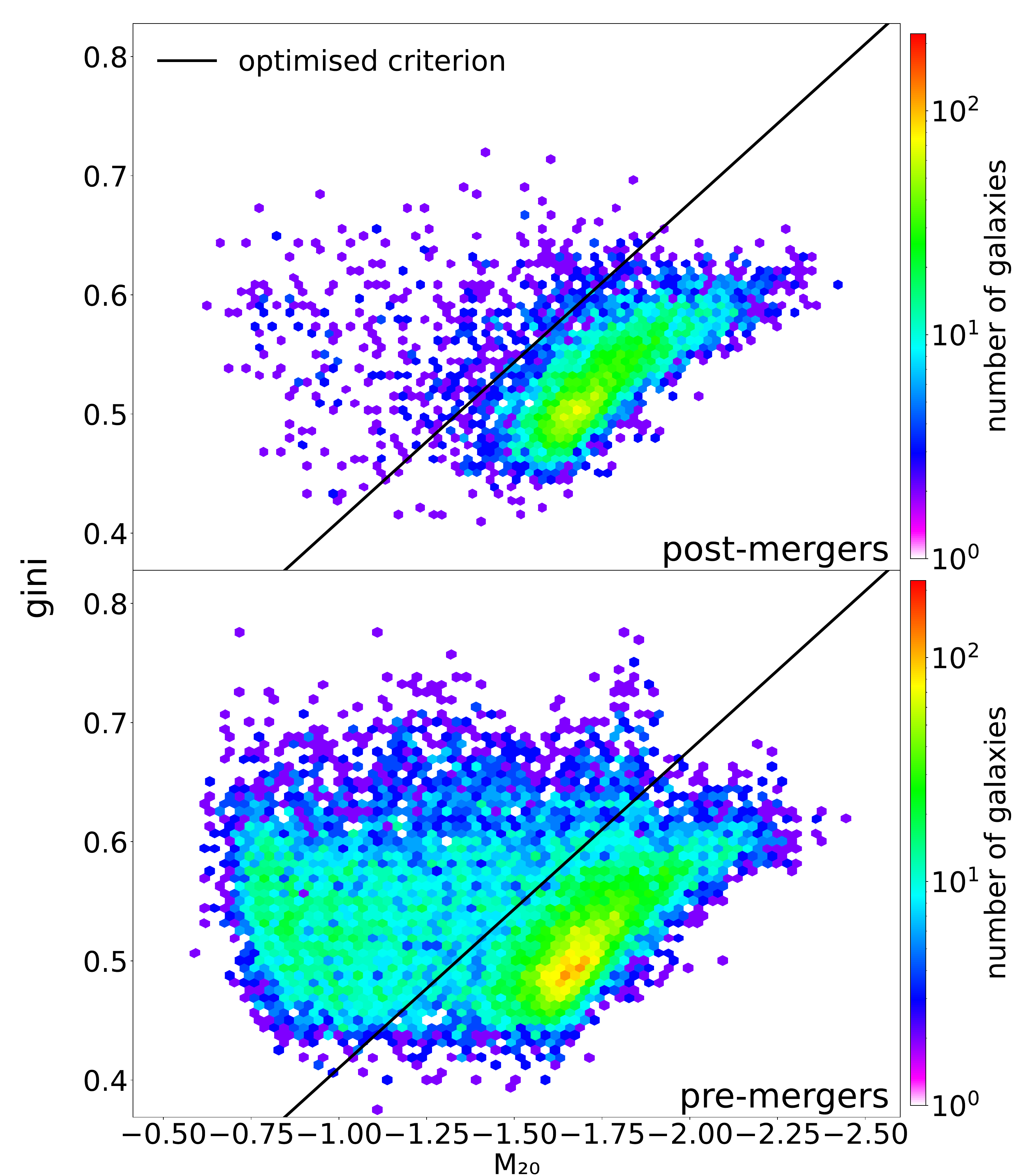}    \includegraphics[width=8.5cm]{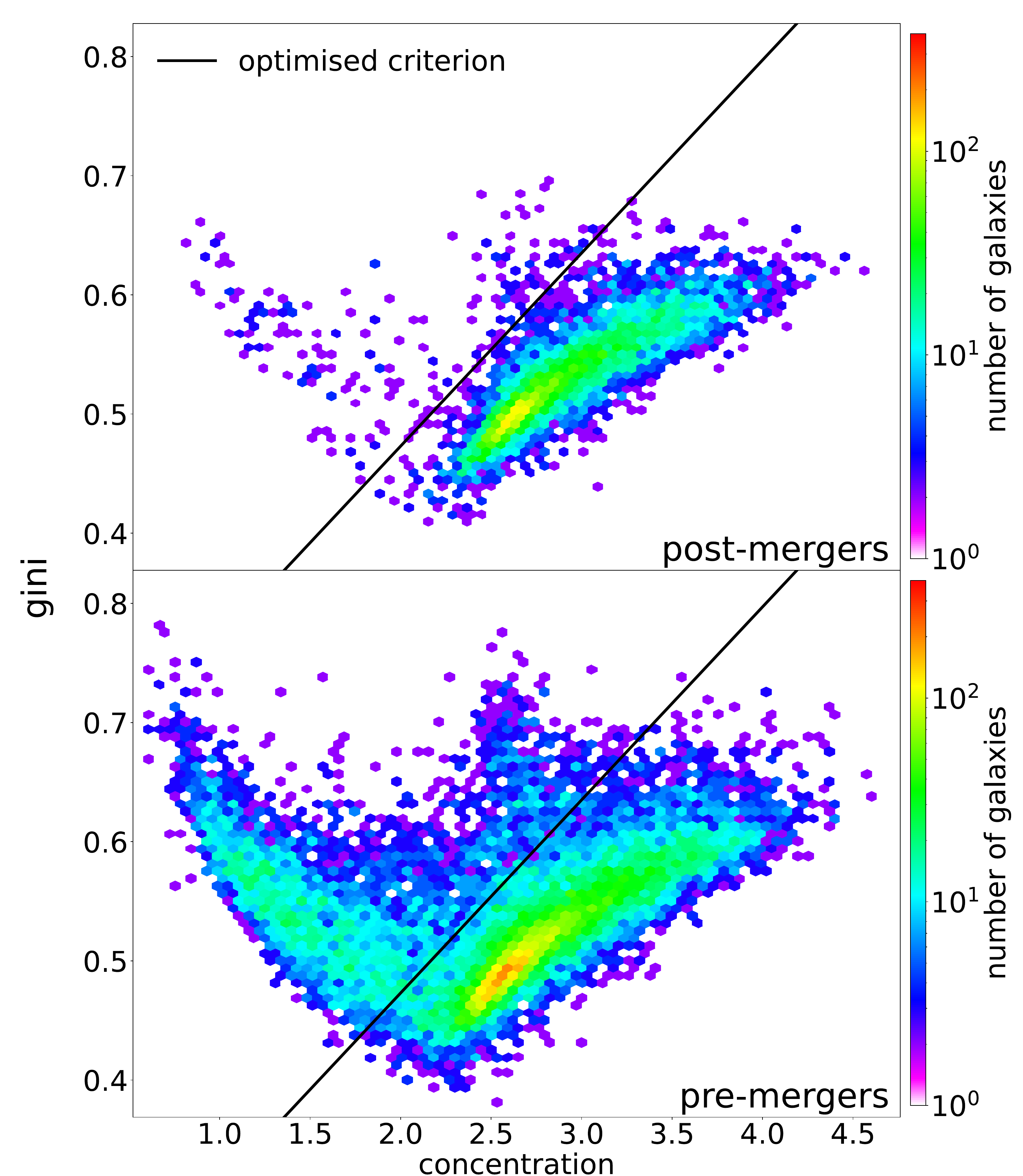}
    \caption{Distribution of merger subclasses in the testing data in $G$-$M_{20}$ (left) and $G-C$ (right). Data is separated into post-mergers (this class also includes the ongoing mergers) in the top panels and pre-mergers in the bottom panels. Post-mergers are more concentrated below the merger criteria in both cases compared to pre-mergers so pre-mergers can be selected with high precision. Colouring shows the number of galaxies per cell in morphology space from low (in purple) to high (in red).}
    \label{fig:morph_dist_prepost}
\end{figure*}

\begin{figure}
    \centering
    \resizebox{\hsize}{!}{\includegraphics{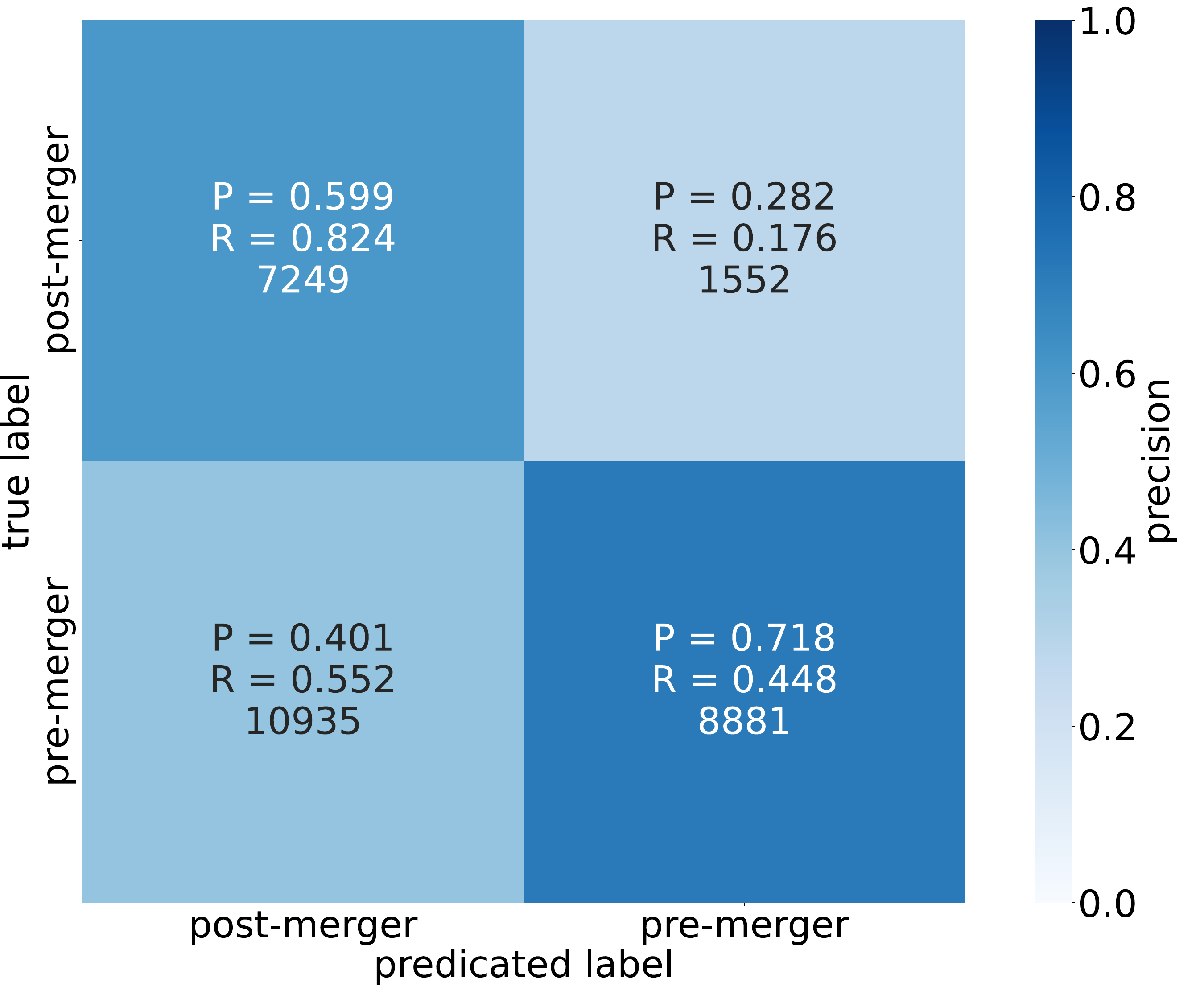}}
    \resizebox{\hsize}{!}{\includegraphics{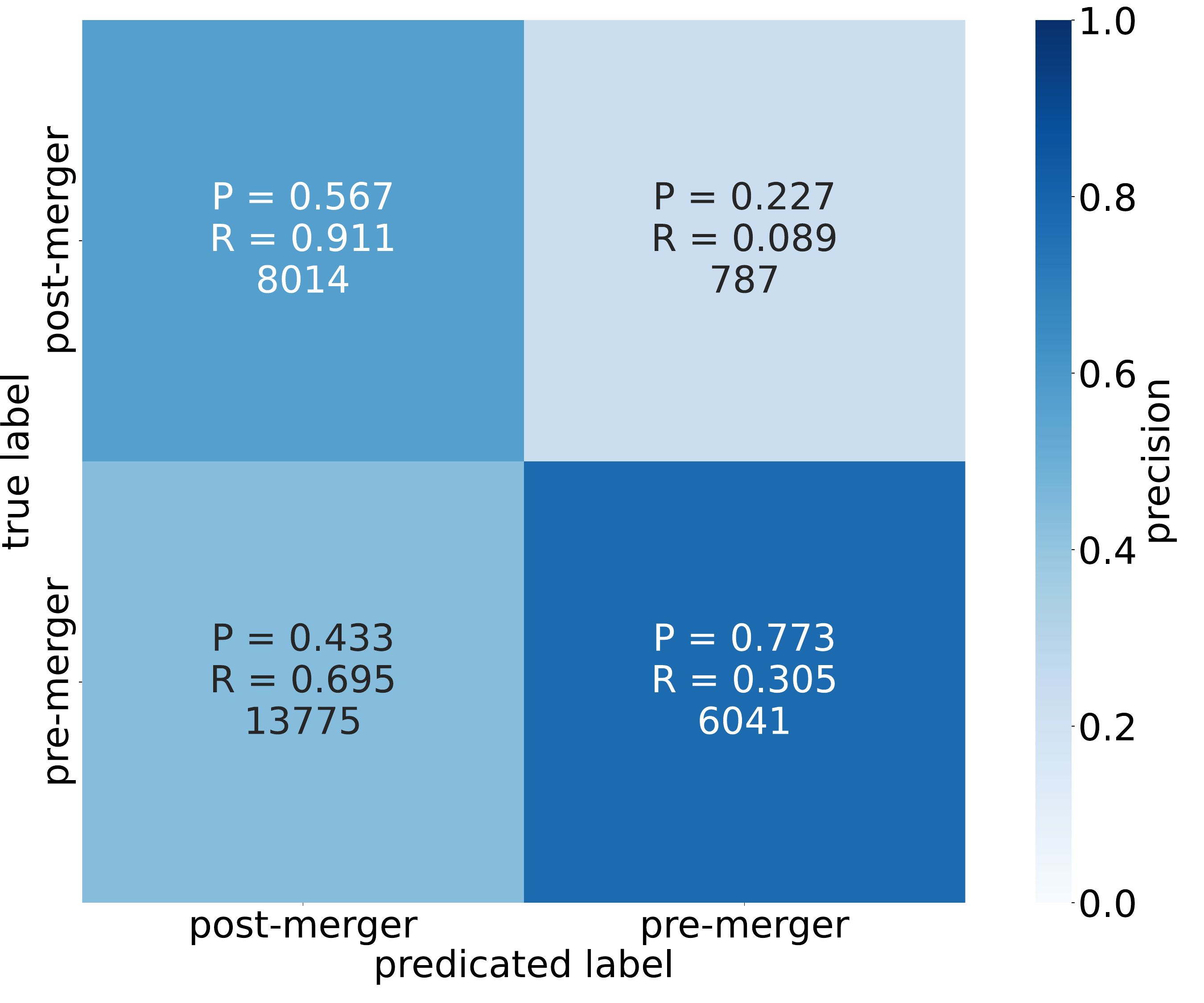}}
    \caption{Confusion matrices displaying the performance of the $G-M{20}$ (left) and $G-C$ (right) merger classifiers on the pre and post merger task. The merger criterion used is from the MCMC routine. The values shown in the cells are the precision, recall, and number of classifications, as in Figure \ref{fig:mcmc_conf}. Ongoing mergers are included in the post-merger class.}
    \label{fig:prepost_conf}
\end{figure}

From a pre-prepared sample of mergers, the $G$-$M_{20}$ classifier is able to distinguish pre-mergers from post-mergers and ongoing mergers with a precision of 71.7\% and recall of 44.8\%. This capability is demonstrated in the left panels of Figure \ref{fig:morph_dist_prepost} using the test dataset; for this task the ongoing mergers and post-mergers are combined into a single class. The lower-left panel of Figure \ref{fig:morph_dist_prepost} shows the distribution of pre-mergers and the upper-left panel shows the distribution of post-mergers and ongoing mergers. This classification task achieves an accuracy of 63.6\%. The full results of this classification task can be seen in the confusion matrix in the top panel of Figure \ref{fig:prepost_conf}.

The $G$-$C$ classifier is able to extract pre-mergers from a sample of mergers with a precision of 77.7\% and a recall of 30.5\%. This classification is shown in the right panels of Figure \ref{fig:morph_dist_prepost} in which the pre-merger distribution is shown in the lower panel and the post-mergers are shown in the upper panel. For this task, the classifier achieved an accuracy of 60.8\%. The full results of this classification task can be seen in the confusion matrix in the top panel of Figure \ref{fig:prepost_conf}.

\subsection{Dependence on redshift and stellar mass}\label{subsec:Dependence on redshift and stellar mass}
\begin{figure*}
    \centering
    \includegraphics[width=17cm]{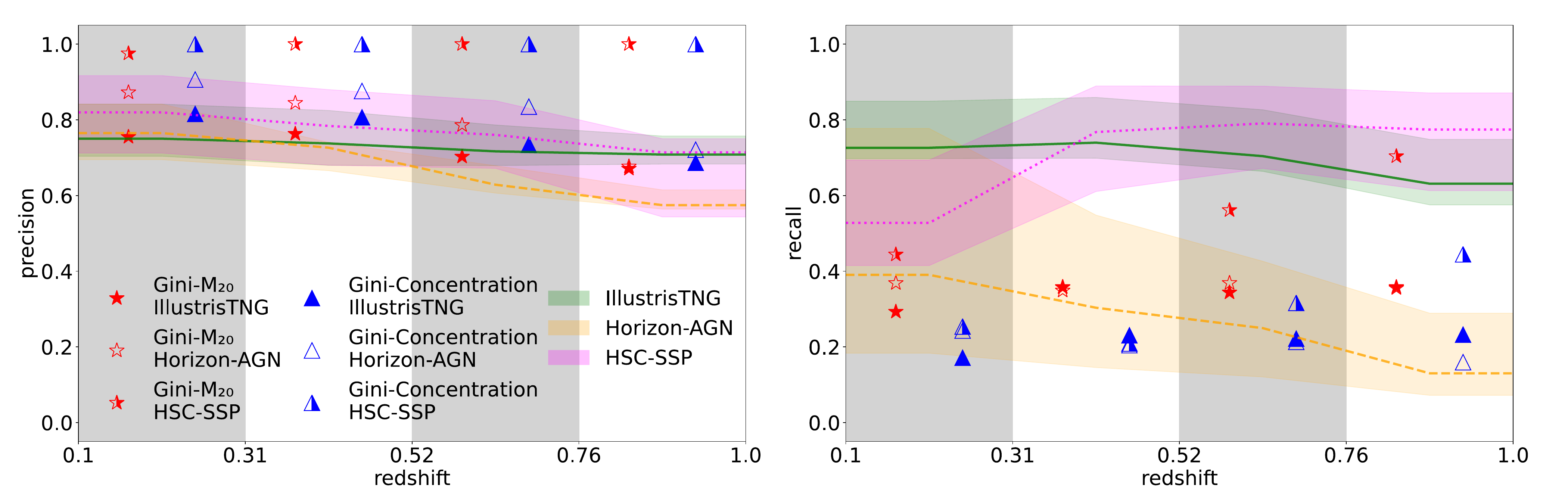}
    \caption{Performance of the merger classifiers with binned redshift in terms of precision (left) and recall (right). Red stars represent the performance of the $G$-$M_{20}$ and blue triangles represent the performance of the $G$-$C$ classifier. Filled symbols show the results from classifying the IllustrisTNG testing data, empty symbols show the results of classifying the Horizon-AGN dataset, and the half filled symbols show the performance when classifying the HSC-SSP observations. The lines show the median performance of the classifiers in \citetalias{Margalef-Bentabol2024} for each redshift bin and the surrounding shaded areas show the range of performance of those models. The solid green line and green shaded region summarise the classification of the IllustrisTNG testing data using the \citetalias{Margalef-Bentabol2024} models, the orange dashed line and orange region are for the classification of the Horizon-AGN dataset, and the magenta dotted line and magenta region are for the HSC-SSP classifications.}
    \label{fig:z evolve}
\end{figure*}

\begin{figure*}
    \centering
    \includegraphics[width=17cm]{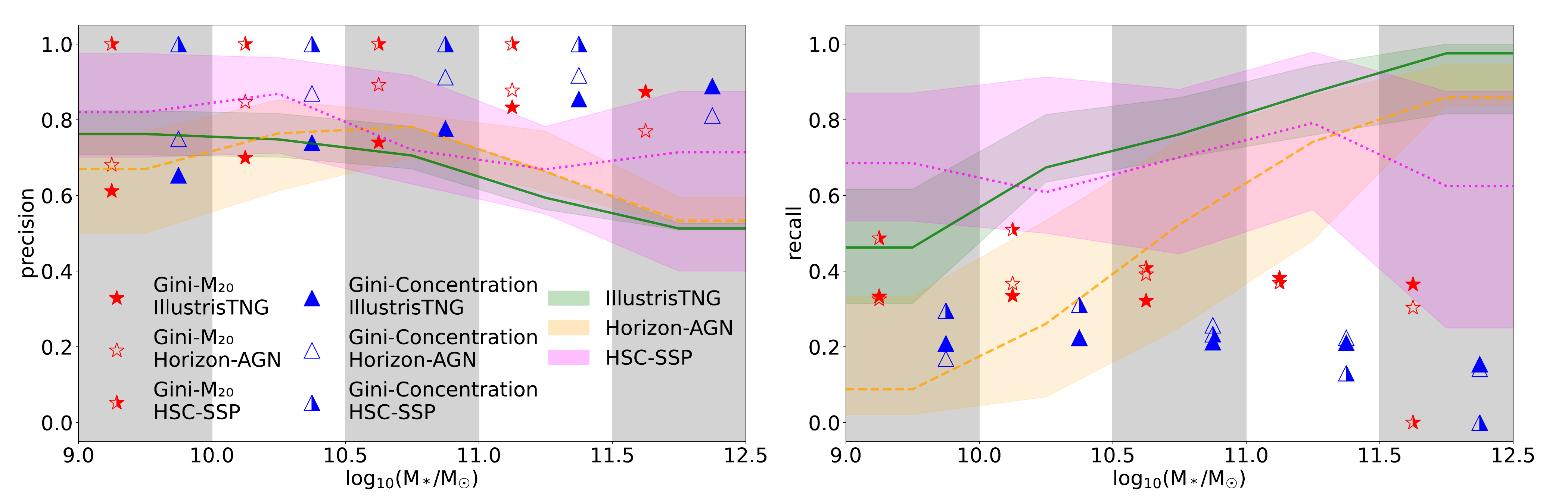}
    \caption{Performance of the merger classifiers with binned stellar mass in terms of precision (left) and recall (right). The symbols and lines follow the same formatting as Figure \ref{fig:z evolve}.}
    \label{fig:m evolve}
\end{figure*}

The dependence of the performance of the morphological classifiers on redshift and stellar mass was investigated. The testing dataset, Horizon-AGN dataset, and HSC-SSP dataset were binned by redshift and by stellar mass. The bin ranges used follow \citetalias{Margalef-Bentabol2024} in which the redshift bins are the same as those used in creating the mock images (see Section \ref{subsec:Mock images}) and the stellar mass bins are 9.0 < $\log_{10}(M_1/M_\odot)$ < 10.0, 10.0 < $\log_{10}(M_2/M_\odot)$ < 10.5, 10.5 < $\log_{10}(M_3/M_\odot)$ < 11.0, 11 < $\log_{10}(M_4/M_\odot)$ < 11.5, and 11.5 < $\log_{10}(M_5/M_\odot)$ < 12.5. The precision and recall for the $G$-$M_{20}$ and $G$-$C$ classifiers are displayed in terms of redshift in Figures \ref{fig:z evolve} and in terms of stellar mass in Figure \ref{fig:m evolve}. The results from the testing data are shown with solid symbols, the Horizon-AGN results are shown with empty symbols, and the results from the HSC-SSP are shown with half-filled symbols. The results from the $G$-$M_{20}$ classifier are displayed with red stars and the $G$-$C$ classification results are shown with blue triangles, the results are offset from the centre of each bin for ease of reading. These results are shown alongside the classifications from \citetalias{Margalef-Bentabol2024} which are represented by the median precision and recall in each redshift and mass bin displayed as the solid, dashed and dotted liens for the IllustrisTNG, Horizon-AGN and HSC-SSP results, respectively. The range of the \citetalias{Margalef-Bentabol2024} results are shown as the corresponding shaded regions.

The precision of the morphological classifiers decreases with redshift. When applied to the IllustrisTNG test set and Horizon-AGN sample, the precision decreases for both classifiers after z = 0.52; $G$-$M_{20}$ decreases by 9.2\% for IllustrisTNG and by 19.8\% for Horizon-AGN across the redshift range, the $G$-$C$ classifier drops by 12.9\% for the IllustrisTNG data and 18.5\% for Horizon-AGN. With the HSC-SSP dataset, precision remains approximately constant at 100\%. The unrealistically high precision seen in the HSC-SSP sample is discussed in Section \ref{subsec:Reliability of the morphological classifiers}.

Recall increases slightly for both classifiers on the IllustrisTNG dataset across redshift with an increase of 6.5\% and 6.1\% for $G$-$M_{20}$ and $G$-$C$, respectively. With the Horizon-AGN dataset, a decrease of 1.3\% and an increase of 8.4\% for $G$-$M_{20}$ and $G$-$C$, respectively were found across the redshift range. Recall increases substantially after z = 0.52 with redshift when applied to HSC-SSP for both classifiers with an increase of 26.0\% for $G$-$M_{20}$ and 19.0\% for the $G$-$C$ classifier.

The precision of the morphological classifiers increases with stellar mass. Precision increases when applied to the IllustrisTNG dataset by 26.2\% for $G$-$M_{20}$ and by 23.6\% for $G$-$C$. The precision of the $G$-$M_{20}$ classifier increases when applied to the Horizon-AGN dataset from 68.1\% up to a maximum of 89.2\% at $\log_{10}(M_*/M_\odot)=11.0$ before decreasing slightly to 77.0\% over the two most massive bins. Similarly, the $G$-$C$ classifier increases up to $\log_{10}(M_*/M_\odot)=11.0$ from 74.9\% to a maximum of 91.7\% and then decreases to 81.1\%. For the HSC-SSP dataset it remains constant at 100\%. No data is available for the precision on the HSC-SSP data in the highest mass bin.

The recall of the $G$-$M_{20}$ classifier remains approximately constant with mass for all three datasets with the exception of the most massive bin for the HSC-SSP observations where no mergers were correctly classified. The recall of the $G$-$C$ classifier decreases above the mass of $\log_{10}(M_*/M_\odot)=11.0$ for all three datasets. The recall of the $G$-$C$ classifier decreases by 5.5\% across the mass range when applied to the IllustrisTNG testing set, for the Horizon-AGN dataset the recall decreases by 2.7\%, and when applied to the HSC-SSP observations the recall decreases by 16.6\% before dropping to 0\% in the most massive bin.

\section{Discussion}\label{sec:Discussion}
\subsection{Merger criteria quality}\label{subsec:Merger criteria quality}
The MCMC optimised criterion for $G$-$M_{20}$ better traces the difference in the distribution of mergers and non-mergers compared to \citet{Lotz2008} and \citet{Lotz2004}. The number of mergers above the criteria in Figure \ref{fig:morph_dist_main} is significantly higher than the number of non-mergers above the criteria across the whole morphology space. Both mergers and non-mergers have their highest density at $G\sim0.5$ and $M_{20}\sim-1.7$. The distribution of mergers is more extended, with the density of mergers an order of magnitude higher than that of non-mergers for $M_{20}>-1.3$.

\begin{figure}
    \centering
    \resizebox{\hsize}{!}{\includegraphics{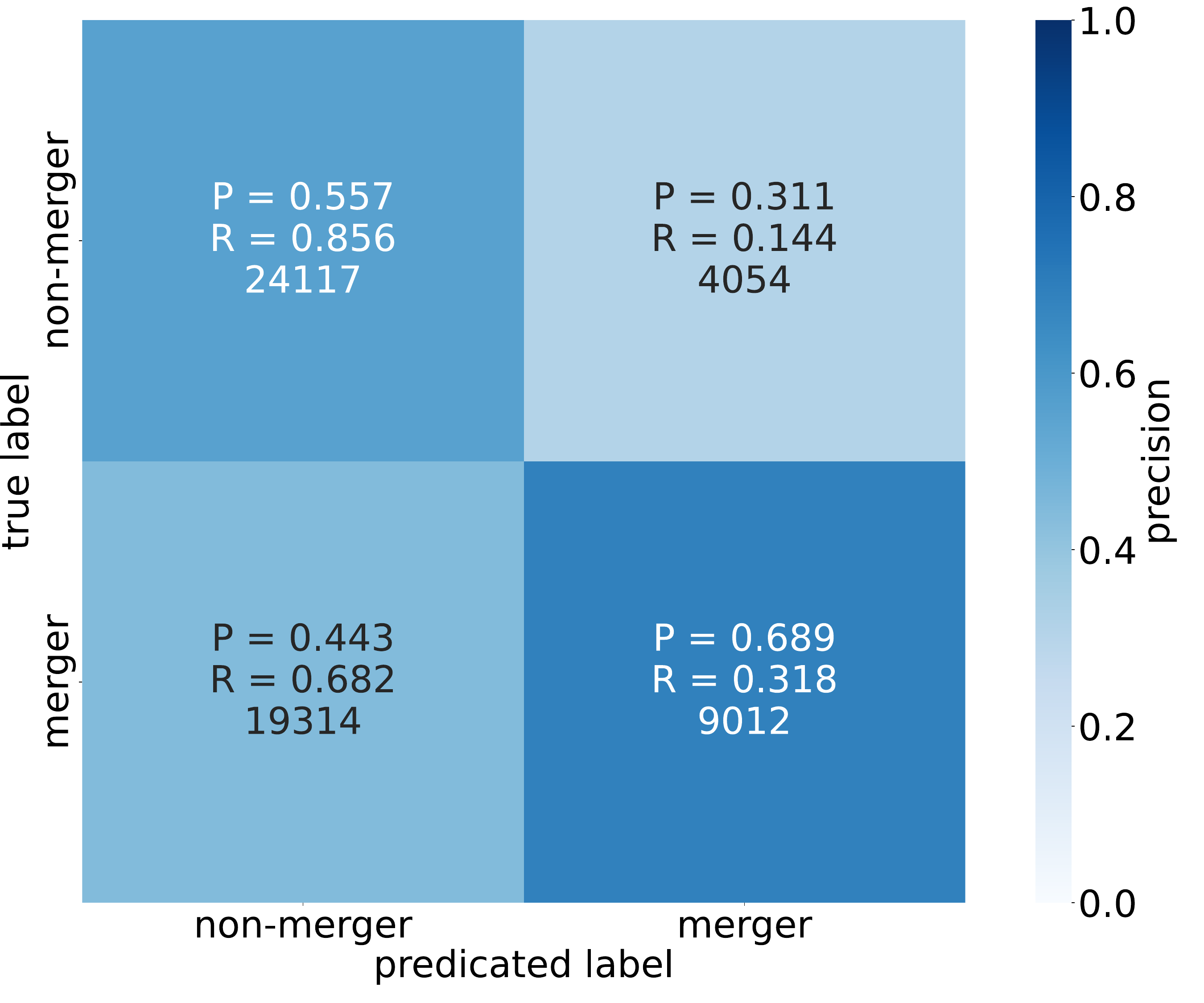}}
    \resizebox{\hsize}{!}{\includegraphics{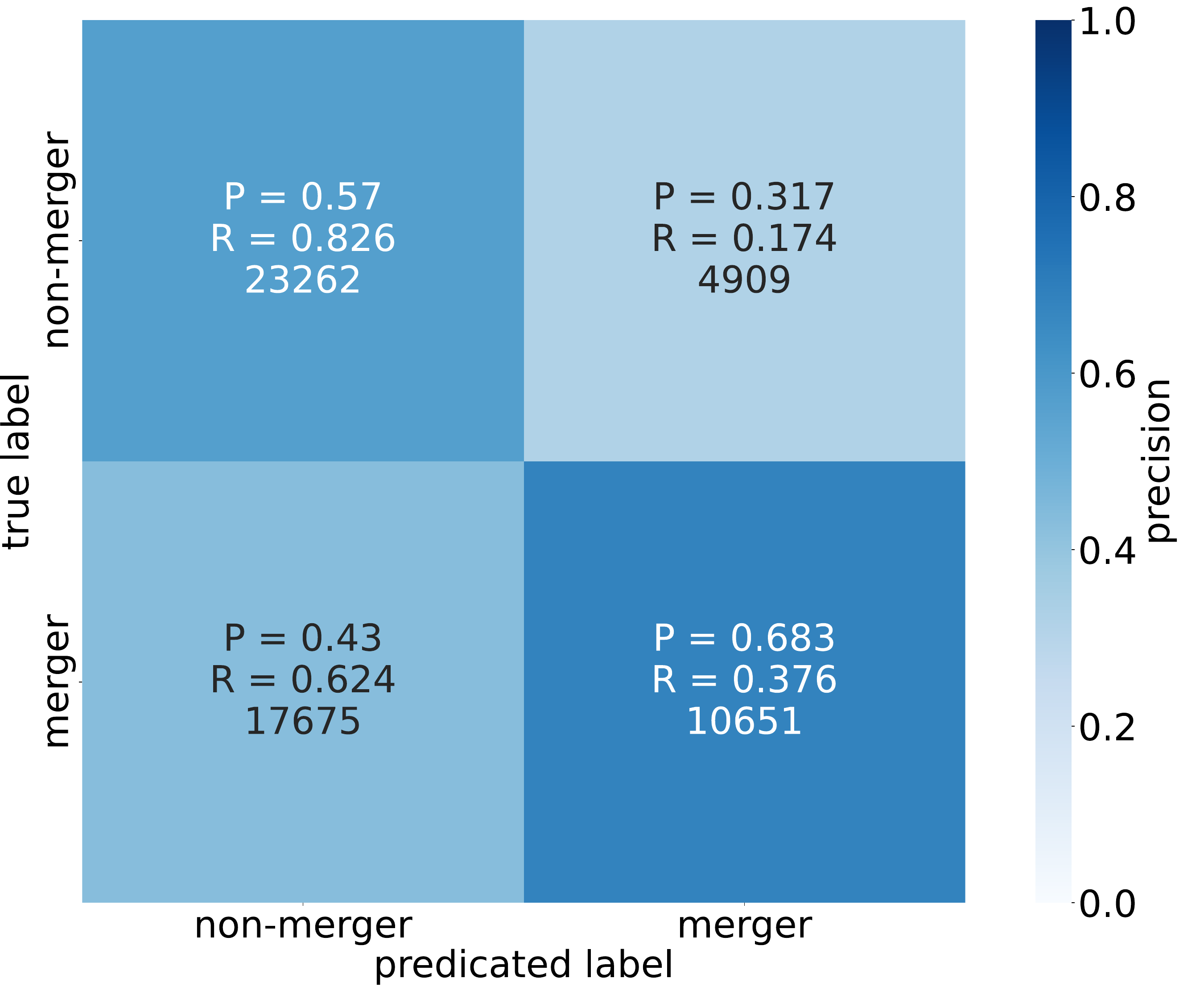}}
    \caption{Confusion matrices displaying the performance of $G$-$M_{20}$ merger criteria from \citet{Lotz2004} (top) and \citet{Lotz2008} (bottom). The values shown in the cells are the precision, recall, and number of classifications, in Figure \ref{fig:mcmc_conf}}
    \label{fig:Lotz gm20 conf}
\end{figure}

Figure \ref{fig:Lotz gm20 conf} shows the confusion matrices of the two literature cuts for $G$-$M_{20}$. The criteria from \citet{Lotz2004} and \citet{Lotz2008} have approximately the same precision at 68.9\% and 68.3\%, respectively; this is 0.6\% and 1.2\% lower than the updated $G$-$M_{20}$ criterion, and 3.4\% and 4\% lower than the $G$-$C$ criterion. \citet{Lotz2004} reaches an accuracy of 58.7\% which is 1.5 \% lower than the updated $G$-$M_{20}$ classifier and 1.3\% higher than the $G$-$C$ classifier, the recall of the \citet{Lotz2004} criterion was 31.8\% which is 4.6\% lower than the $G$-$M_{20}$ and 8.2\% higher than the $G$-$C$ classifier. The criterion from \citet{Lotz2008} has an accuracy approximately the same as the updated criteria at 60.1\%, this is 0.1\% lower than the updated $G$-$M_{20}$ classifier and 2.7\% higher than the $G$-$C$ classifier. The recall of the \citet{Lotz2008} criterion was 37.6\% which is approximately the same as the updated $G$-$M_{20}$ classifier but substantially higher, a difference of 14\%, than the $G$-$C$ classifier.

As in the case of the $G$-$M_{20}$ distribution, the mergers and non-mergers have their peak densities in approximately the same position in the $G-C$ plane. The peak density is at $G\sim0.5$ and $C\sim2.6$. The mergers which are successfully classified have typical values of $G$ and a lower $C$. The density of merger above the criterion is an order of magnitude higher compared to non-mergers, resulting in a highly precise classification.

The cuts presented in this work also trace the distribution of physical and morphological statistics not considered by the MCMC routine. We find that the cuts closely trace the distributions of stellar mass, time before merger, and Petrossian radius in the morphological space. In Appendix \ref{app:Distributions of additional parameters in morphological space} we discuss the relation between the morphological cuts and asymmetry, time after merger, mass ratio, and redshift.

\begin{figure*}
    \centering
    \includegraphics[width=8.5cm]{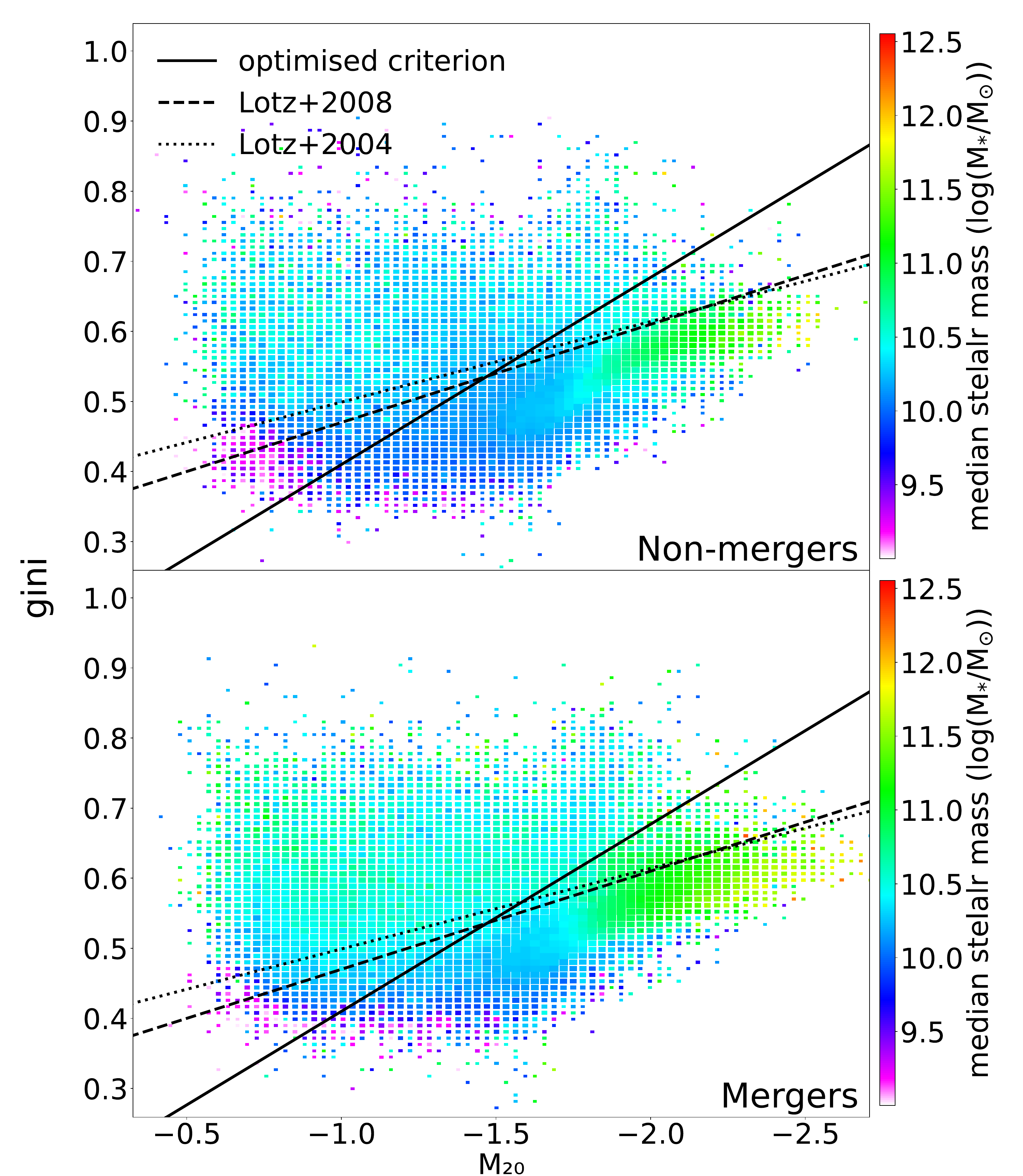}
    \includegraphics[width=8.5cm]{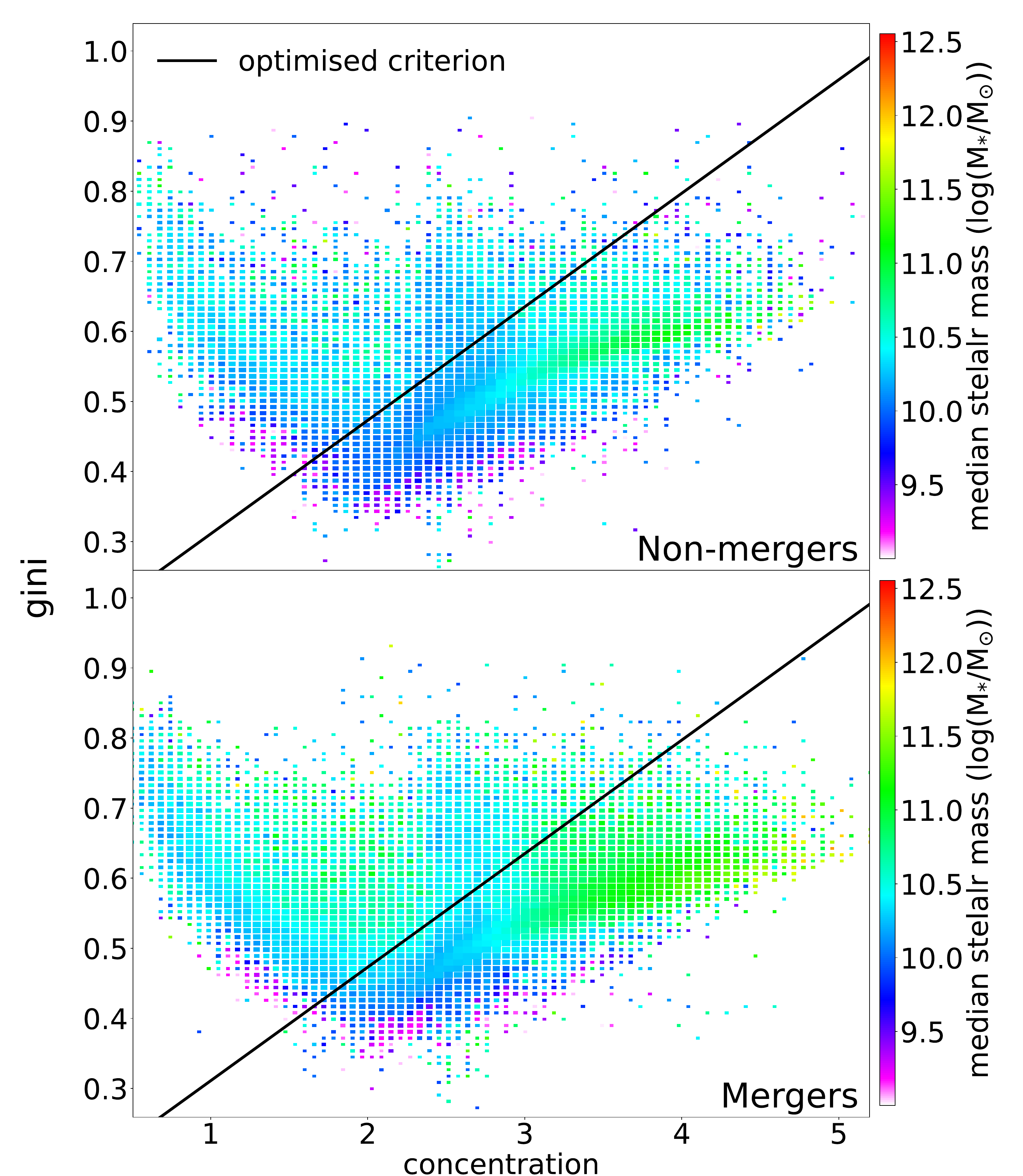}
    \caption{2D histogram of the $G-M_{20}$ (left) and $G-C$ (right) distributions of the training sample from \citetalias{Margalef-Bentabol2024}. Top panels include non-merging galaxies and the bottom panels include merging galaxies. The uniformly sized bins are displayed by rectangles scaled logarithmically by their population, the colouring shows the median stellar mass in each bin from low (in purple) to high (in red). The optimised criterion is included as the solid black line and the literature cuts from \citet{Lotz2004} and \citet{Lotz2008} are shown in dashed black lines.}
    \label{fig:morph_dist_mass}
\end{figure*}

In Figure \ref{fig:morph_dist_mass}, the distribution of stellar mass of IllustrisTNG galaxies is shown in the $G-M_{20}$ and $G-C$ planes. The stellar mass of galaxies classified as non-mergers show a bimodal distribution in both morphological planes. Galaxies with $G>0.5$ and $C>3$ or $M_{20}<-1.8$ have a median stellar mass of $\log_{10}(M_*/M_\odot)\sim11.0$ and galaxies with $G<0.5$ and $C<3$ or $M_{20}>-1.8$ have a median stellar mass of $\log_{10}(M_*/M_\odot)\sim10.0$. This bimodal distribution in stellar mass is consistent with the separation between early type and late type galaxies identified by \citet{Lotz2008}. 

The median stellar mass distribution above the $G$-$M_{20}$ merger criterion in Figure \ref{fig:morph_dist_mass} is bimodal but less pronounced compared to the distribution below the criterion. Additionally, the distribution of median stellar mass above the $G$-$M_{20}$ criterion does not follow the early type and late type classification from \citet{Lotz2008}. Above the merger criterion, two groups can be seen with a higher median stellar mass at $G>0.5$ and a lower median stellar mass at $G<0.5$. In the case of mergers, the stellar masses at $G>0.5$ reach $\log_{10}(M_*/M_\odot)\sim11$ and in the region of $G<0.5$ the median stellar mass reaches $\log_{10}(M_*/M_\odot)\sim9$. For the non-mergers the median stellar mass for $G>0.5$ reaches a lower value of $\log_{10}(M_*/M_\odot)\sim10.5$ while the median stellar mass for $G<0.5$ remains the same as for the mergers at $\log_{10}(M_*/M_\odot)\sim9$. The criteria from \citet{Lotz2004} and \citet{Lotz2008} are not able to classify mergers in this low mass region, these cuts only reach $G=0.4$ in the data range.

In the $G$-$C$ plane seen in the right panels of Figure \ref{fig:morph_dist_mass}, the galaxies above the merger criterion have a mostly homogeneous median stellar mass. Mergers in the region at $C\sim2$ and $G\sim0.5$ have a higher median stellar mass of up to $\log_{10}(M_*/M_\odot)\sim11.0$ however, this region has a comparatively low population. In Figure \ref{fig:morph_dist_main} this region has around ten times fewer galaxies in it compared to the two regions described in Section \ref{subsec: optimised merger criteria}.

\begin{figure*}
    \centering
    \includegraphics[width=8.5cm]{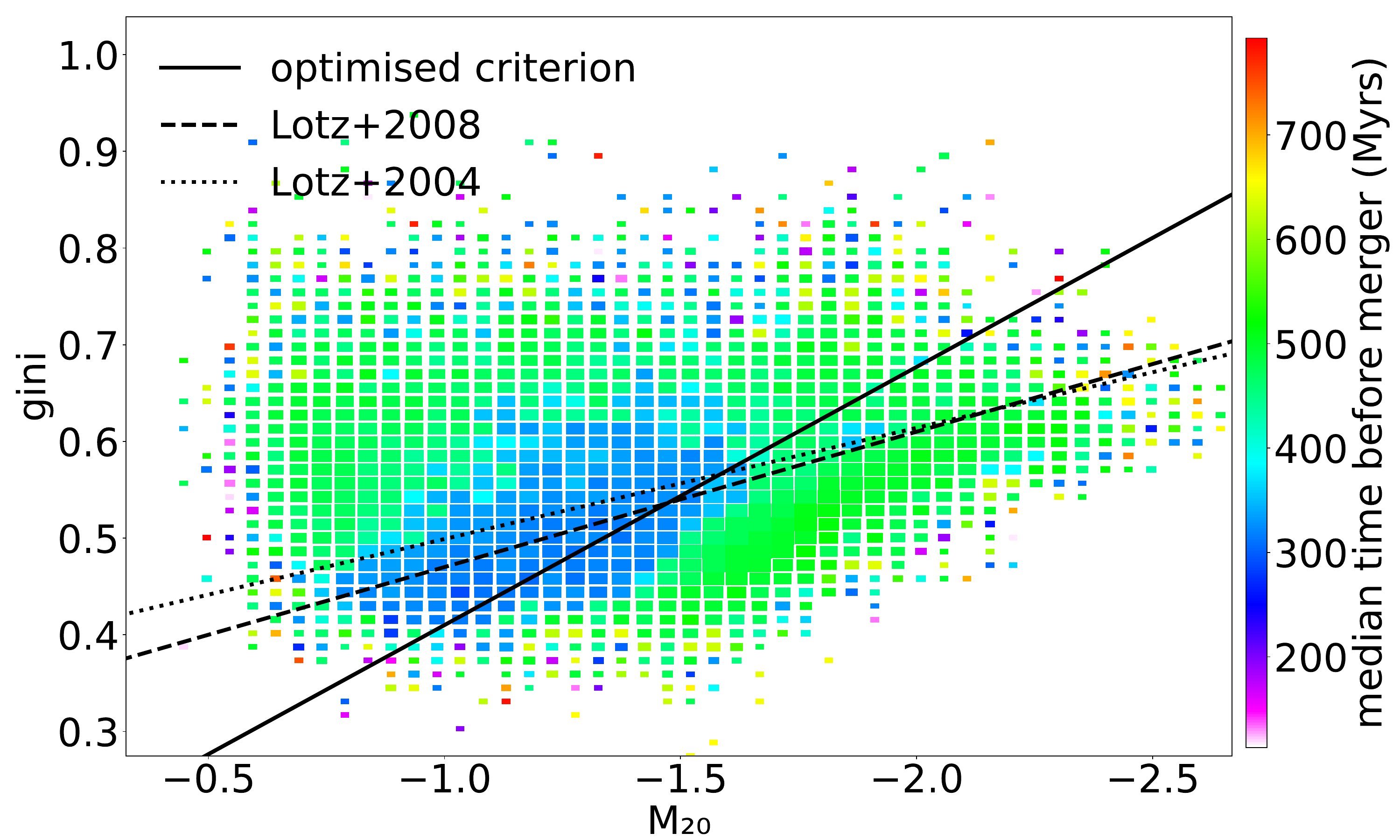}
    \includegraphics[width=8.5cm]{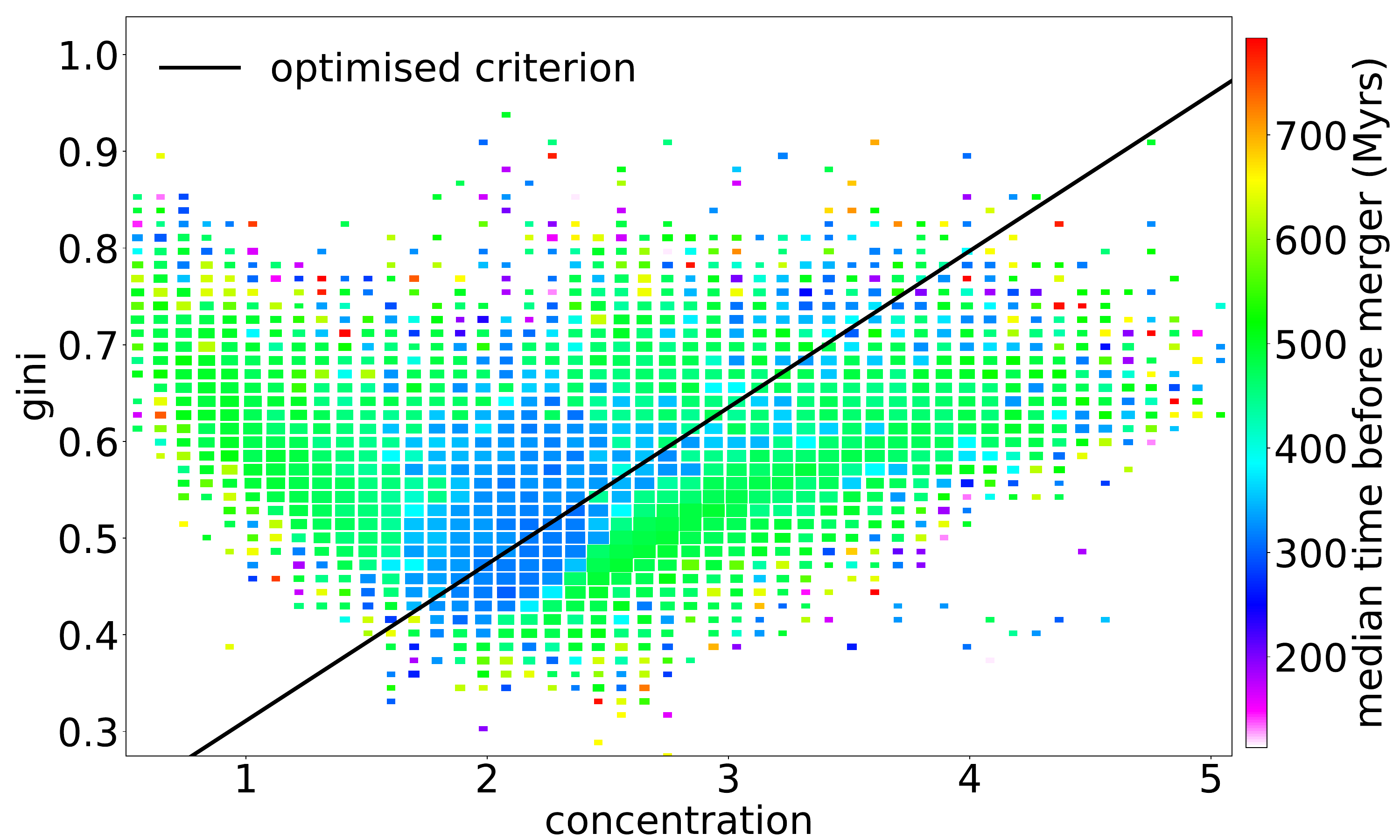}
    \caption{2D histogram of the $G-M_{20}$ (left) and $G-C$ (right) distributions of the mergers in the training sample from \citetalias{Margalef-Bentabol2024}. The uniformly sized bins are displayed by rectangles scaled logarithmically by their population, the colouring shows the median time before the next merger in each bin from low (in purple) to high (in red). The black lines represent the same criteria as in Figure \ref{fig:morph_dist_mass}.}
    \label{fig:morph_dist_time}
\end{figure*}

The distribution of time before the merger is shown for merging galaxies within 800Myrs of coalescence in Figure \ref{fig:morph_dist_time}. The merger criterion in each plane closely traces the distribution of time before the merger however the variation in the distribution is exaggerated due to the discrete time steps between the snapshots in IllustrisTNG. Both distributions include a region mainly above the merger criterion with a median time before the merger of $\sim300$Myrs and a region below the criterion (where the majority of galaxies are located) with a median of $\sim500$Myrs before the merger. This indicates that the morphology of pre-mergers only becomes distinct from non-mergers within 400Myrs of coalescence in agreement with \citet{deGraaff2025}, this is discussed further in Section \ref{subsec:Classification of subclasses}.

\begin{figure*}
    \sidecaption
        \includegraphics[width=6cm]{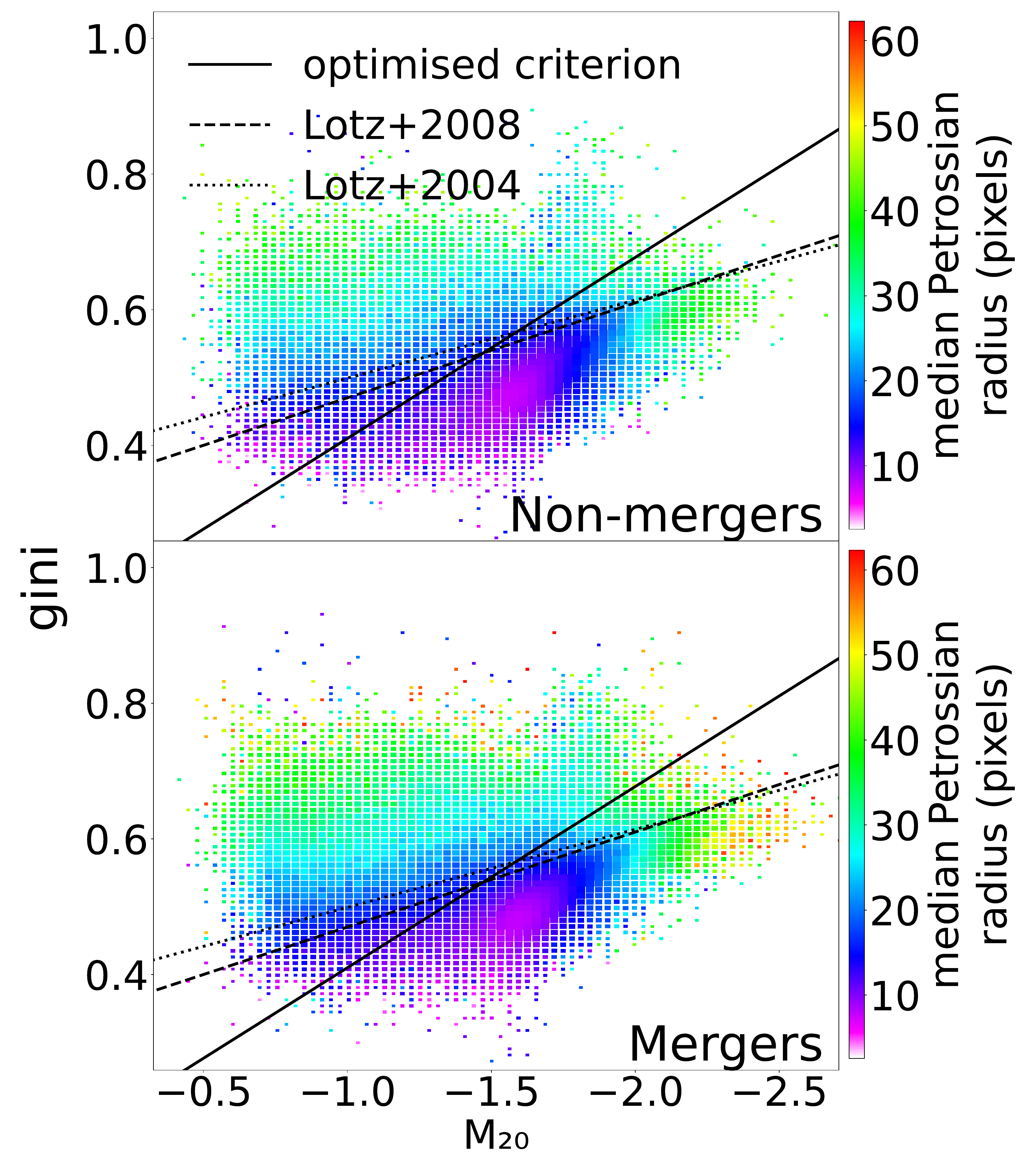}
        \includegraphics[width=6cm]{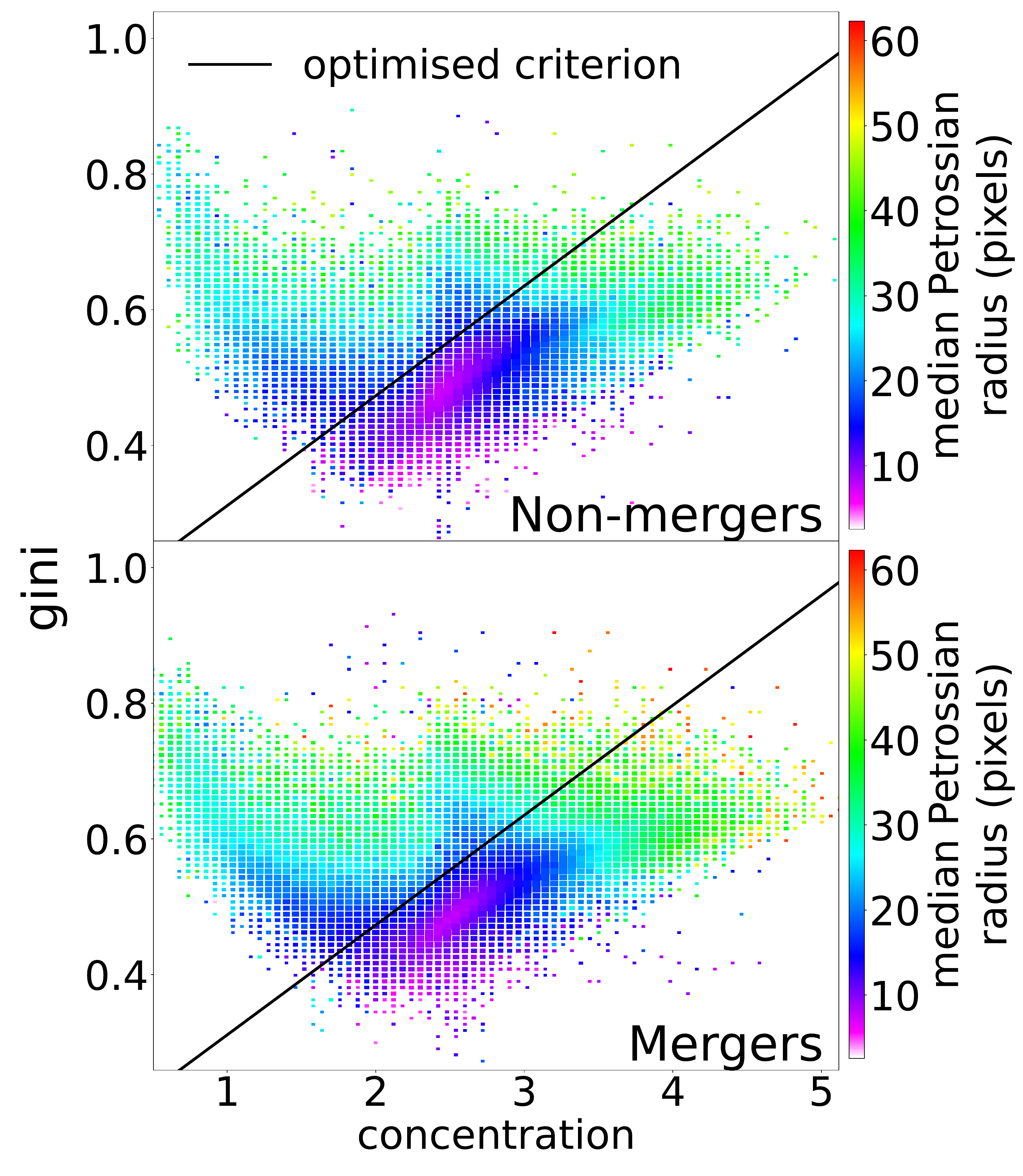}
            \caption{2D histogram of the $G-M_{20}$ (left) and $G-C$ (right) distributions of the training sample from \citetalias{Margalef-Bentabol2024}. Top panels include non-merging galaxies and the bottom panels include merging galaxies. The uniformly sized bins are displayed by rectangles scaled logarithmically by their population, the colouring shows the median stellar mass in each bin from low (in purple) to high (in red). The black lines represent the same criteria as in Figure \ref{fig:morph_dist_mass}.}
            \label{fig:morph_dist_rpetro}
\end{figure*}

In Figure \ref{fig:morph_dist_rpetro}, the distribution of the Petrossian radius \citep{Petrosian1976} of galaxies in IllustrisTNG is shown in the $G-M_{20}$ and $G-C$ planes, respectively. The Petrossian radius is included in the statistics calculated by \texttt{statmorph} and is ideal for analysis of the size of galaxies because it is robust to limiting surface brightness and signal to noise ratio \citep{Rodriguez-Gomez2019}. For galaxies below the merger criterion, the Petrossian radius does not vary strongly with $G$ so appears to correlate with the $C$ and $M_{20}$ statistics. At $G$ values above the merger criterion, the median Petrossian radius does not follow the $C$ or $M_{20}$ values; the Petrossian radius at $G\sim0.4$ reaches below 10 pixels in the $G$-$M_{20}$ plane and $\sim15$ pixels in the $G$-$C$ plane. The cuts from \citet{Lotz2004} and \citet{Lotz2008} are not able to classify galaxies with the smallest Petrossian radii, only reaching a size of $\sim15$ pixels. When calculating the Petrossian radius, an elliptical shape is assumed and so for galaxies classified as mergers this method may cause unexpected values producing the distribution seen in Figure \ref{fig:morph_dist_rpetro}.

\subsection{Classification of subclasses}\label{subsec:Classification of subclasses}
\begin{figure*}
    \sidecaption
        \includegraphics[width=6cm]{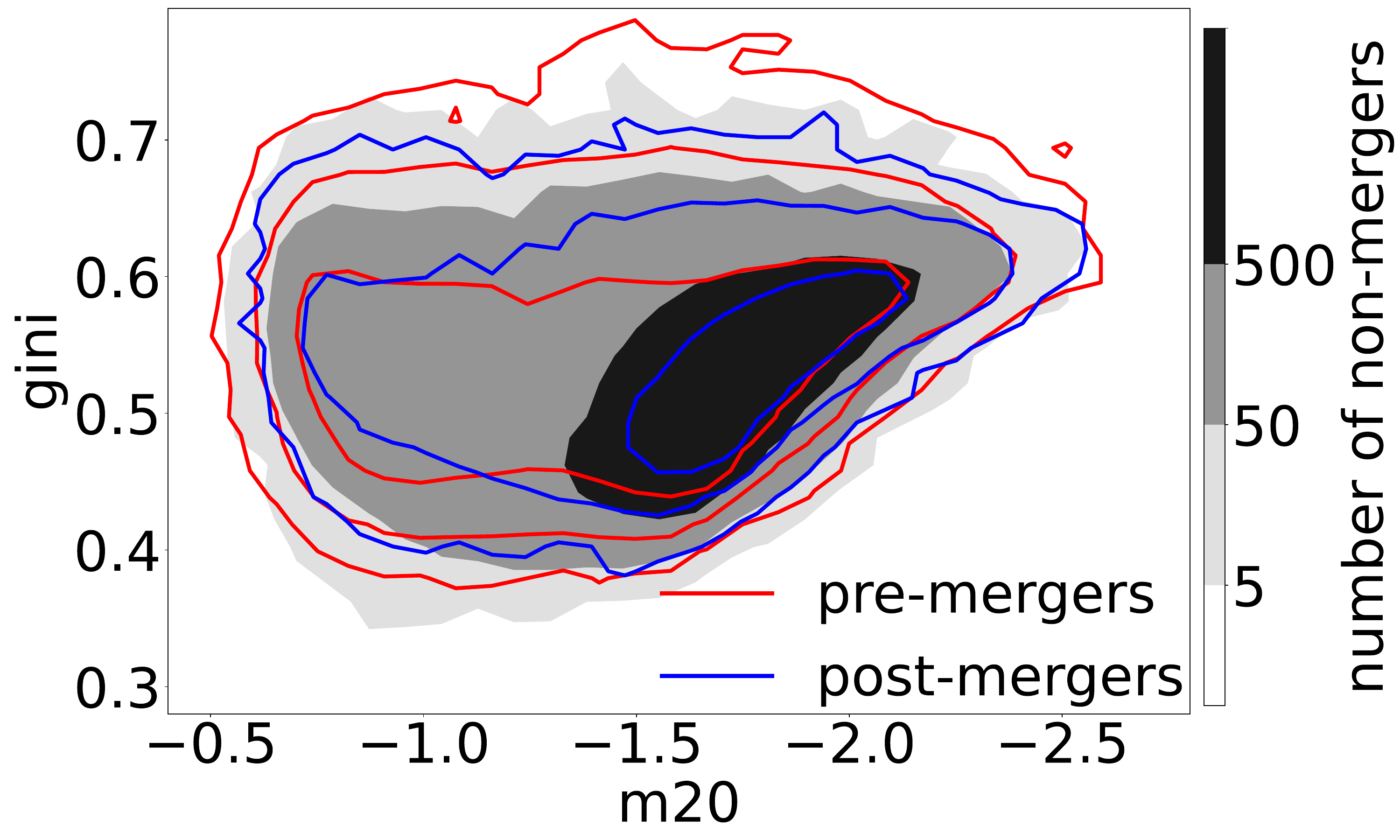}
        \includegraphics[width=6cm]{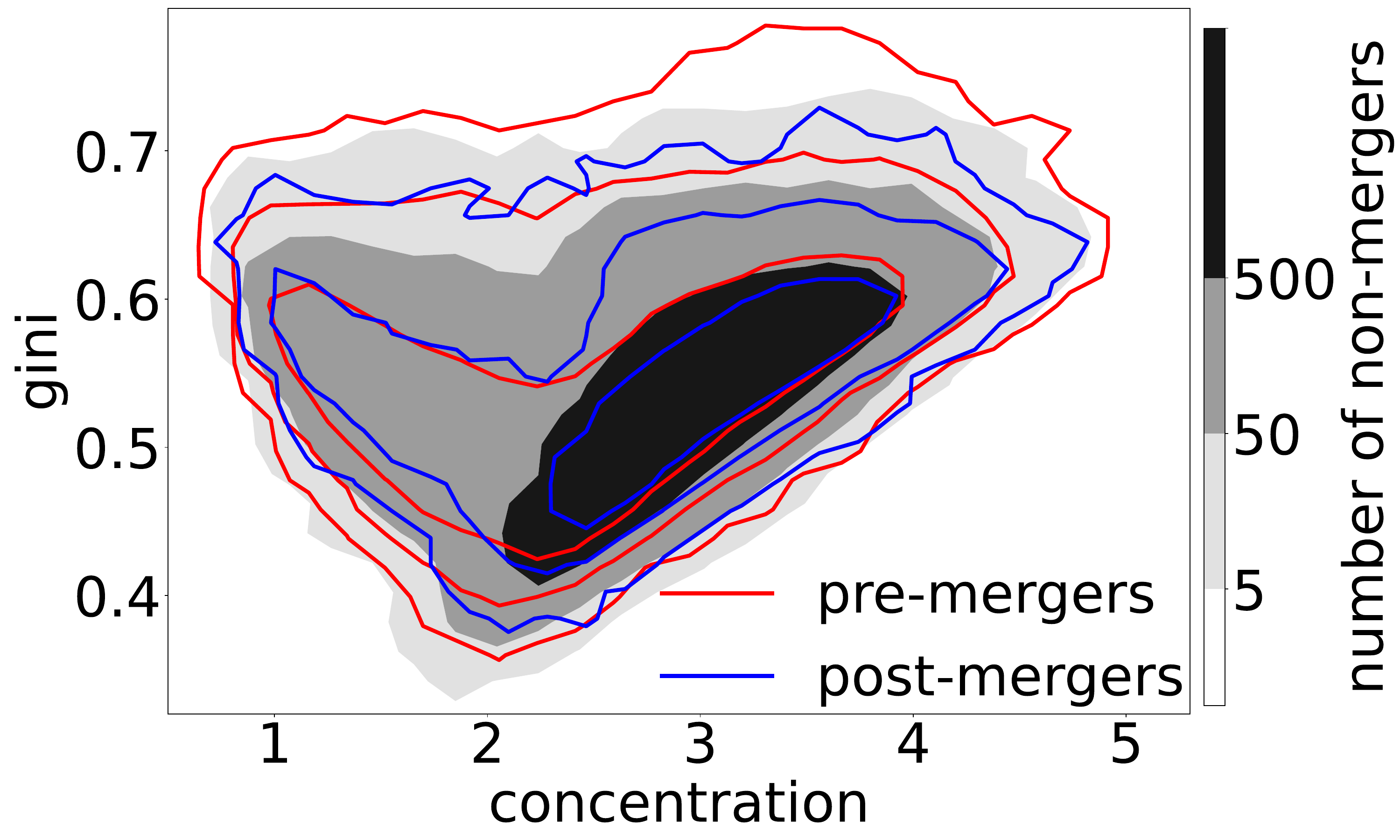}
            \caption{Contours showing the distribution of pre-mergers (in red) and post-mergers (in blue) from the training sample in the $G-M_{20}$ (left) and $G-C$ (right) planes at thresholds of 5, 50, and 500 galaxies per cell. The post-merger class is combined with the ongoing mergers. Grey contours show the distribution of non-mergers with the same thresholds.}
            \label{fig:contours}
\end{figure*}

The low recall of the morphological classifiers and their poor precision for non-mergers can be traced back to how the subclasses of mergers are distributed. The distribution of these subclasses in morphology space can be seen in the contours of Figures \ref{fig:contours}.

\begin{figure*}
    \sidecaption
        \includegraphics[width=6cm]{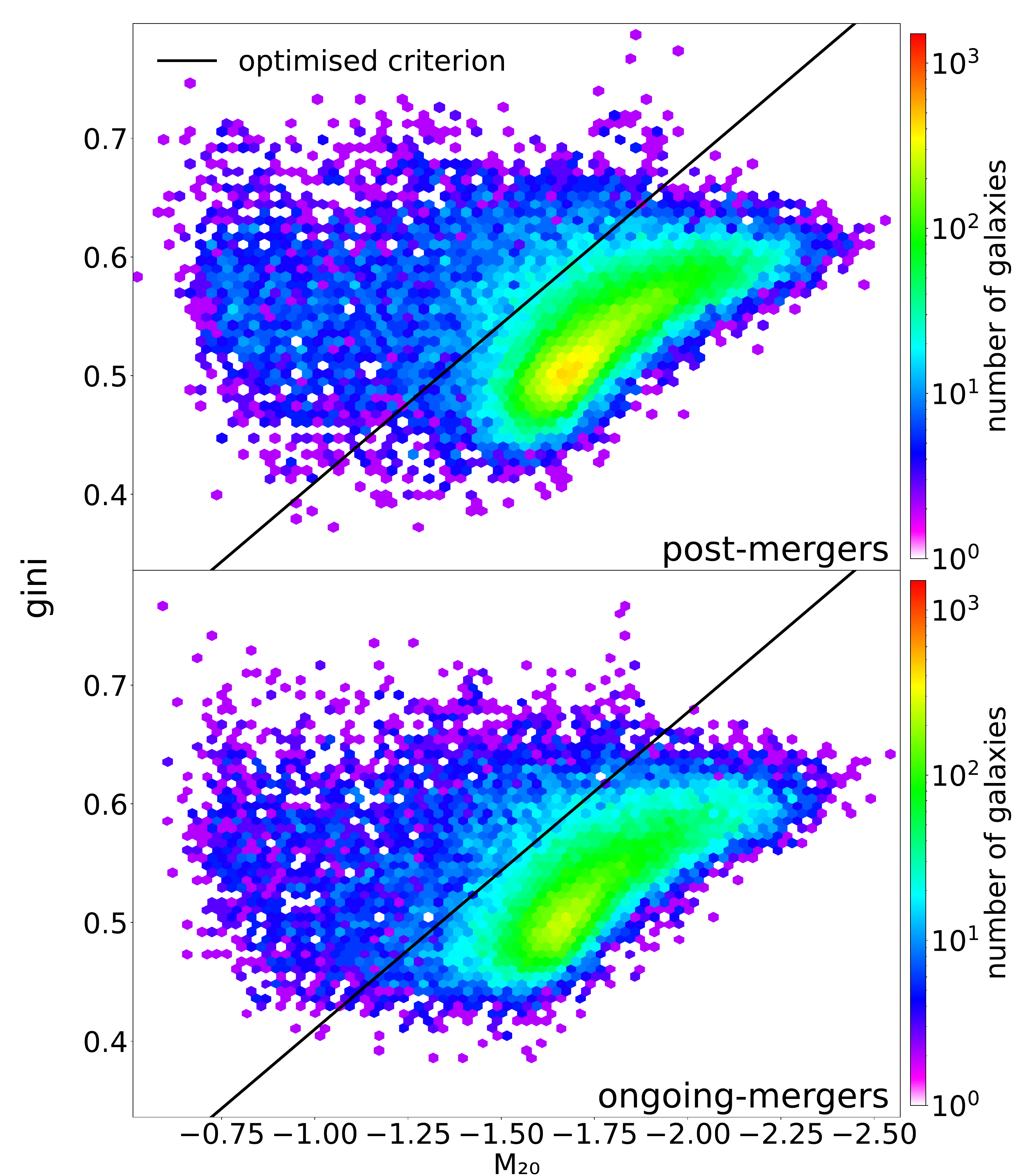}
        \includegraphics[width=6cm]{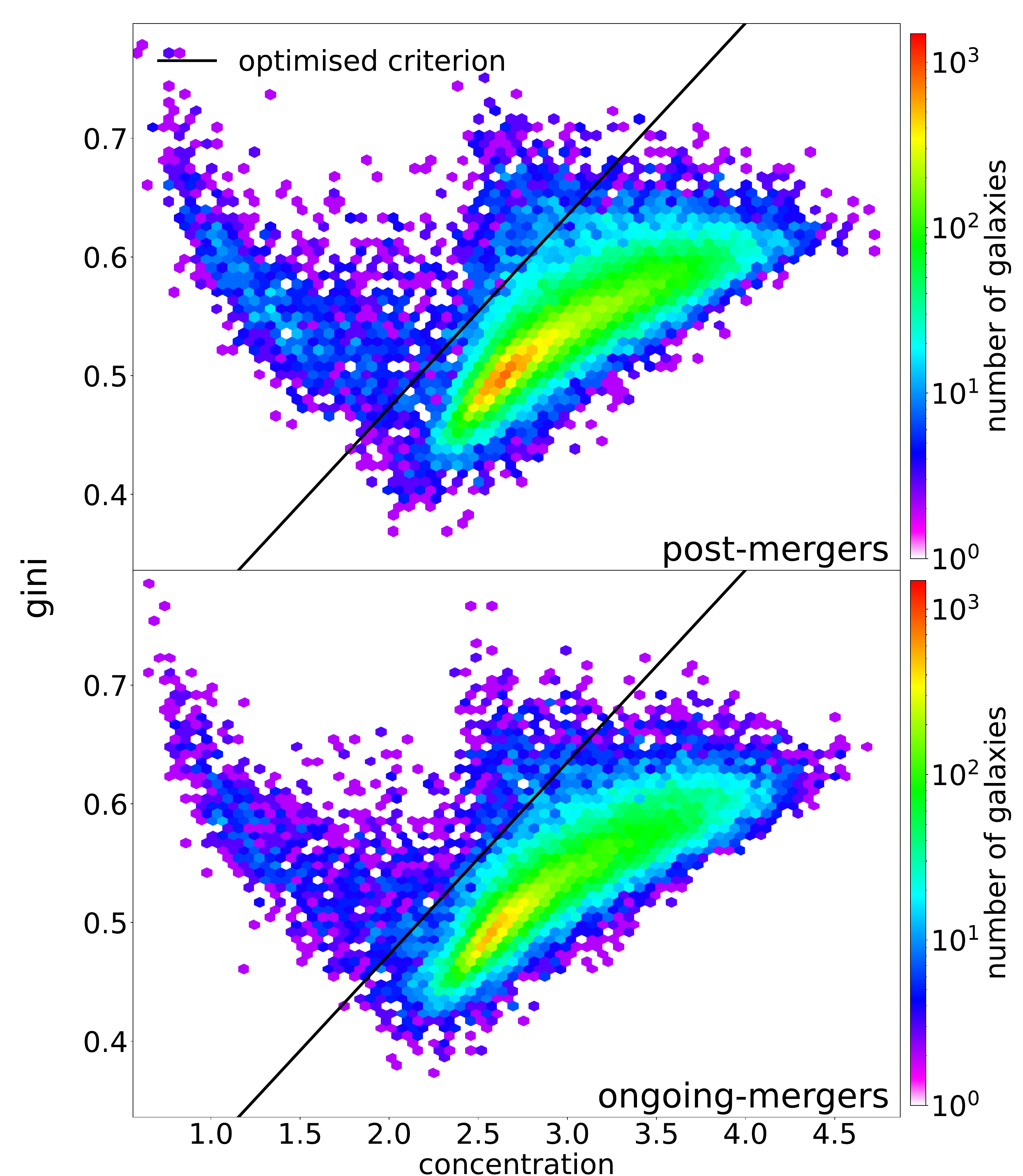}
            \caption{Distribution of the ongoing mergers (bottom panels) and post mergers (top panels) from the training sample in the $G-M_{20}$ (left) and $G-C$ (right) planes. Overall, 83.0\% of the ongoing mergers and post-mergers are below the $G-M_{20}$ merger criterion and 91.6\% are below the $G$-$C$ merger criterion. The colouring shows the population in each bin from low (in purple) to high (in red).}
            \label{fig:morph_dist_ongoing-post}
\end{figure*}

In both the $G$-$C$ and $G$-$M_{20}$ distributions, the ongoing and post-mergers occupy the same region as demonstrated with the training sample in Figures \ref{fig:morph_dist_ongoing-post}. Post-mergers show a slightly more extended distribution above the merger criterion but over 90\% of the ongoing mergers and post mergers are found below this criterion with a similar distribution which reach their maximum density in the same region at $G\sim0.5$, $M_{20}\sim1.7$, and $C\sim2.6$. This region is at the centre of the non-merger distribution in both cases and so ongoing and post-mergers cannot be distinguished from non-mergers using their $C$, $M_{20}$, and $G$ statistics. This is one cause of the poor recall in identifying mergers with this methods.

Only a small part of the pre-merger distribution is separated from the non-mergers. In both cases, these pre-mergers have typical $G$ but the $C$ statistic and the absolute value of $M_{20}$ are much lower than a normal galaxy. The $M_{20}$ statistic traces the bright features of a galaxy, a more negative value indicates that these bright regions are less centralised which is consistent with features such as a clumpy disk or tidal tails. A low $C$ value indicates a more extended central bright region consistent with tidal distortions.

\begin{figure}
    \centering
    \resizebox{\hsize}{!}{\includegraphics{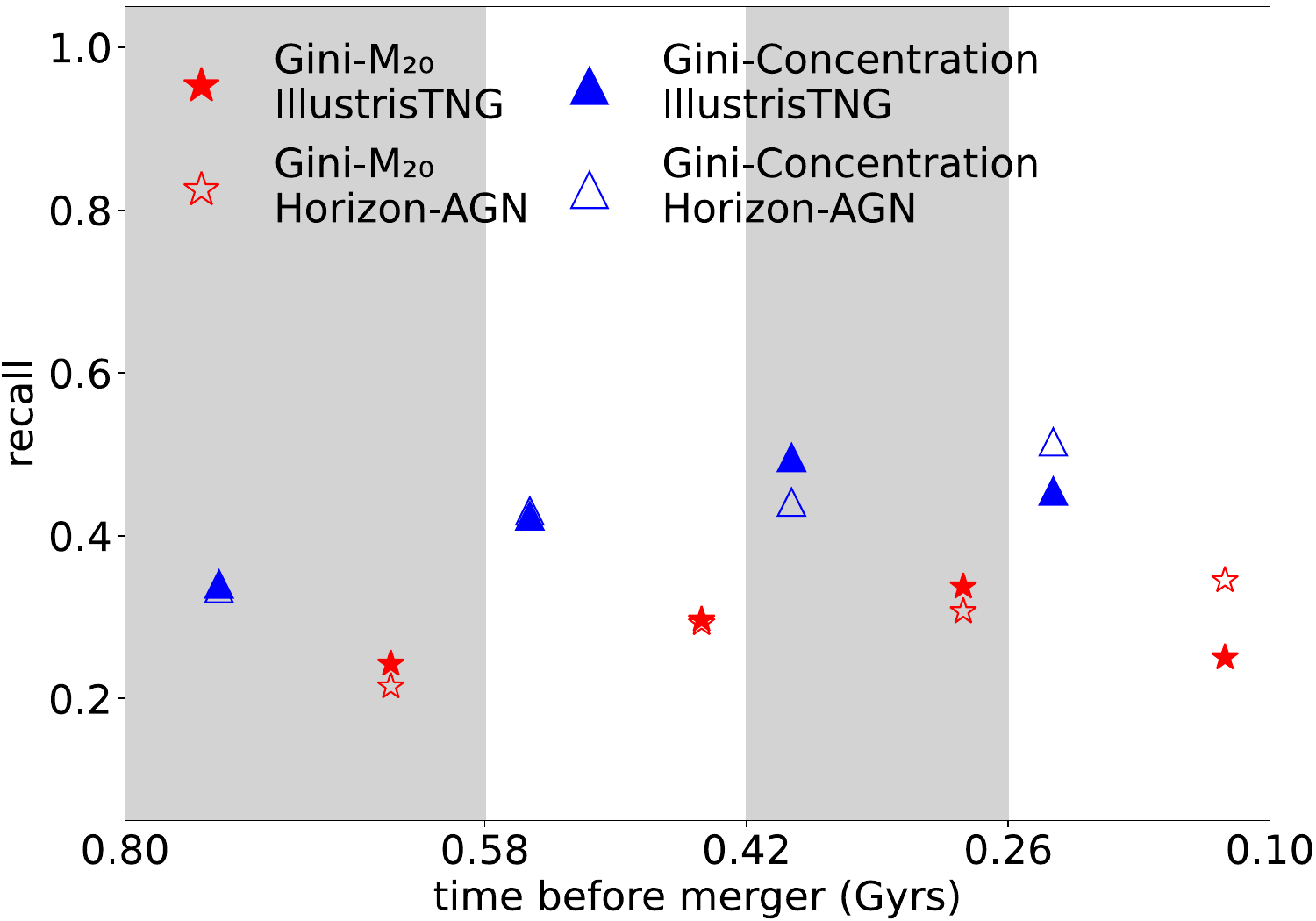}}
    \caption{Recall of the morphological classifiers as a function of time before coalescence. The $G$-$M_{20}$ classifications are shown with red stars and the $G$-$C$ classifications are shown with blue triangles. The filled symbols represent the recall when applied to the testing dataset and the empty symbols represent the recall when applied to the Horizon-AGN mock images.}
    \label{fig:t evolve}
\end{figure}

The pre-merger images can further be binned by the time before coalescence using the bins 0.8Gyrs < t$_1$ < 0.58Gyrs, 0.58Gyrs < t$_2$ < 0.42Gyrs, 0.42Gyrs < t$_3$ < 0.26Gyrs, and 0.26Gyrs < t$_4$ < 0.1Gyrs. These bins were chosen to reflect the time resolution of the IllustrisTNG simulation. Figure \ref{fig:t evolve} shows that the recall of the morphological classifiers on the testing dataset increases up to $\sim48\%$ at $\sim0.4$Gyrs before coalescence after which it is constant. Recall traces the completeness of the classification and so after 0.4Gyrs the classifier is identifying as many mergers as possible, this reinforces the result from \citet{deGraaff2025} which showed that 0.4Gyrs is the optimal time to define pre-mergers for mock IllustrisTNG images. However for Horizon-AGN, the recall of the morphological classifiers continues to increase throughout the four time bins and so the 0.4Gyrs definition may not be universal. 

\subsection{Extending the merger challenge}\label{subsec:Extending the merger challenge}
The updated cuts presented in this work are valuable for drawing comparisons to the ML methods in \citetalias{Margalef-Bentabol2024} because they are optimised and evaluated with the same data as was used to train and test the ML methods. The morphological classifiers using $G$, $M_{20}$ statistic, and $C$ achieve similar precisions to the ML methods tested in \citetalias{Margalef-Bentabol2024} when applied to the testing data from IllustrisTNG. The precisions of the \citetalias{Margalef-Bentabol2024} models range in $69.1\%<P<80.1\%$ with a median value of 72.65\%. Thus, both morphological classifiers achieve precisions within the range of the ML models with both the $G$-$M_{20}$ and $G$-$C$ classifiers close to the median at 69.5\% and 72.3\%, respectively. The full comparative results are included in Table \ref{tab:results}. Therefore, for the task of producing high purity merger samples morphological classifiers are well suited since they produce similarly pure merger samples to ML methods, while being much simpler to implement.

\begin{table*}
    \caption{\label{tab:results}Evaluation metrics of the three datasets using each merger classifier from this work (top) and \citetalias{Margalef-Bentabol2024} (bottom).}
    \centering
    \begin{tabular}{l|cccc|cccc|cccc}
        \hline\hline
        \rule{0pt}{2.5ex}\multirow{2}{4em}{Method} & \multicolumn{4}{c|}{IllustrisTNG} & \multicolumn{4}{c|}{Horizon-AGN} & \multicolumn{4}{c}{HSC-SSP} \\
         & A & P & R & F1 & A & P & R & F1 & A & P & R & F1 \\
        \hline
        \rule{0pt}{2.5ex}$G$-$M_{20}$ & 60.2 & 69.5 & 36.4 & 47.8 & 57.8 & 63.5 & 36.7 & 46.5 & 72.6 & 99.2 & 45.5 & 62.4 \\
        $G$-$C$ & 57.4 & 72.3 & 23.6 & 35.9 & 55.1 & 68.7 & 18.5 & 29.2 & 62.6 & \textbf{100.0} & 25.1 & 40.2 \\
        \hline
        \rule{0pt}{2.5ex}Method-1 (RF) & 70.5 & 71.6 & 68.1 & 69.8 & 57.7 & 71.2 & 26.5 & 38.7 & 74.4 & 74.1 & \textbf{74.9} & 74.5 \\
        Method-2 (Swin) & 72.6 & 69.1 & \textbf{81.9} & 74.9 & \textbf{62.5} & 68.4 & 46.6 & \textbf{55.4} & 71.3 & 71.4 & 71.2 & 71.3 \\
        Method-3 (Zoobot) & \textbf{77.7} & \textbf{80.1} & 73.8 & \textbf{76.8} & 57.6 & \textbf{72.0} & 24.8 & 36.9 & 74.1 & 84.6 & 58.8 & 69.4 \\
        Method-4 (CNN1) & 74.1 & 76.6 & 69.5 & 72.8 & 57.1 & 71.7 & 23.5 & 35.4 & \textbf{75.5} & 83.0 & 64.2 & \textbf{72.4} \\
        Method-5 (CNN2) & 70.6 & 72.0 & 67.2 & 69.5 & 61.2 & 65.4 & \textbf{47.3} & 54.9 & 69.1 & 74.4 & 58.4 & 65.4 \\
        Method-6 (CNN3) & 71.5 & 73.3 & 67.6 & 70.4 & 53.2 & 67.5 & 12.4 & 21.0 & 72.6 & 79.9 & 60.5 & 68.9 \\
        \hline
    \end{tabular}
    \tablefoot{Left column includes results for the testing data from IllustrisTNG, centre column is for the results of classifying the Horizon-AGN mock images, and the right most column is for the HSC-SSP observations. Sub-columns include the accuracy (A), precision (P), recall (R), and F1-score (F1). The highest value in each metric is shown in bold for each dataset.}
\end{table*}

In all other metrics, the morphological classifiers cannot match the performance of the ML techniques in \citetalias{Margalef-Bentabol2024} when applied to the same testing dataset. In terms of recall, accuracy and F-1 score the two morphological classifiers presented in this work perform worse than all six models in \citetalias{Margalef-Bentabol2024}. The $G$-$M_{20}$ classifier is 11.85\% less accurate than the median accuracy of the ML models (the median is 72.05\%) and the $G$-$C$ is 14.65\% less accurate than the median accuracy. In terms of recall, the median performance of the ML methods from \citetalias{Margalef-Bentabol2024} is 68.8\% which is 29.4\% higher than that of the $G$-$M_{20}$ classifier and 45.2\% higher than the recall of the $G$-$C$ classifier. The median F-1 score for the ML models is 71.2\% which is 23.8\% higher than the F-1 score of the $G$-$M_{20}$ classifier and 35.7\% higher than that of the $G$-$C$ classifier. Therefore, for tasks requiring a highly complete sample of mergers, morphological classifiers are poorly suited.

Additionally, the morphological classifiers are ineffective at classifying non-mergers in the testing data. This can be seen in Figures \ref{fig:mcmc_conf} and \ref{fig:Lotz gm20 conf} where the number of correctly classified non-mergers does not exceed 60\% of the total number of classified non-mergers for either the optimised criteria or the criteria from \citet{Lotz2004} and \citet{Lotz2008}.

The RF classifier employed in \citetalias{Margalef-Bentabol2024} is most similar to the morphological classifiers in this work. The $G$-$M_{20}$ and $G$-$C$ classifiers achieved a similar precision for the testing data than the RF, with differences of 2.1\% and 0.7\%, respectively, but their recalls are much lower than the RF causing the accuracies to be lower by 10.3\% in the case of the $G$-$M_{20}$ classifier and 13.1\% for the $G$-$C$ classifier. The RF had a higher accuracy , recall, and F-1 score because it is able to consider more morphological statistics, but we note that a larger morphology space does not result in a proportionally better classifier especially in terms of precision.

When the classifiers are applied to the Horizon-AGN dataset, the performance of the morphological classifiers across all metrics become consistent with the ML models in \citetalias{Margalef-Bentabol2024}. The accuracy, precision, recall and F-1 score of the two morphological classifiers can be seen alongside the ML methods in the second section of Table \ref{tab:results}. When transferred to the Horizon-AGN mock images, the accuracies of the \citetalias{Margalef-Bentabol2024} models have a range of $53.2\%<A<62.5\%$, the precisions have a range of $65.4\%<P<72.0\%$, the recalls have a range of $12.4\%<R<49.3$, and the F-1 scores range in $21.0<F1<55.4\%$. In each metric, both the $G$-$M_{20}$ and $G$-$C$ classifiers fall into these ranges with accuracies of 57.8\% and 55.1\%, precisions of 63.5\% and 68.1\%, recalls of 36.7\% and 18.5\%, and F-1 scores of 46.5\% and 29.2\%. Additionally, the $G$-$M_{20}$ classifier reaches an accuracy, recall, and F-1 score higher than the median for the ML methods; the median accuracy from \citetalias{Margalef-Bentabol2024} on the Horizon-AGN dataset was 57.65\% which is 0.15\% lower than the $G$-$M_{20}$ accuracy, the median recall was 25.65\% which the $G$-$M_{20}$ classifier exceeds by 11.05\%, and the median F-1 score was 37.8\% which is 8.7\% lower than the F-1 score of the $G$-$M_{20}$ classifier. Despite performing better compared to the ML classifiers, the morphological classifiers maintain similar performance metrics, which indicates they are more robust than the ML models.

The final section of Table \ref{tab:results} shows the performance metrics of all eight classifiers form this work and from \citetalias{Margalef-Bentabol2024} when used on the HSC-SSP observations. Both of the morphological classifiers classify the HSC-SSP sample more precisely than any ML method presented in \citetalias{Margalef-Bentabol2024} with precisions of 98.3\% for $G$-$M_{20}$ and 96.6\% for the $G$-$C$ classifier. The $G$-$M_{20}$ classifier achieves the best overall accuracy at 79.2\%. Both morphological classifiers reach precision $>95\%$ despite having recalls $<50\%$, this suggests that there is a bias in the visual classifications which provide the ground truths for this sample. The bias originates from the addition of the $\sim1000$ clear non-mergers in \citetalias{Margalef-Bentabol2024}. The high precision values indicates that almost no non-mergers occupied the merger region of the $G$-$M_{20}$ and $G$-$C$ planes while a low recall indicates that mergers were spread more regularly across the morphology space as in Figure \ref{fig:morph_dist_main}. These results demonstrate that the morphological classifiers recover the visually classified HSC-SSP sample well but are not indicative of the true performance of these methods.

The high precision of these simple morphological classifiers across all three datasets indicates that high performing, high efficiency merger classifiers are possible. The performance of the morphological classifiers is consistent with ML methods when transferred to new data showing that simple classifiers are as effective for finding mergers in new data as the state-of-the-art ML models.

\subsection{Reliability of the morphological classifiers}\label{subsec:Reliability of the morphological classifiers}
A reliable merger classifier should perform at a high level for all redshifts and stellar masses. Such a classifier would be able to produce complete and unbiased merger catalogues. To further test the reliability, the redshift and mass dependent performance was found for the Horizon-AGN and HSC-SSP images. Since the training data was exclusively from IllustrisTNG, testing on Horizon-AGN and HSC-SSP shows how the performance changes when the classifier is applied to previously unseen types of data.

Figure \ref{fig:z evolve} shows that as the simulated images go to higher redshifts, the precision decreases. A decrease in precision of 8.5\% for the $G$-$M_{20}$ classifier and 12.7\% $G$-$C$ is seen from $z=0.1$ to $z=1.0$ which is in line with the trend seen for the ML methods in \citetalias{Margalef-Bentabol2024}. Such a similar drop in precision shows that this is a limitation from the imaging rather than the classifiers themselves, this is reinforced by the precision in each bin being similar between IllustrisTNG and Horizon-AGN. 

On the real images from HSC-SSP (shown in Figure \ref{fig:z evolve}), the precision is close to, or exactly, 100\% for both classifiers and all redshift bins. This is caused by the bias in the dataset which is discussed in Section \ref{subsec:Extending the merger challenge}.

The recall in Figure \ref{fig:z evolve} remains roughly constant across the redshift bins and between simulations in contrast to the ML methods in \citetalias{Margalef-Bentabol2024}. Across both simulations, the $G$-$M_{20}$ classifier varies in recall by only 7.6\% and the $G$-$C$ classifier varies by a recall of 8.4\%. In all bins the recall is similar between Horizon-AGN and IllustrisTNG showing that the performance remains consistent on previously unseen types of data. The recall in \citetalias{Margalef-Bentabol2024} decreases by between 7.8$\%$ and 57.6$\%$ when the models are tested on Horizon-AGN. This drop is not seen in the morphological classifiers, when applied to Horizon-AGN. In Figure \ref{fig:z evolve}, the recall for HSC-SSP increases with redshift in a similar pattern to the methods in \citetalias{Margalef-Bentabol2024} but at a substantially lower recall.

The precision of the morphological classifiers in Figure \ref{fig:z evolve} is slightly higher in each redshift bin when applied to the Horizon-AGN mock images compared to the IllustrisTNG mock images. In every redshift bin and for both morphological classifiers the Horizon-AGN precisions are higher than the IllustrisTNG results, despite this the overall precision is lower by 6\% for the $G$-$M_{20}$ classifier and 3.6\% for $G$-$C$. After $z=0.52$, the precision of ML methods from \citetalias{Margalef-Bentabol2024} on the Horizon-AGN dataset become inconsistent with that of IllustrisTNG; the best Ml method in the $0.52<z<0.76$ bin has a precision of 67.9\% on the Horizon-AGN mock images which is the same as the worst precision on the IllustrisTNG mock images, in the highest redshift bin the difference then reaches 6.8\%. This divergence does not occur for the morphological classifiers and the precisions in fact converge as redshift increases. The difference between the precisions on the IllustrisTNG and Horizon-AGN mock images for the $G$-$M_{20}$ classifier decreases from 11.7\% in the lowest redshift bin to 0.6\% in the highest redshift bin. For the $G$-$C$ classifier, the difference between the precisions on the IllustrisTNG and Horizon-AGN mock images decreases from 9\% in the lowest redshift bin to 3.2\% in the highest redshift bin.

\begin{figure}
    \centering
    \resizebox{\hsize}{!}{\includegraphics{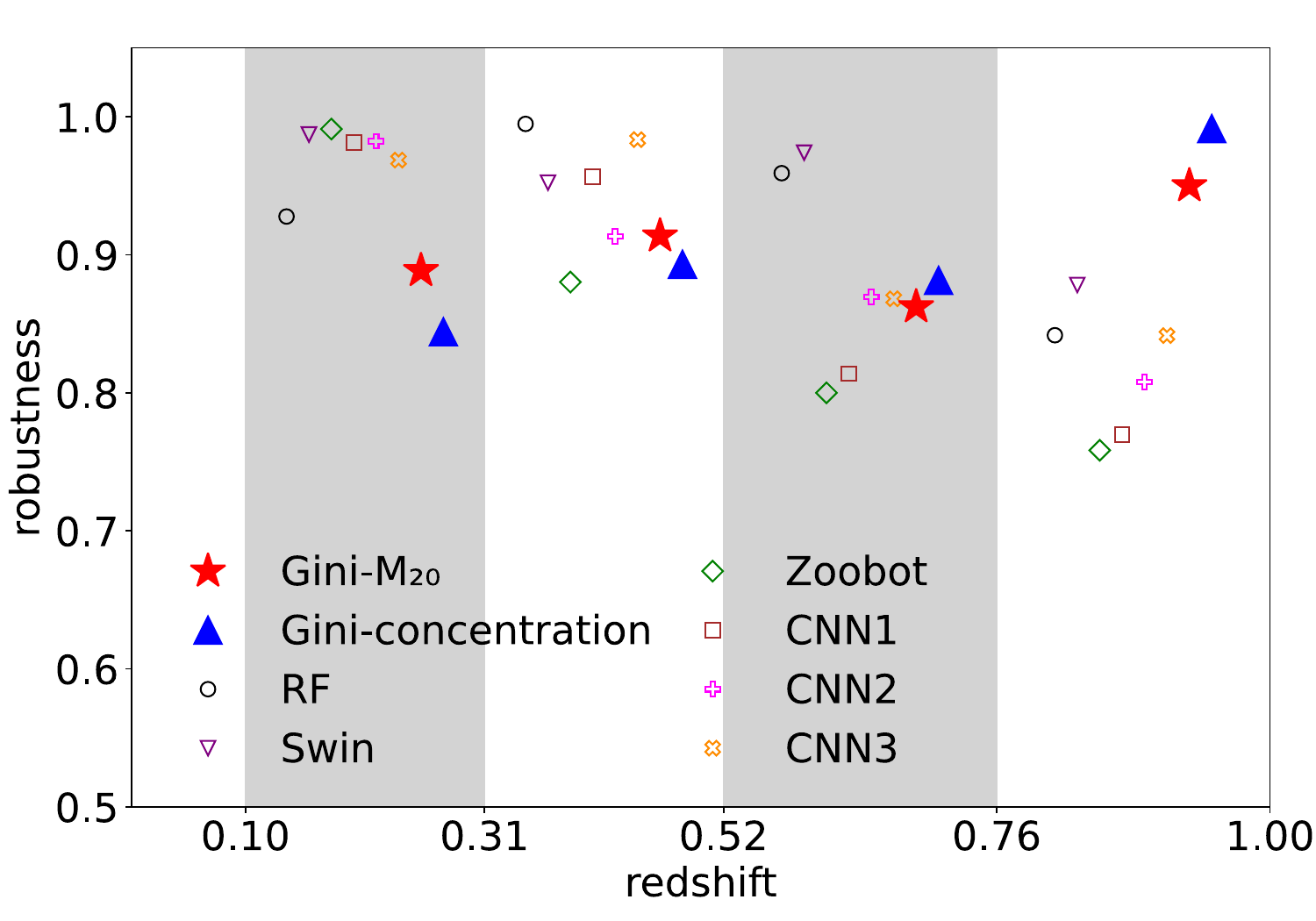}}
    \caption{Robustness of the $G$-$M_{20}$ and $G$-$C$ classifiers (in red and blue, respectively) compared to the 6 ML classifiers from \citetalias{Margalef-Bentabol2024}. Robustness is calculated from the precision on the testing data and the Horizon-AGN data.}
    \label{fig:robustness}
\end{figure}

To evaluate how well the morphological classifiers transfer to new data, we define a robustness metric as
\begin{equation}
    \rho = 1 - \Bigg|1 - \frac{P'}{P}\Bigg|
    \label{eq:robustness},
\end{equation}
where P' is the precision on a new dataset and P is the precision on the testing dataset. This robustness metric measures the change in the purity of a merger sample produced by a given classifier. Robustness was calculated for both morphological classifiers across each redshift bins. The robustness of the classifiers in \citetalias{Margalef-Bentabol2024} were also calculated and are compared to the morphological classifiers in Figure \ref{fig:robustness}.

\begin{table}
    \caption{\label{tab:robustness}Robustness of the morphological classifiers (top) and ML classifiers from \citetalias{Margalef-Bentabol2024} (bottom).}
    \centering
    \begin{tabular}{l|cc}
        \hline\hline
        \rule{0pt}{2.5ex}\multirow{2}{4em}{Method} & \multicolumn{2}{c}{$\rho$} \\
         & Horizon-AGN & HSC-SSP\\
        \hline
        \rule{0pt}{2.5ex}$G$-$M_{20}$ & 91.4 & 57.3 \\
        $G$-$C$ & 95.0 & 61.7 \\
        \hline
        \rule{0pt}{2.5ex}Method-1 (RF) & \textbf{99.4} & 96.5 \\
        Method-2 (Swin) & 99.0 & 96.7 \\
        Method-3 (Zoobot) & 89.9 & \textbf{98.1} \\
        Method-4 (CNN1) & 93.6 & 91.6 \\
        Method-5 (CNN2) & 90.8 & 96.7 \\
        Method-6 (CNN3) & 92.1 & 91.0 \\
        \hline
    \end{tabular}
    \tablefoot{For each method, the robustness is shown in terms of the precision on the Horizon-AGN dataset (column two) and the HSC-SSP dataset (column three) compared to the test dataset. The highest robustness for each dataset is highlighted in bold.}
\end{table}

Table \ref{tab:robustness} shows the robustness values for the two classifiers presented in this work and the six methods from \citetalias{Margalef-Bentabol2024}. Since the HSC-SSP classification is unrepresentative for the morphological classifiers, only the robustness based on the Horizon-AGN dataset is considered in Figure \ref{fig:robustness}. The RF from \citetalias{Margalef-Bentabol2024} has the best robustness at 99.4\%, closely followed by Swin transformer at 99.0\%. The range of robustness for the ML methods is $89.9\%<\rho<99.4\%$ with a median of 92.85\% and so the morphological classifiers, with robustness of 91.4\% for $G$-$M_{20}$ and 95.0\% for $G$-$C$, are consistent with the ML methods. In Figure \ref{fig:robustness}, the dependency of robustness on redshift is plotted. The methods follow two paths in Figure \ref{fig:robustness} with the morphological classifiers, the RF classifier, and the Swin transformer maintaining a high robustness up to $z\sim0.76$, while Zoobot and the three CNN methods decrease in robustness. The morphological classifiers have the highest robustness above $z\sim0.76$. 

Figure \ref{fig:m evolve} shows that the morphological classifiers increase in precision when applied to the testing data across the entire mass range. An increase of 26.2\% is seen for the precision of the $G$-$M_{20}$ classifier in the range $9.0<\log_{10}(M_*/M_\odot)<12.5$, in the same range the $G$-$C$ classifier increases by 23.6\% in precision. In the case of the Horizon-AGN mock images, the precisions increases over the same mass range by 8.9\% and 6.2\% for the $G$-$M_{20}$ and $G$-$C$ classifiers, respectively; both classifiers see their maximum precision at lower masses however with $G$-$M_{20}$ peaking at 89.2\% for $\log_{10}(M_*/M_\odot)=10.75$ and $G$-$C$ peaking at 91.7\% for $\log_{10}(M_*/M_\odot)=11.25$. This increase over the mass range is due to more massive galaxies have more easily resolved morphologies allowing the most massive mergers to be more precisely selected. This is the opposite trend to the the ML models in \citetalias{Margalef-Bentabol2024} which decrease in precision above a stellar mass of $\log_{10}(M_*/M_\odot)\sim11.0$. The increase also exceeds the range of the ML models in all three datasets as illustrated in the left panel of Figure \ref{fig:m evolve} and so this difference in trend is significant. Once again, the precisions from HSC-SSP are too biased to make any meaningful conclusions, as seen in Figure \ref{fig:m evolve}, where the precision is 100\% in all bins except the most massive in which no classifications were made.

When the morphological classifiers are applied to the IllustrisTNG images the recall, shown in Figure \ref{fig:m evolve}, remains approximately constant with increasing stellar mass. For the $G$-$M_{20}$ classifier we see a variation of 3.2\% over the mass range and for $G$-$C$ the recall varies by 5.5\%. However, in the case of Horizon-AGN and HSC-SSP, the recall initially increases before decreasing at large stellar masses. The $G$-$M_{20}$ classifier reaches its highest recall for the Horizon-AGN mock images at $\log_{10}(M_*/M_\odot)=10.75$ with a recall of 39.1\%, over the entire mass range the recall remains approximately constant with a decrease of 2\%. Also on the Horizon-AGN mock images, the $G$-$C$ classifier reaches a maximum recall of 25.7\% at a mass of $\log_{10}(M_*/M_\odot)=11.25$, over the entire mass range the recall remains approximately constant with a decrease of by 2.8\%. When used to classify the HSC-SSP images, the recall of the morphological classifiers remains approximately constant, at $\sim50\%$ for $G$-$M_{20}$ and $\sim30\%$ for $G$-$C$, until a mass of $\log_{10}(M_*/M_\odot)=10.5$ after which both classifiers decrease steadily to a recall of zero in the highest mass bin. This is in contrast to the ML methods in \citetalias{Margalef-Bentabol2024}, which all reveal an increase in terms of recall at higher stellar masses. As with the redshift dependence, the recall binned by stellar mass is more consistent between IllustrisTNG and Horizon-AGN for the morphological classifiers compared to the ML classifiers in \citetalias{Margalef-Bentabol2024}, which reinforces that the methods shown in this work are more robust to new data.

A possible explanation of the decrease in recall at higher mass is that more massive mergers are able to retain Hubble type morphologies for a longer time during the pre-merger phase, before tidal forces distort them and more massive post-merger systems then form more rapidly into Hubble type morphologies. In Section \ref{subsec:Classification of subclasses}, it was shown that morphological classifiers only detect mergers during a short period when their morphologies are significantly disturbed; therefore, for higher mass mergers, this period might be shortened resulting in lower recall. 

\section{Conclusions}\label{sec:Conclusions}
The morphological statistics of galaxies can be used to produce high precision merger classifiers that are highly robust to new data. In this work, the \texttt{statmorph} package \citep{Rodriguez-Gomez2019} was used to generate morphological statistics for mock images from the IllustrisTNG and Horizon-AGN simulations prepared by \citetalias{Margalef-Bentabol2024} along with a set of observations from HSC-SSP selected by \citetalias{Margalef-Bentabol2024} from a visually classified dataset from \citet{Goulding2018}. 

Two morphological classifiers for merging galaxies were produced using $G$ and the $M_{20}$ statistic, and $G$ and C. An MCMC was used to find optimised merger criteria using the reciprocal accuracy as a likelihood function. The training and testing data from the merger challenge in \citetalias{Margalef-Bentabol2024} was used for the MCMC routine, which eliminated the biases from the original merger criteria from \citet{Lotz2008} and \citet{Lotz2004}. The resulting criteria are:
\begin{equation}
    \begin{split}
        &G>(-0.267\pm0.081)M_{20}+(0.143\pm0.012);\\
        &G>(0.162\pm0.048)C+(0.149\pm0.12),
    \end{split}
    \label{eq:optimised cuts}
\end{equation}
where galaxies above these lines are classified as mergers.

The main conclusions from this work are as follows:
\begin{itemize}
    \item Morphological classifiers can achieve a similar precision to ML techniques. When used to classify the testing data, the $G$-$M_{20}$ classifier achieved 69.5\% precision and the $G$-$C$ classifier achieved 72.3\%. When compared to the classifiers in \citetalias{Margalef-Bentabol2024}, the $G$-$M_{20}$ classifier is more precise than one of the six models and the $G$-$C$ classifier is more precise than three of the ML models.
    \item The recall of morphological classifiers is much lower than ML methods and they are only sensitive to pre-mergers when used to classify the testing data. Using $G$-$M_{20}$ classification on the testing data, a recall of 36.4\% is achieved. With $G$-$C$ on the testing dataset, the recall is 23.6\%. Post-mergers and ongoing mergers are indistinguishable from non-mergers in terms of their $G$, $C$, and $M_{20}$ statistics contributing to the poor recall.
    \item When applied to the Horizon-AGN dataset and HSC-SSP observations, the accuracy of the morphological classifiers is consistent with the ML methods in \citetalias{Margalef-Bentabol2024}. On the Horizon-AGN dataset, the $G$-$M_{20}$ classifier is more accurate than four of the six ML models and the $G$-$C$ classifier is more accurate than one of the ML models in \citetalias{Margalef-Bentabol2024}. The accuracy of the $G$-$M_{20}$ classifier when applied to the HSC-SSP observations is higher than three of the six ML methods from \citetalias{Margalef-Bentabol2024} and the $G$-$C$ classifier reaches an accuracy 6.5\% lower than the least accurate ML method.
    \item Pre-mergers can be extracted from samples of mergers to a precision of 77.3\% using the $G$-$C$ classifier, when using $G$-$M_{20}$ the pre-mergers are extracted with a precision of 71.8\%.
    \item The morphological classifiers reach their maximum precision at low redshift and high masses. Recall is approximately constant with redshift and maximised at low mass. 
    \item The robustness of morphological classifiers is comparable to that of ML classifiers and less sensitive to redshift. The $G$-$M_{20}$ classifier had a robustness of 91.4\% when transferred to the Horizon-AGN images, the $G$-$C$ classifier has a 95.0\% robustness. By comparison, the ML techniques in \citetalias{Margalef-Bentabol2024} have robustness values ranging from 89.0\% to 99.4\%.
\end{itemize}

In future works, the accuracy of morphological merger classifiers could be improved upon by developing new morphological statistics that are sensitive to post-mergers. The precision might also be improved in the future by analysing a higher dimensional morphology space with unsupervised learning methods. 

\begin{acknowledgements}
    The authors would like to thank Krzysiek Lisiecki for their software contribution. A.C., W.J.P., and S.D. have been supported by the Polish National Science Center project UMO-2023/51/D/ST9/00147. This work made use of Astropy:\footnote{\href{http://www.astropy.org}{www.astropy.org}} a community-developed core Python package and an ecosystem of tools and resources for astronomy \citep{astropy:2013, astropy:2018, astropy:2022}. This work made use of Scipy\footnote{\href{https://www.scipy.org}{www.scipy.org}} \citep{2020SciPy-NMeth}.
\end{acknowledgements}

\bibliographystyle{bibtex/aa} 
\bibliography{references} 

\begin{appendix}

\FloatBarrier
\section{Histograms of the sample morphologies}\label{app:Histograms of the sample morphologies}
\begin{figure}
    \centering
    \resizebox{\hsize}{!}{\includegraphics{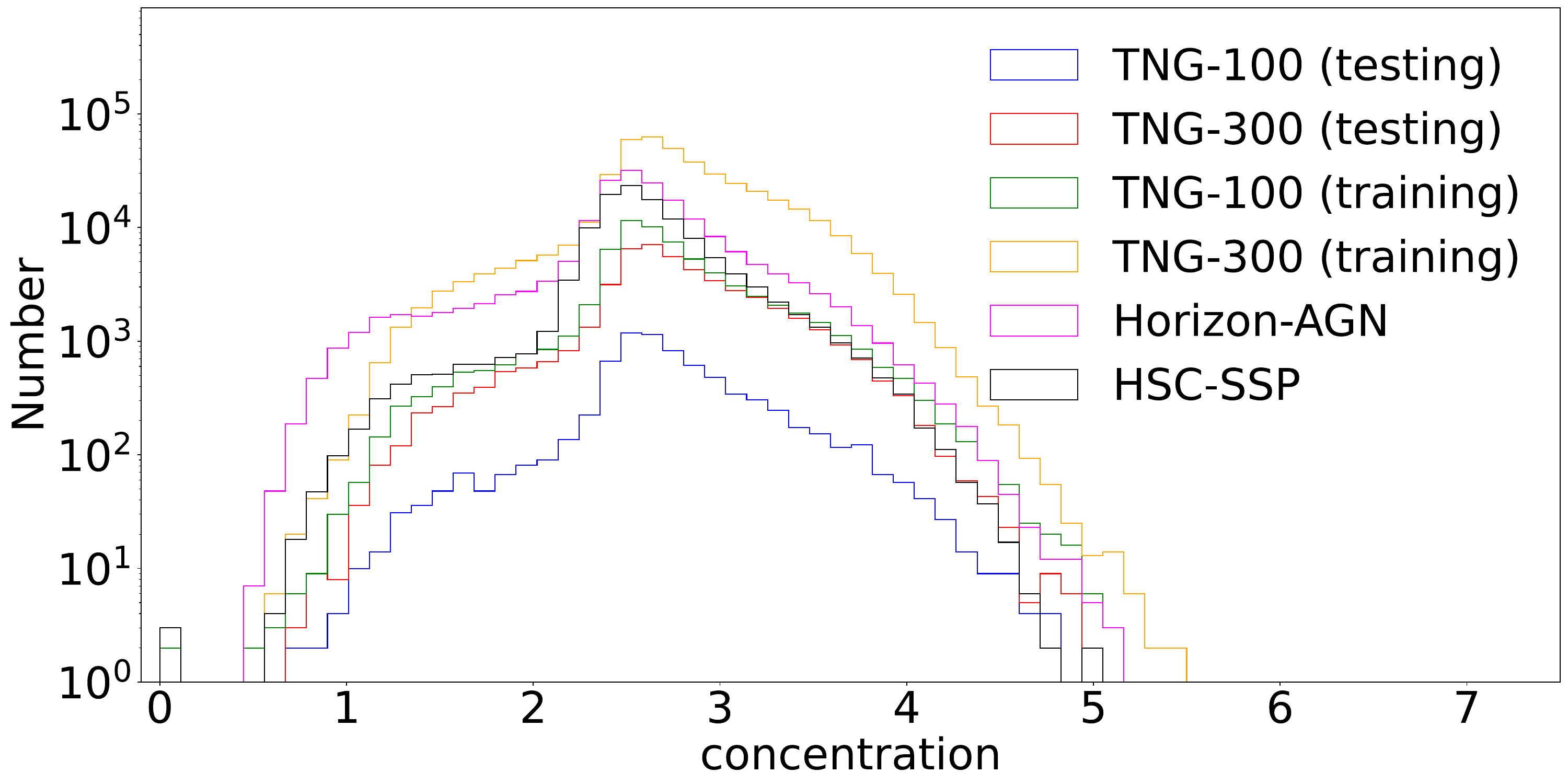}}
    \resizebox{\hsize}{!}{\includegraphics{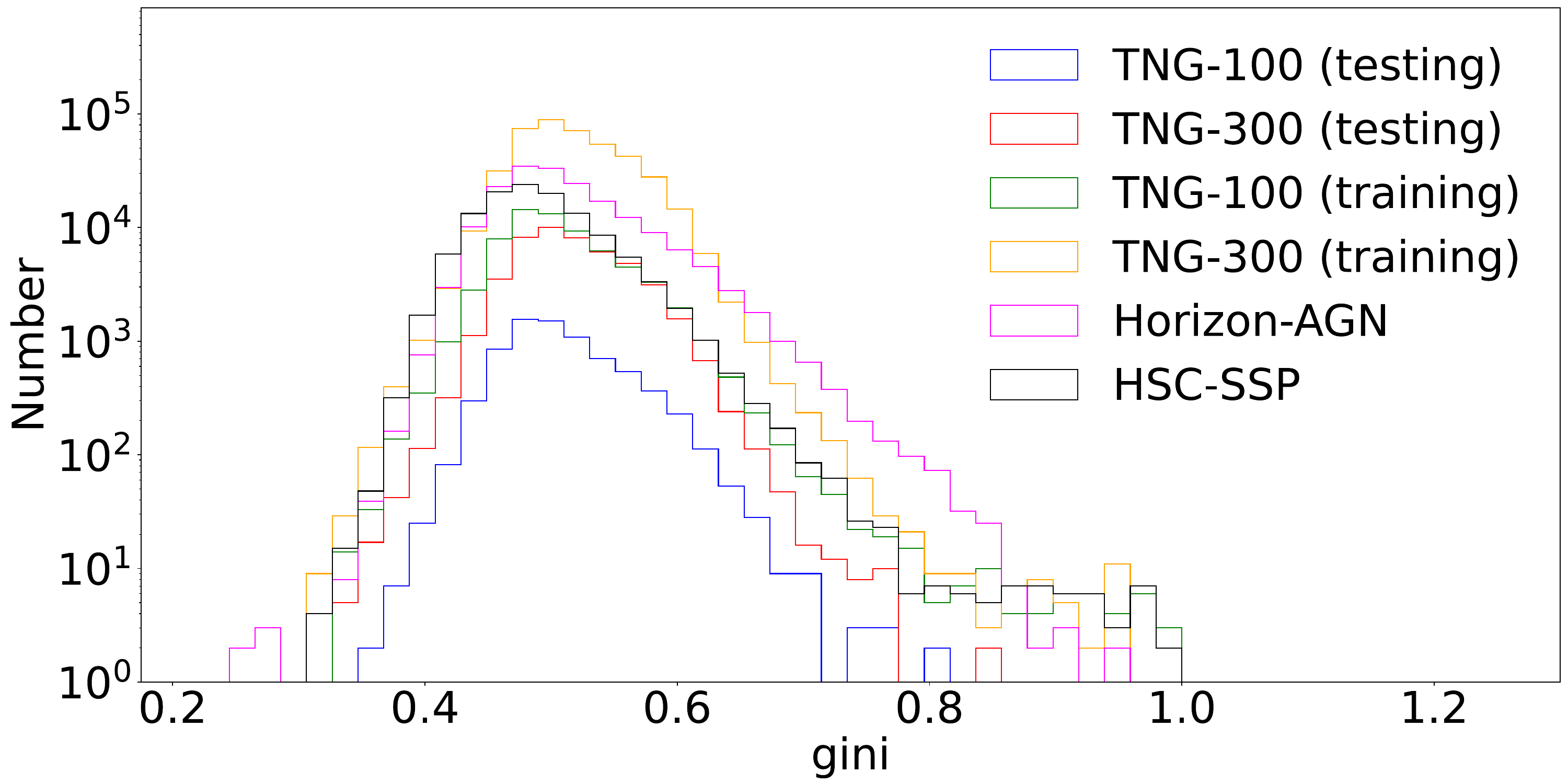}}
    \resizebox{\hsize}{!}{\includegraphics{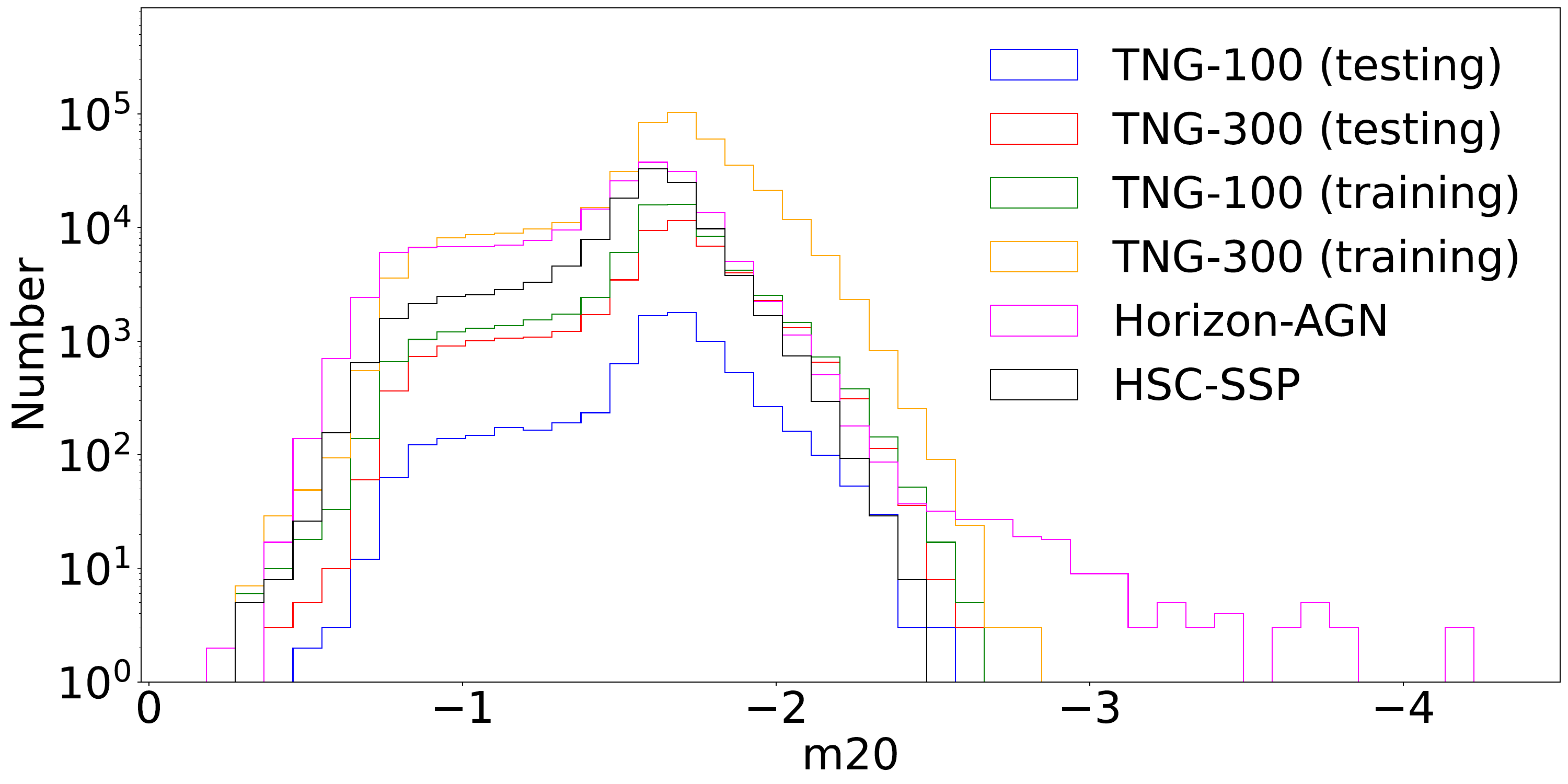}}
    \caption{Non-parametric morphological statistics used in this work. Top panel shows the distribution of $C$, the middle panel shows the $G$ distribution and the lower panel shows $M_{20}$. The IllustrisTNG mock images are shown in four histograms: TNG-100 testing sample in blue, TNG-300 testing sample in red, TNG-100 training sample in green, and TNG-300 training sample in yellow. The Horizon-AGN mock images are shown in the magenta histogram and the HSC-SSP images are shown in black.}
    \label{fig:morphology histograms}
\end{figure}

The $G$, $M_{20}$, and $C$ statistics used in this work are summarised in Figure \ref{fig:morphology histograms}. The IllustrisTNG testing and training datasets from \citetalias{Margalef-Bentabol2024} are divided into TNG-100 aand TNG-300. The distributions are consistent with those shown in \citet{Pearson2022} and \citet{Guzman-Ortega2023}.

For both TNG-100 and TNG-300, the testing and training datasets show nearly identical distributions scaled by an order of magnitude indicating that the testing data is representative. The morphology of TNG-100 galaxies have a slightly lower $C$ and $G$ and slightly higher $M_{20}$ compared to TNG-300, this is expected as TNG-100 has a superior mass resolution leading to more featured galaxies. The TNG-100 morphological statistics are most similar to HSC-SSP.

The morphological statistics of the Horizon-AGN galaxies closely follow the HSC-SSP statistics in the most populated regions of the morphology space. A small number ($\sim100$) of the mock images in the Horizon-AGN dataset show an exceptionally low value of $M_{20}$ meaning that these galaxies have highly extended bright features not seen in the IllustrisTNG mock images or the HSC-SSP observations.

\FloatBarrier
\section{MCMC initial condition and prior selection}\label{app:MCMC initial condition and prior selection}
\begin{figure}
    \centering
    \resizebox{\hsize}{!}{\includegraphics{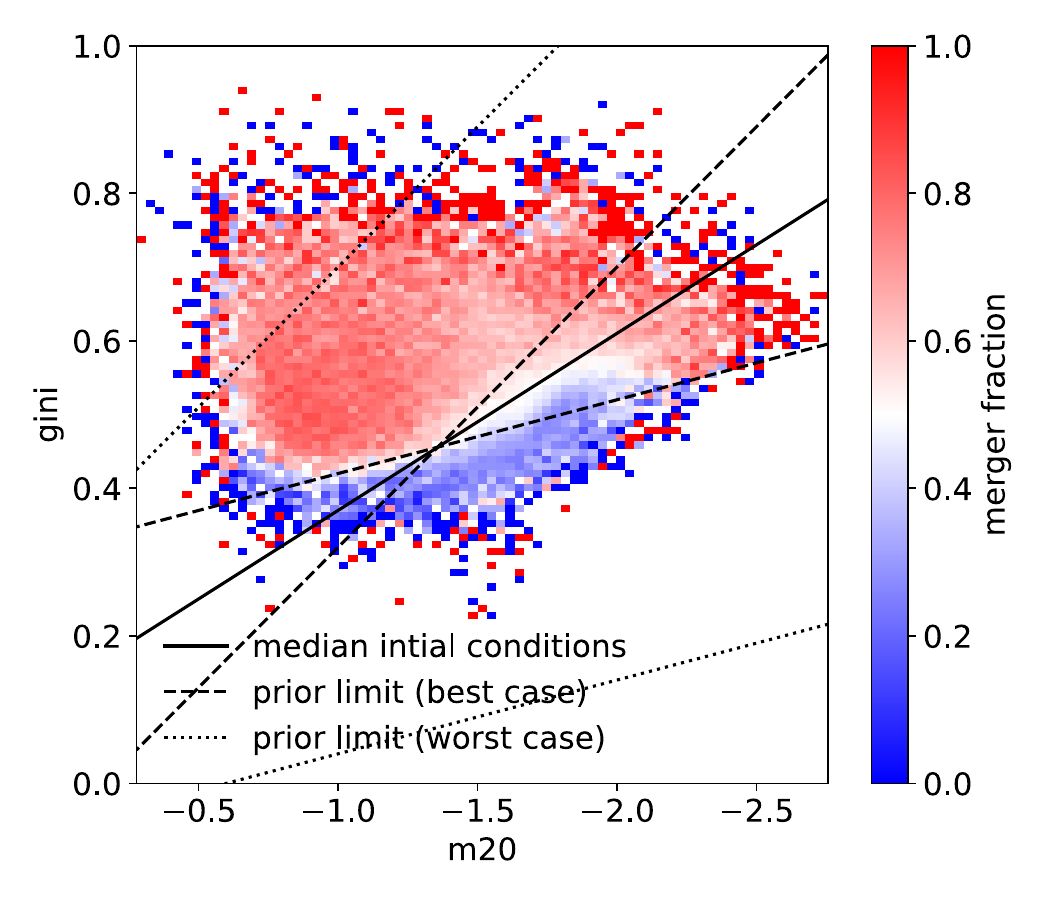}}
    \resizebox{\hsize}{!}{\includegraphics{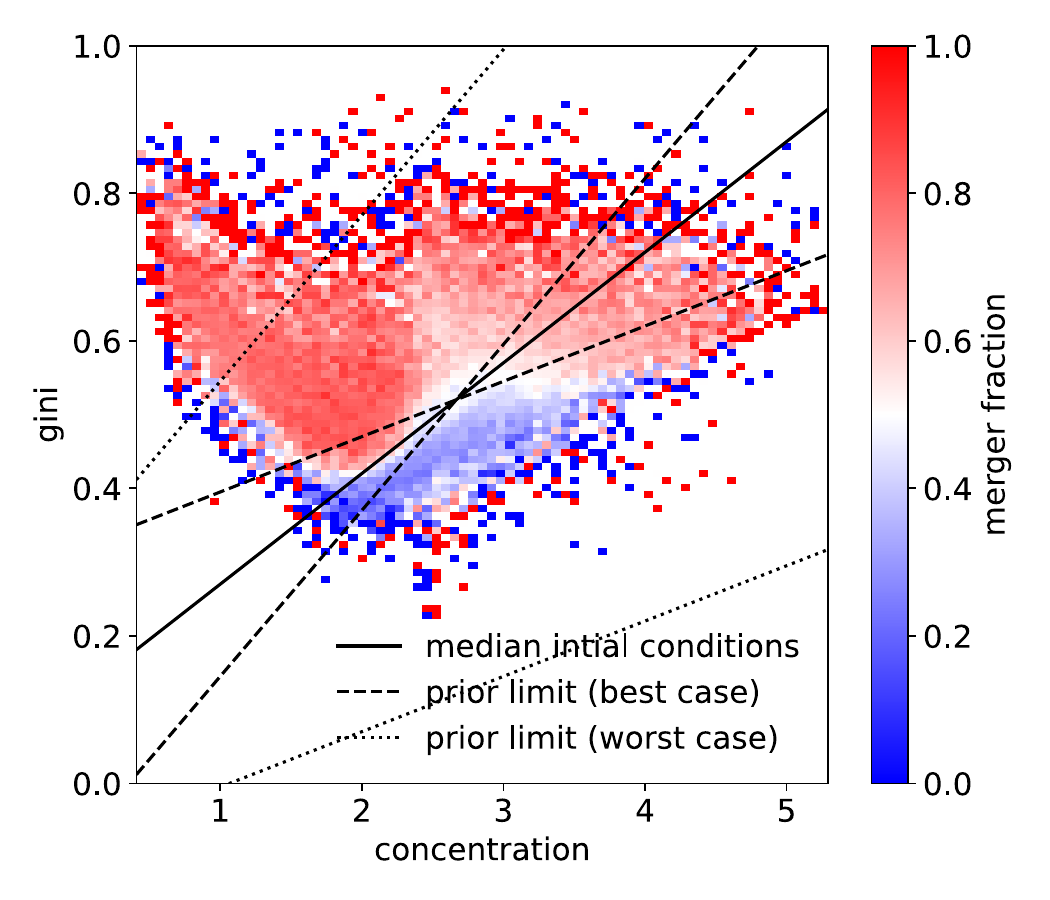}}
    \caption{Histograms based on the number of mergers as a fraction of total galaxies per cell in the $G-M_{20}$ (top) and $G-C$ (bottom) planes. Colouring shows the merger fraction with blue cells being dominated by non-mergers, red cells dominated by mergers, and white cells having approximately equal numbers of mergers and non-mergers. The solid black lines are fitted to the white regions and are used as the median initial conditions for walkers in the MCMC routines in Section \ref{subsec:Classifieroptimisation}. The dashed black lines are the worst acceptable fit to the white region, these are mirrored in the solid black liens to give upper and lower limits. The dotted black liens are formed by combining the parameters of the dashed black lines to make the largest region possible around the solid black lines, these are used as the priors for the MCMC routines in Section \ref{subsec:Classifieroptimisation}.}
    \label{fig:prior_picking}
\end{figure}

The initial conditions and priors used in Section \ref{subsec:Classifieroptimisation} were selected by analysing the difference in distributions of mergers and non-mergers in the $G-M_{20}$ and $G-C$ planes. By making histograms of the mergers and non-mergers in the same bins on the morphological planes, the ratio of mergers to total galaxies (hereafter referred to as the merger fraction) can be calculated for each cell. This merger fraction per cell is shown in the colouring of Figure \ref{fig:prior_picking}, where the red cells contain more mergers than non-mergers and blue cells contain more non-mergers than mergers.

In both panels of Figure \ref{fig:prior_picking}, a certain region can be seen at $0.4 < G < 0.6$, where the merger fraction is $\sim0.5$, as indicated by  white-coloured cells. In these cells, the mergers and non-mergers contain approximately equal populations. This region (hereafter referred to as the turnover region) bisects the morphological planes. With the exception of outlying points (where cells contain few galaxies so the merger fraction is close to 0 or 1), all cells below the turnover region have more non-mergers than mergers and all cells above the turnover region have more mergers than non-mergers.

The optimal cut should be expected to pass through the turnover region and, thus, we used it to inform the initial conditions for the MCMC routine described in Section \ref{subsec:Classifieroptimisation}. The turnover region in both $G-M_{20}$ and $G-C$ is well approximated by a linear fit which is shown as the solid black line on Figure \ref{fig:prior_picking}. By taking these visually fitted cuts as the initial conditions for the MCMC routine, we ensure that the walkers begin near the minima in the likelihood function thereby constraining the optimised criterion.

The dashed black lines in Figure \ref{fig:prior_picking} represent the priors selected for the MCMC routine. The shallower dashed line is the worst acceptable linear fit to the turnover region and, again, this is visually fitted. The steeper dashed line is then made by mirroring the worst acceptable fit about the solid black line. The dotted black lines show the worst possible combination of the parameters of the dashed lines so that the top dotted line is the steepest allowed gradient for a walker combined with the largest allowed intercept value and the lower dotted line is the shallowest allowed gradient and the lowest allowed intercept.

The parameters of all the visually fitted lines in this appendix are summarised in Table \ref{tab:mcmc setup}. Because an MCMC routine is applied to these cuts, high accuracy in these visually fitted cuts is not necessary.

\section{Distributions of additional parameters in morphological space}\label{app:Distributions of additional parameters in morphological space}
Further to Section \ref{subsec:Merger criteria quality}, the distribution of several additional parameters in the $G-M_{20}$ and $G-C$ planes. We find that $A$ and redshift are traced by the merger criteria presented in this work but do not link as strongly to the criteria as the parameters presented in Section \ref{subsec:Merger criteria quality}. Time after the most recent merger and mass ratio were found to remain constant over the morphological space and so were not traced by the merger criteria.

\begin{figure}
    \centering
    \resizebox{\hsize}{!}{\includegraphics{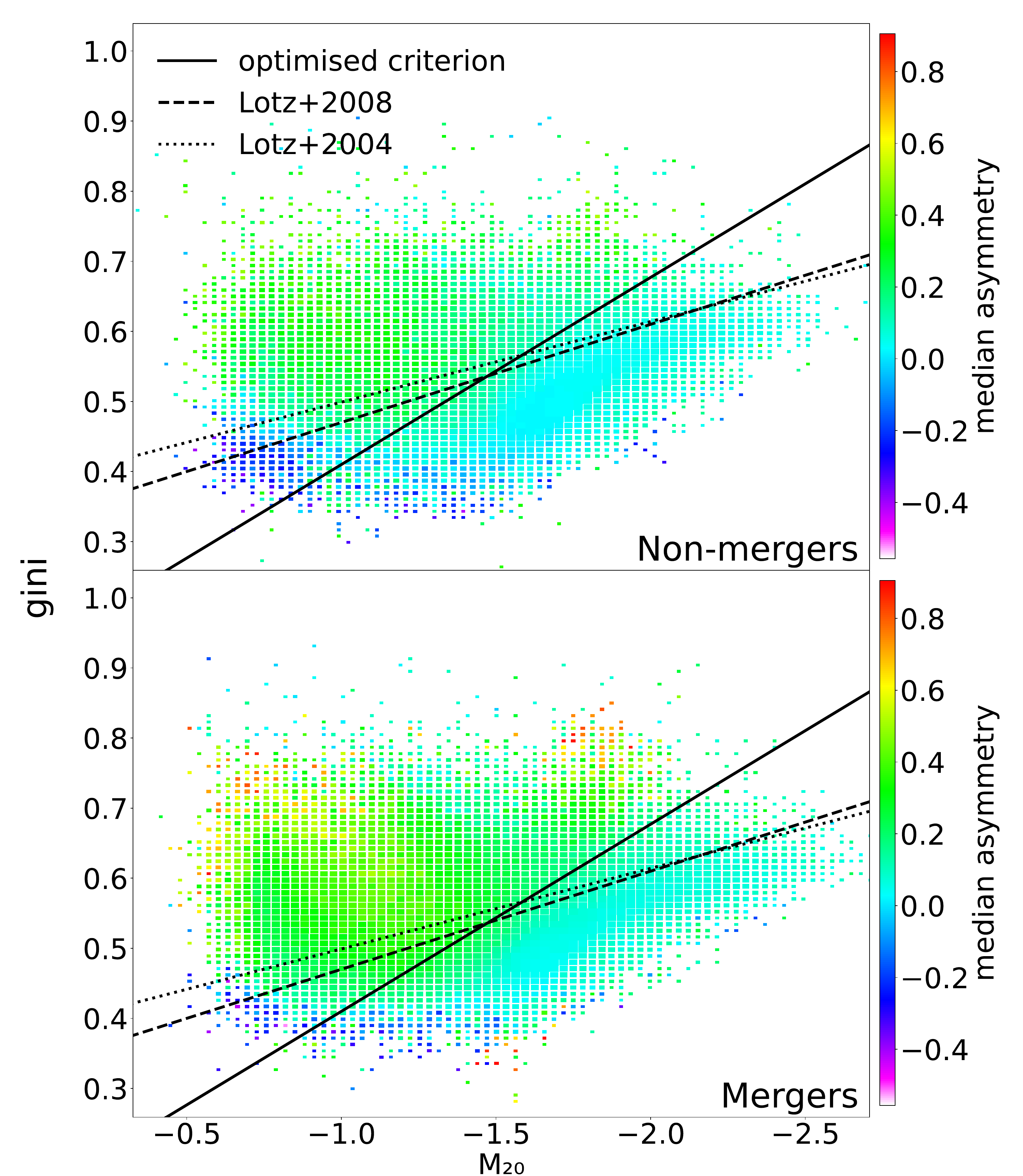}}
    \caption{2D histogram of the $G-M_{20}$ distribution of the training sample from \citetalias{Margalef-Bentabol2024}. Top panel includes non-merging galaxies and the bottom panel includes merging galaxies. The uniformly sized bins are displayed by rectangles scaled logarithmically by their population, the colouring shows the median $A$ in each bin from low (in purple) to high (in red). The optimised criterion is included as the solid black line and the literature cuts from \citet{Lotz2004} and \citet{Lotz2008} are shown in dashed black lines.}
    \label{fig:gm20 asymmetry}
\end{figure}

\begin{figure}
    \centering
    \resizebox{\hsize}{!}{\includegraphics{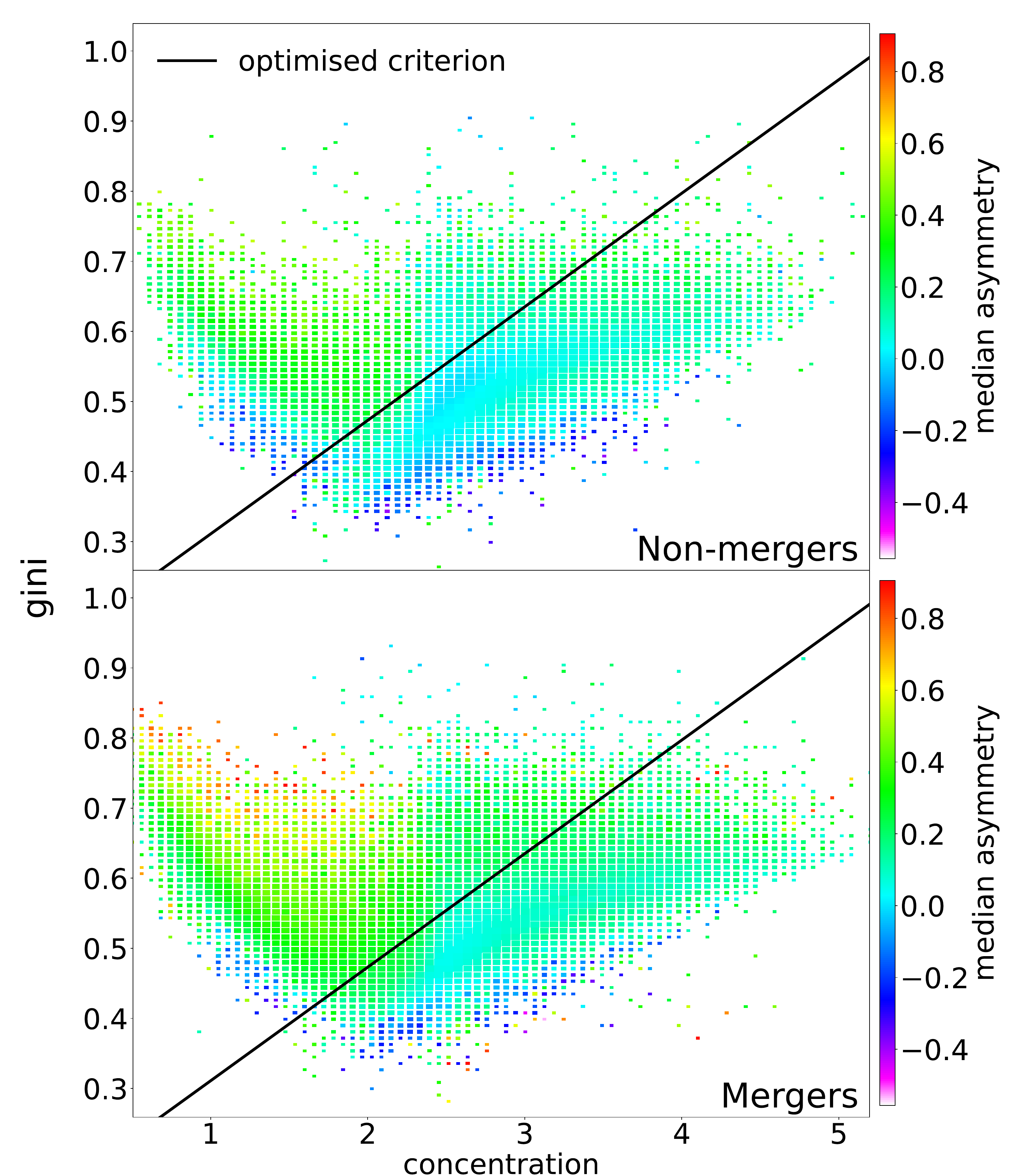}}
    \caption{2D histogram of the $G-C$ distribution of the training sample from \citetalias{Margalef-Bentabol2024} with colouring to represent median binned asymmetry. Formatting is the same as in Figure \ref{fig:gm20 asymmetry}.}
    \label{fig:gc asymmetry}
\end{figure}

The distribution of $A$ for the training sample form IllustrisTNG is shown in the $G-M_{20}$ plane in Figure \ref{fig:gm20 asymmetry} and in the $G-C$ plane in Figure \ref{fig:gc asymmetry}. In both cases, $A$ has a median value of $<$0.2 below the merger criterion and the median value of $A$ increases further from the criterion for classified mergers. This indicates that a classification based on $A$ is implicitly included in the $G-M_{20}$ classification, when the \citet{Conselice2003} merger criterion ($A>0.35$) is applied we find that using $A$ achieves an accuracy of 55.5\%, a precision of 75.6\% and recall of 16,3\%.

\begin{figure}
    \centering
    \resizebox{\hsize}{!}{\includegraphics{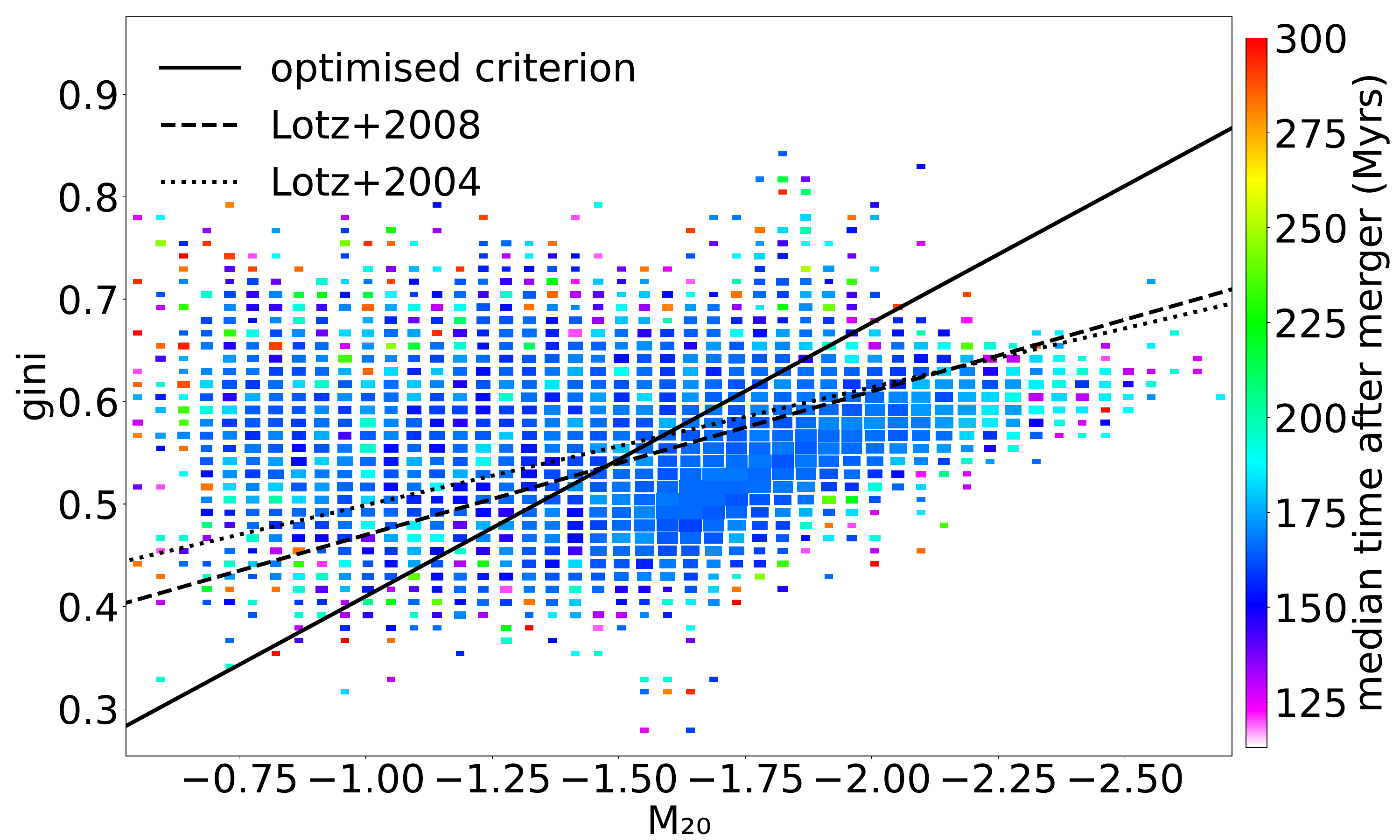}}
    \caption{2D histogram of the $G-M_{20}$ distribution of the training sample from \citetalias{Margalef-Bentabol2024} with colouring to represent median binned time after the most recent merger. Formatting is the same as in Figure \ref{fig:gm20 asymmetry}.}
    \label{fig:gm20 time after merger}
\end{figure}

\begin{figure}
    \centering
    \resizebox{\hsize}{!}{\includegraphics{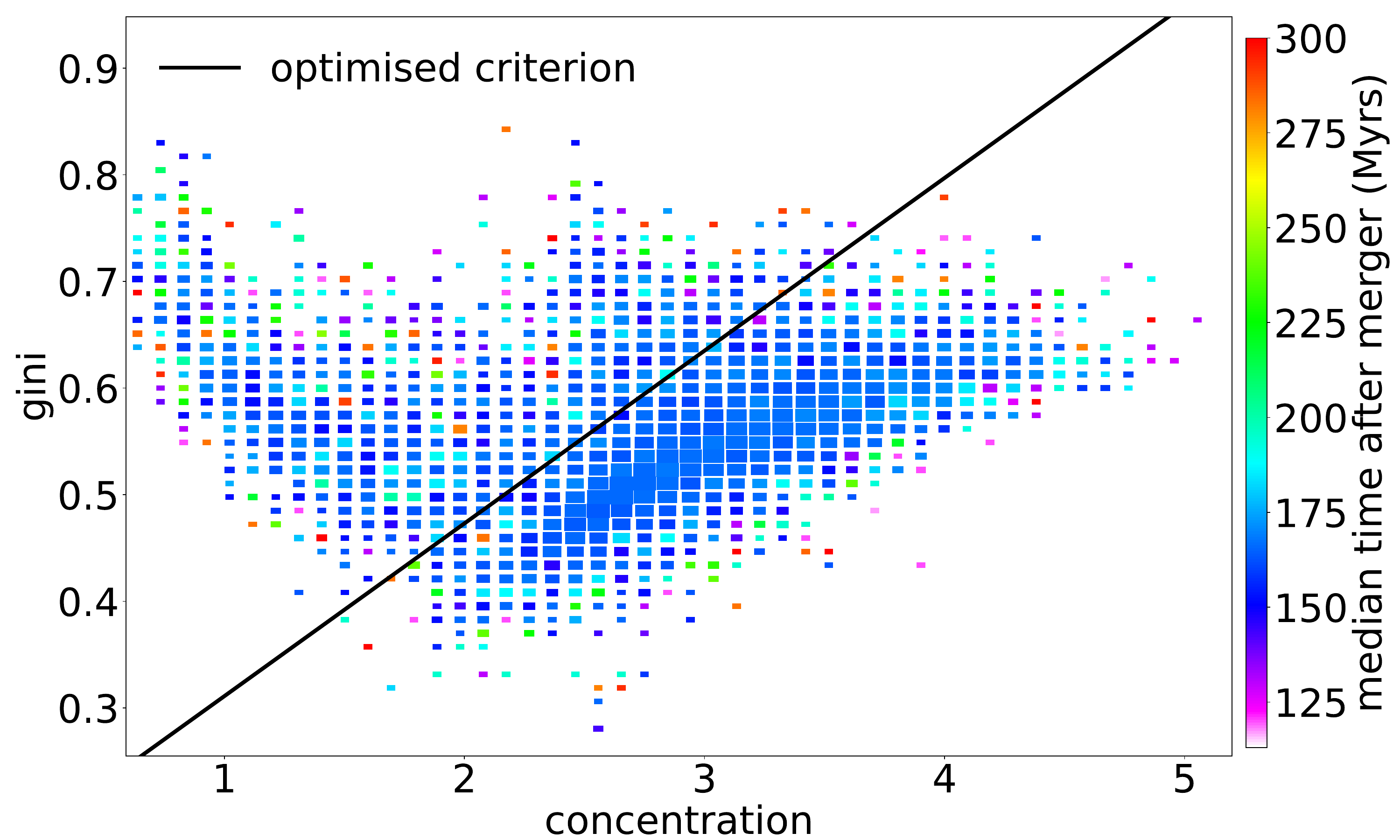}}
    \caption{2D histogram of the $G-C$ distribution of the training sample from \citetalias{Margalef-Bentabol2024} with colouring to represent median binned time after the most recent merger. Formatting is the same as in Figure \ref{fig:gm20 asymmetry}.}
    \label{fig:gc time after merger}
\end{figure}

The time after the last merger for galaxies up to 300Myrs after coalescence in the training dataset are shown in Figures \ref{fig:gm20 time after merger} and \ref{fig:gc time after merger} in the $G-M_{20}$ and $G-C$ planes, respectively. The median time after the merger is $\sim150$Myrs for the majority of bins across the morphological space. As in \ref{subsec:Classification of subclasses}, post-mergers do not show any morphological distinction in $G$, $C$ or $M_{20}$.

\begin{figure}
    \centering
    \resizebox{\hsize}{!}{\includegraphics{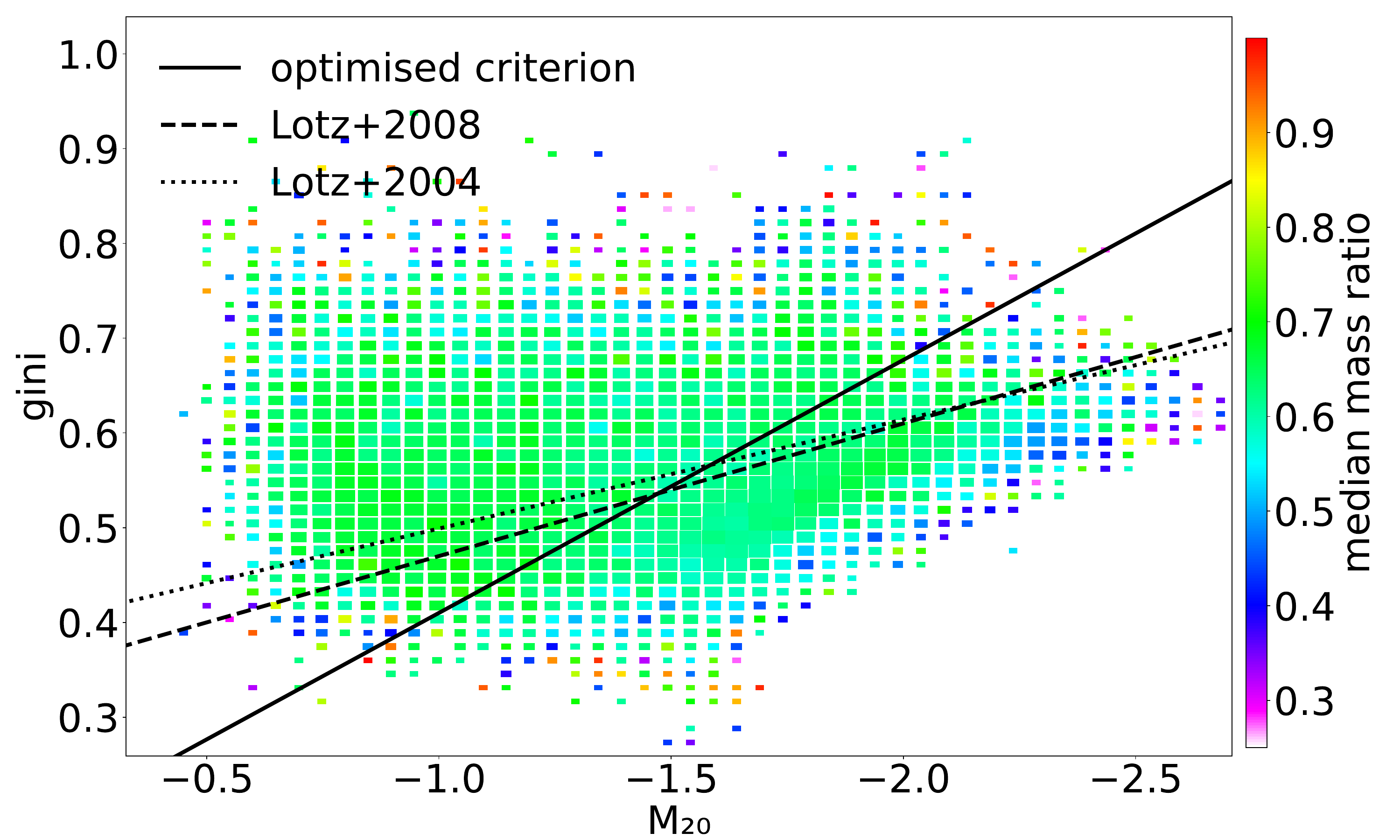}}
    \caption{2D histogram of the $G-M_{20}$ distribution of the training sample from \citetalias{Margalef-Bentabol2024} with colouring to represent median binned mass ratio. Formatting is the same as in Figure \ref{fig:gm20 asymmetry}.}
    \label{fig:gm20 mass ratio}
\end{figure}

\begin{figure}
    \centering
    \resizebox{\hsize}{!}{\includegraphics{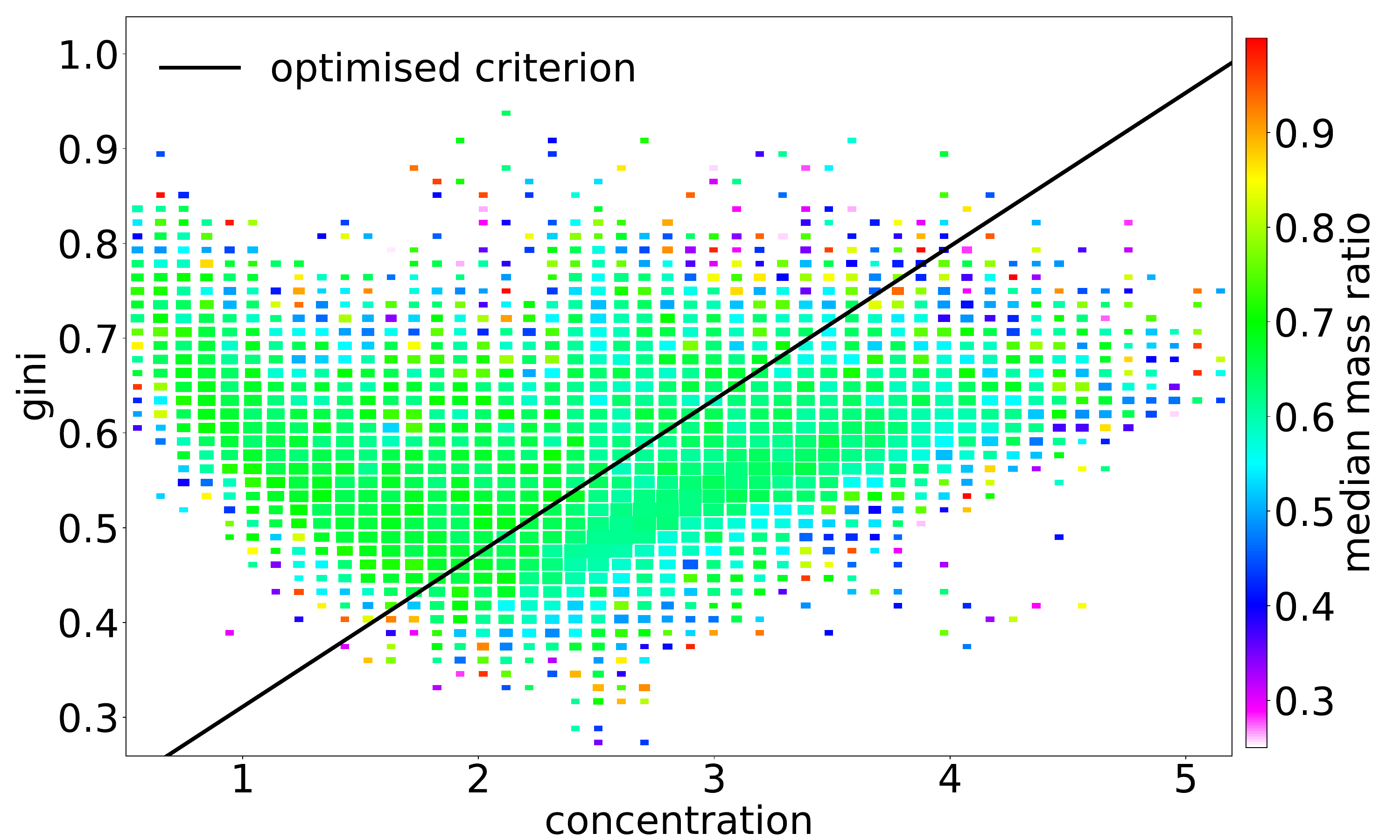}}
    \caption{2D histogram of the $G-C$ distribution of the training sample from \citetalias{Margalef-Bentabol2024} with colouring to represent median binned mass ratio. Formatting is the same as in Figure \ref{fig:gm20 asymmetry}.}
    \label{fig:gc mass ratio}
\end{figure}

In Figures \ref{fig:gm20 mass ratio} and \ref{fig:gc mass ratio}, the distribution of the median mass ratio in the $G-M_{20}$ and $G-C$ planes are shown. The mass ratio has a median value of $\sim0.65$ across the morphological space indicating that $G$, $C$, and $M_{20}$ are robust to mass ratio for major mergers.

\begin{figure}
    \centering
    \resizebox{\hsize}{!}{\includegraphics{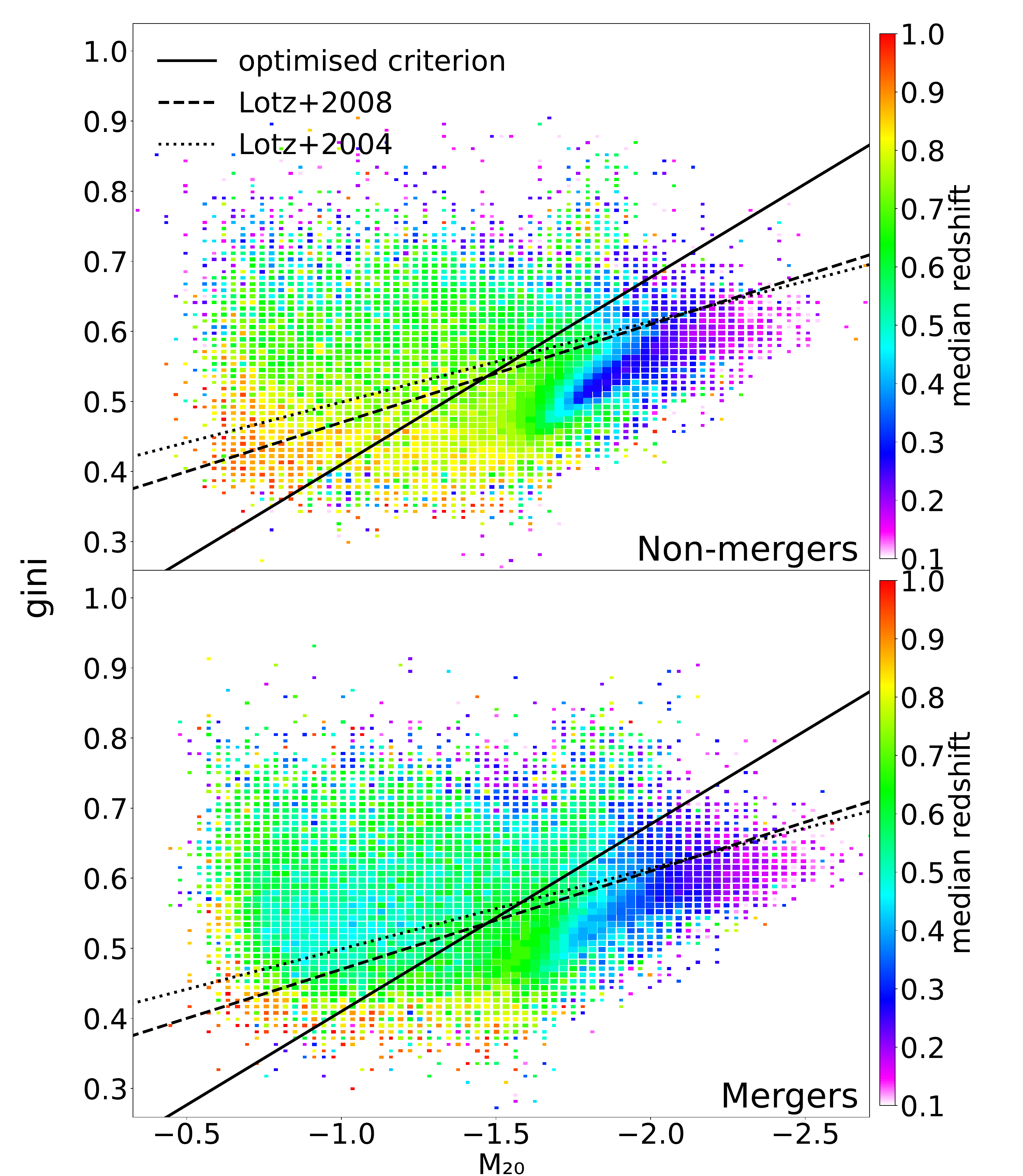}}
    \caption{2D histogram of the $G-M_{20}$ distribution of the training sample from \citetalias{Margalef-Bentabol2024} with colouring to represent median binned redshift. Formatting is the same as in Figure \ref{fig:gm20 asymmetry}.}
    \label{fig:gm20 redshift}
\end{figure}

\begin{figure}
    \centering
    \resizebox{\hsize}{!}{\includegraphics{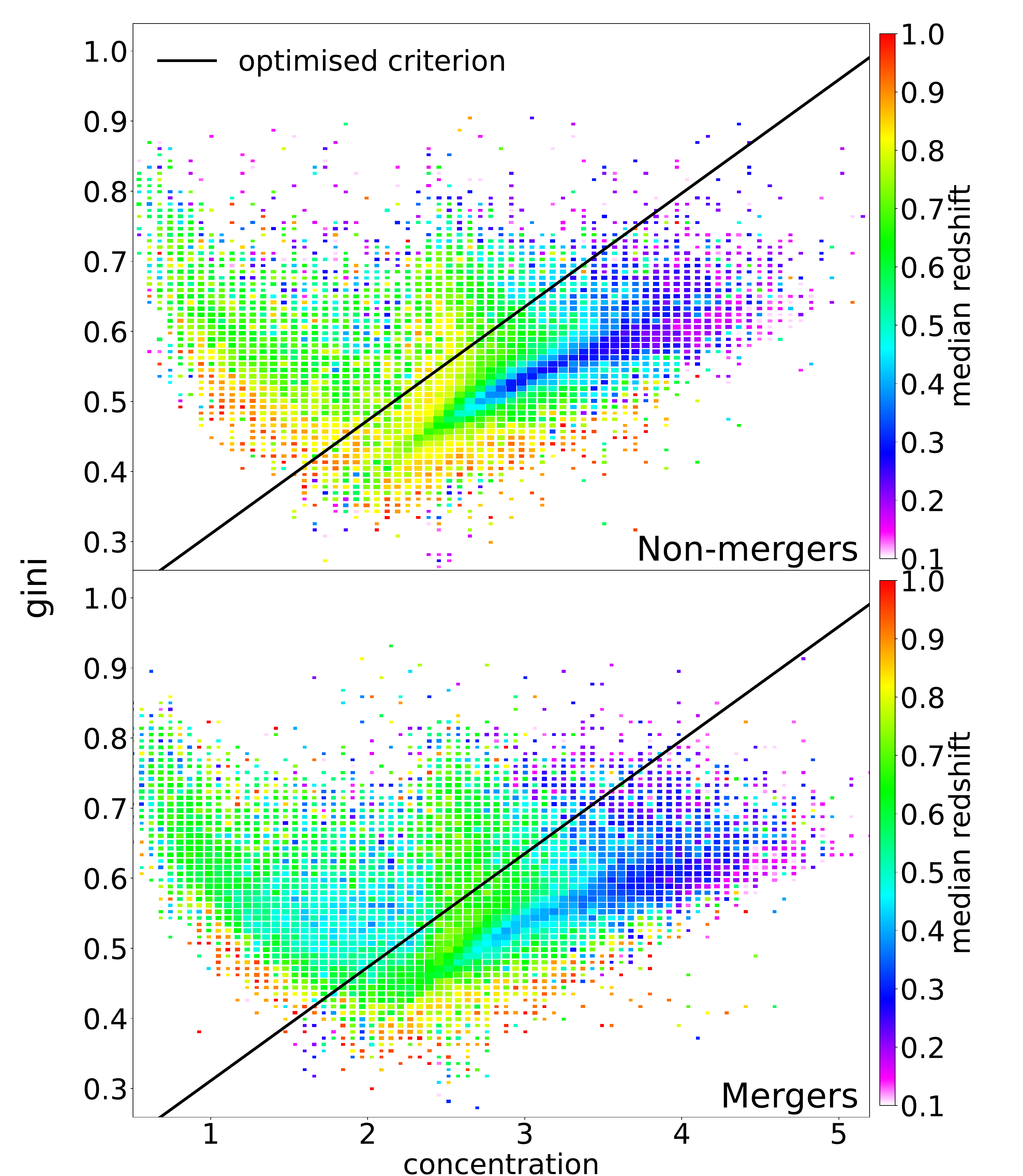}}
    \caption{2D histogram of the $G-C$ distribution of the training sample from \citetalias{Margalef-Bentabol2024} with colouring to represent median binned redshift. Formatting is the same as in Figure \ref{fig:gm20 asymmetry}.}
    \label{fig:gc redshift}
\end{figure}

The distribution of redshift in the $G-M_{20}$ and $G-C$ planes are shown in Figures \ref{fig:gm20 redshift} and \ref{fig:gc redshift}, respectively. $C$ and $M_{20}$ scale with the redshift of the galaxy but $G$ does not. The weak dependence of precision and recall on redshift seen in section \ref{subsec:Dependence on redshift and stellar mass} can be explained by this relation between $C$ and $M_{20}$ and redshift.

\FloatBarrier
\clearpage

\end{appendix}
\end{document}